\documentclass[onecolumn,sort&compress,numbers]{els-mrw} 

\usepackage{amsmath,amssymb,amsfonts,amsthm,makeidx,graphicx}
\usepackage{txfonts}
\usepackage{helvet}
\usepackage[LGR,OT1]{fontenc}

\begin{document}


\chapter{Cosmic Ray Space Experiments}\label{chap9}

\author[1]{Martin Pohl}

\address[1]{\orgname{University of Geneva}, \orgdiv{DPNC}, \orgaddress{1211 Gen\`eve, Switzerland}}

\articletag{\bf Invited article for the Encyclopedia of Particle Physics}

\maketitle

\begin{abstract}[Abstract]
This article describes experiments in space which measure charged cosmic ray particles in the range from $10\,\mathrm{GV}$ to $10^6\,\mathrm{GV}$ of magnetic rigidity $p/(Ze)$. In this energy range, cosmic rays are expected to originate from sources in the Milky Way and be confined to our galaxy. Spectra of nuclei and their chemical composition are discussed. The spectrum of antiprotons and the search for heavier anti-nuclei are covered. All spectra and especially those of electrons and positrons are analysed for indications of unconventional particle sources, acceleration or transport mechanisms. 
\end{abstract}

\begin{keywords}
 	Cosmic rays\sep space experiments\sep PAMELA\sep AMS-02\sep CALET \sep DAMPE\sep HERD \sep ISS-CREAM
\end{keywords}

\begin{figure}[h]
	\centering
	\includegraphics[width=10cm]{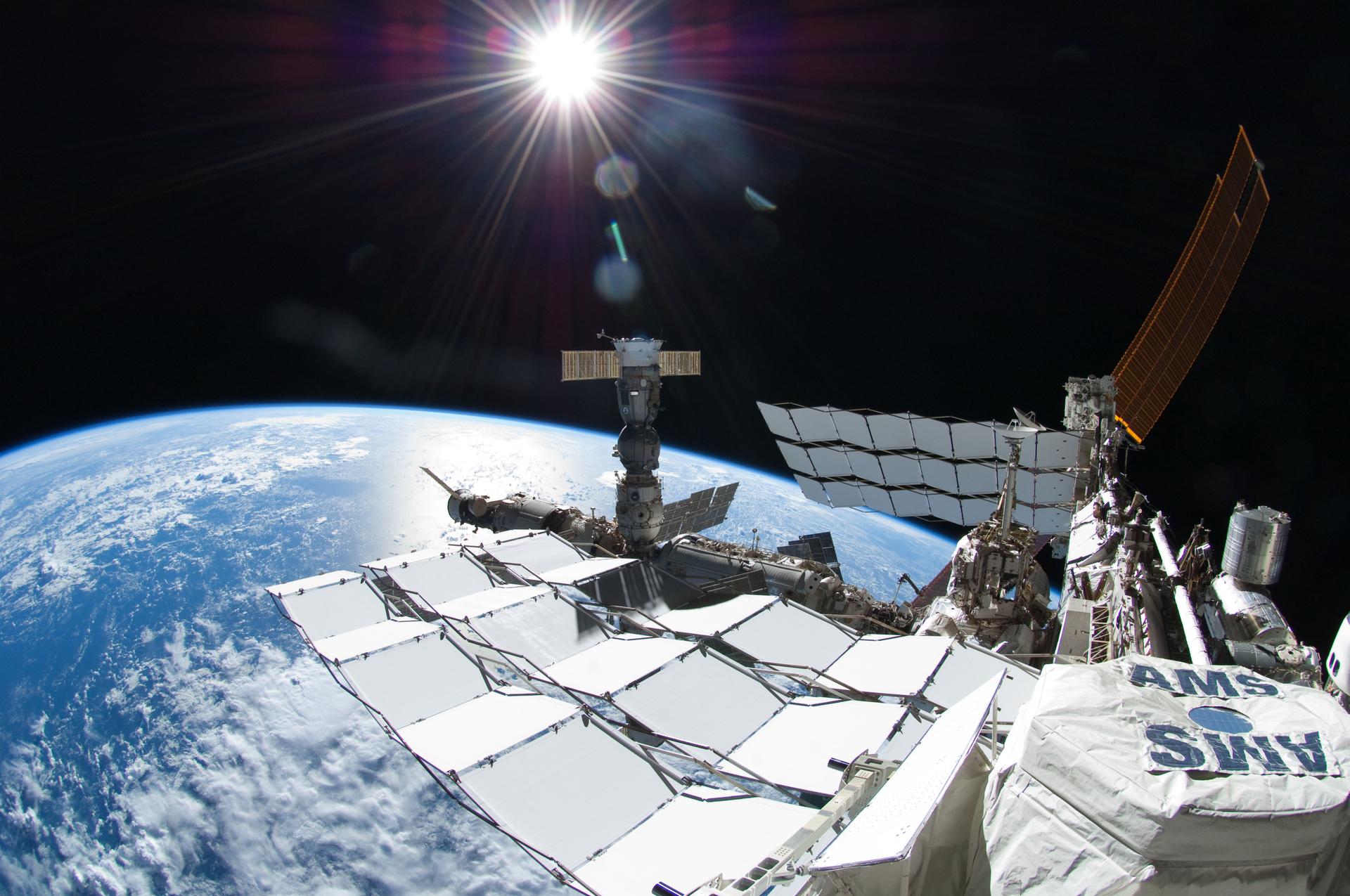}
	\caption{The AMS-02 cosmic ray detector on board the International Space Station}
	\label{fig:titlepage}
\end{figure}

\begin{glossary}[Nomenclature]
	\begin{tabular}{@{}lp{34pc}@{}}
		AMS-02			& Alpha Magnetic Spectrometer		\\
		CALET			& CALorimetric Electron Telescope		\\
		DAMPE			& DArk Matter Particle Explorer			\\
		ISS				& International Space Station			\\
		ISS-CREAM		& International Space Station -- Cosmic Ray Energetics and Mass		\\
		LEO				& Low Earth Orbit					\\
		MDR				& Maximum detectable rigidity			\\
		PAMELA 			& Payload for Antimatter Matter Exploration and Light-nuclei Astrophysics	\\
		HERD			& High Energy cosmic-Radiation Detector
	\end{tabular}
\end{glossary}

\section*{Objectives}
\begin{itemize}
	\item Understanding the main characteristics of astroparticle detectors in space, the function of their components, how they count cosmic rays and identify species; 
	\item Properties of cosmic ray particles in the range from $10\,\mathrm{GV}$ to $10^6\,\mathrm{GV}$ of magnetic rigidity $p/(Ze)$, accessible to current space experiments;
	\item Rigidity spectra of nuclei, light anti-nuclei, electrons and positrons, spectral features in this energy range; 
	\item Chemical composition of galactic cosmic rays;
	\item Astrophysical sources, acceleration and transport of cosmic rays throughout their life cycle; 
	\item Search for unconventional components of galactic cosmic rays, as well as signs for unusual sources, acceleration and transport phenomena.
\end{itemize}

\def\dtod#1#2{\frac{d #1}{d #2}}%
\def\ptop#1#2{\frac{\partial #1}{\partial #2}}%

\section{Introduction}\label{history}

The aim of cosmic ray physics is to understand the complete life cycle of cosmic rays, i.e.~charged particles from outer space. The goal is to describe it from their generation in cosmological and astrophysical events, through their acceleration to energies unreachable by man-made accelerators, to their transport all the way towards Earth. They impinge on the top of the atmosphere at a rate of about ten thousand per $\mathrm{m}^2$ and second. At sea level there is still about one particle per $\mathrm{m}^2$ and minute left. Cosmic rays are thus an important part of our environment and influence human activities and technology.  

In recent time, a basic framework for discussing cosmic rays using known physics has emerged. Its basic ingredients can be summarised as follows:
\begin{itemize}
\item The lightest nuclear components, protons, helium and partially lithium, are to a large extent of cosmological origin. They have been produced during primordial nucleosynthesis seconds after the Big Bang itself. 
\item The rest of the stable elements which make up ordinary matter in the Universe are synthesised in stars through stellar nucleosynthesis, starting from this raw material. The elements up to iron are produced in fusion reactions. Elements beyond iron originate from the intense neutron bombardment of lighter species during explosive events.
\item The chemical composition of cosmic rays thus roughly follows that of stellar matter. Exceptions are the light elements Li, Be, B and F as well as the sub-iron group from Sc to Mn. These are intermediate products consumed in stellar nucleosynthesis and thus rare in stellar material. In cosmic rays, they are more abundant due to spallation reactions of heavier progenitors. 
\item Acceleration to high energies occurs via stochastic mechanisms in the relativistic plasma generated by explosive events like supernovae. The power law shape typical for the spectra of all cosmic ray species is a direct consequence of these multiple acceleration steps. 
\item High energy cosmic rays then travel diffusively through the magnetic fields and particles which fill interstellar space. On the way, their direction and energy is altered, interactions with the interstellar medium also change the particle composition. 
\item Light stable particles like electrons and positrons have a more eventful life cycle. They are easier to produce electromagnetically and subject to more frequent interactions on their way between their source and us.  
\end{itemize}
Cosmic ray physics thus requires input from cosmology and nuclear astrophysics, plasma physics and particle physics. In return, it provides data to all these fields.   

Cosmic ray experiments identify and count single cosmic ray particles as a function of particle type, magnetic rigidity or energy,  direction and time of arrival. These observables are used to measure the differential cosmic ray flux, for single species, groups of particles or all particles. The differential flux $d\Phi/dE$ is defined as the number of particles per unit surface, solid angle, time and energy. It is also often generically called a spectrum, its units are usually chosen as $[d\Phi/dE] = [1/(\mbox{m}^2\mbox{sr}\,\mbox{s}\,\mbox{GeV})]$. 

The choice of the energy variable depends both on the detector properties and the physics problem at hand. The output of magnetic spectrometers is the magnetic rigidity $R = p/(Ze)$  ([R] = [\mbox{GV}]) with momentum $p$ and particle charge $Ze$. Particles of equal rigidity follow the same path in a magnetic field. Rigidity is thus the variable of choice when gathering information about acceleration or transport of cosmic rays. Calorimetric detectors, on the other hand, measure the kinetic energy $E_{kin}$.  When particles are correctly identified and their mass is measured, momentum and energy measurements are of course equivalent. However, identification of nuclear isotopes requires high resolution measurements of both mass and charge, which is often not available. For nuclei, conversion between the two variables then requires assumptions about their isotopic composition.

\begin{figure}[h]
\begin{center}
\includegraphics[width=0.5\textwidth]{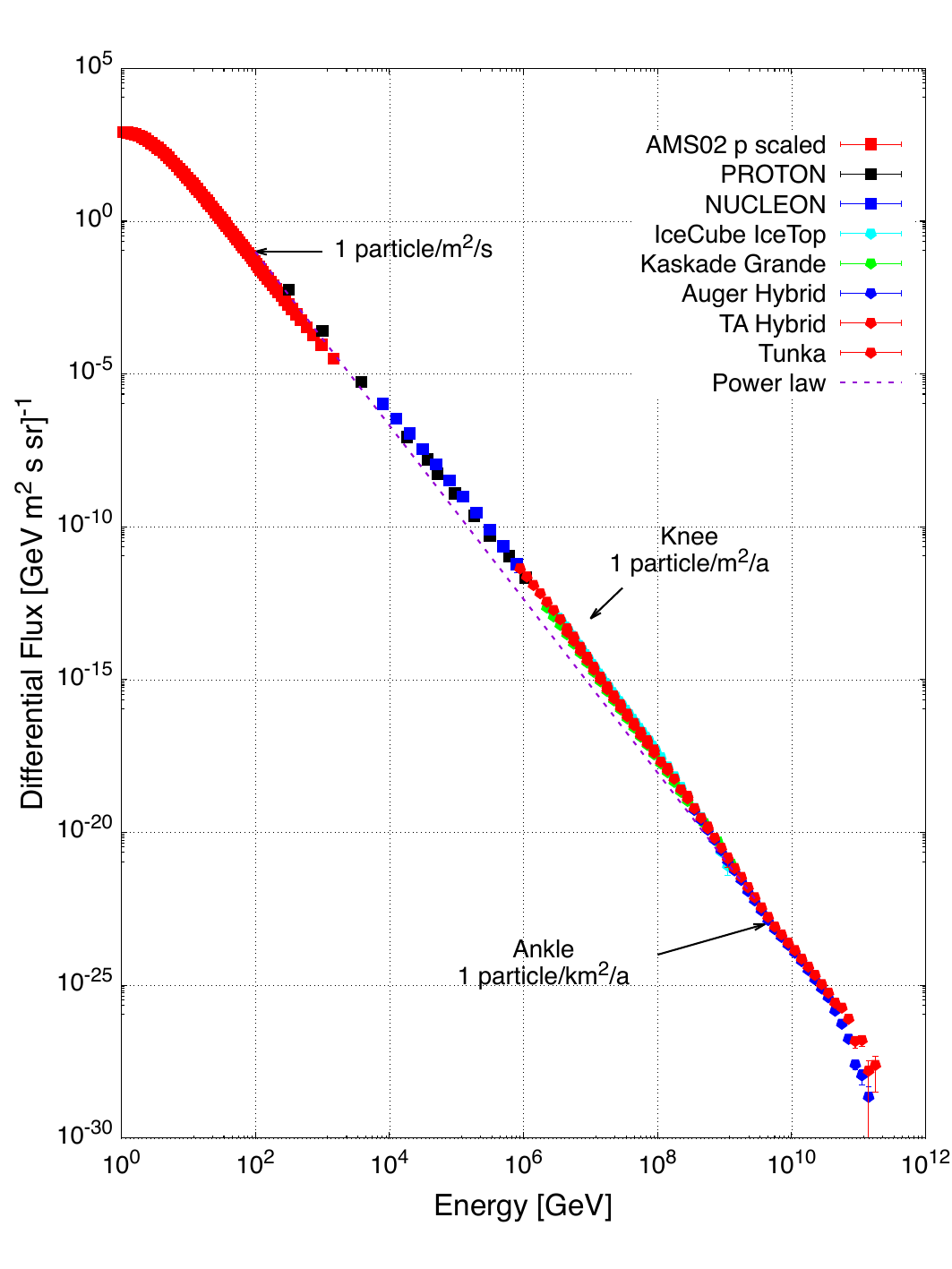}
\end{center}
\caption{\normalsize All-particle spectrum of cosmic rays~\cite{Bindi_2023} from selected space experiments ({\small $\blacksquare$}) and ground arrays ({\Large $\bullet$}). The AMS-02 proton spectrum~\cite{Aguilar_2021} is scaled up to give a rough estimate of an all-particle spectrum at low energies. The dashed line connects the flux at $100\, \mathrm{GeV}$ to the flux at the ankle, $4\times 10^9\, \mathrm{GeV}$; its slope corresponds to a constant spectral index of $-2.84$. 
\label{fig_allspectrum}}
\end{figure}

When all incident cosmic rays are lumped together, the flux as a function of energy indeed roughly follows a power law above $\sim 50\, \mathrm{GeV}$, as seen in Figure~\ref{fig_allspectrum}, $d\Phi/dE \propto E^{-\gamma}$. The exponent $\gamma$ is called spectral index. To make details visible, the gross energy dependence of the flux is often taken out by multiplying with $E^{+\gamma}$. The overall composition is dominated by fully ionised hydrogen (~89\%) and helium (~10\%), heavier nuclei and electrons present a percent-level contribution, antimatter is even rarer. 

The all-particle spectrum has few remarkable features, especially when presented in a log-log plot covering over 30 orders of magnitude in flux and close to 13 orders of magnitude in energy. The first feature marked in the plot is the so-called knee between $10^6$ and $10^7\, \mathrm{GeV}$,  where the fall-off steepens a bit. The second feature is called the ankle of the spectrum between $10^9$ and $10^{10}\, \mathrm{GeV}$, where it flattens a bit. However, the spectral index is not constant between these features, nor is it the same for all species. For energies up to a few TeV, this is shown in Section~\ref{physics} with the high-precision data from recent space experiments. At a few times $10^{20}\,\mathrm{eV}$, the Greisen-Zatsepin-Kuzmin then cuts off the spectrum rather sharply.\footnote{The Greisen-Zatsepin-Kuzmin (GKZ) cut-off ~\cite{Greisen_1966,Zatsepin_1966} is due to the resonant production of pions by the interaction of protons with photons from the microwave background. This reaction has a large cross section, degrades ultra-high cosmic ray energies and limits their range.}  The highest energy particle observed so far had of the order of $300\,\mathrm{EeV}$~\cite{Bird_1995}, corresponding to about $50\,\mathrm{J}$, seven orders of magnitude beyond the highest energies reached by man-made accelerators. 

Since charged particles are deviated by magnetic fields at every scale -- galaxies, stars, even planet Earth --, they only point back to their origin at extreme energies. In the energy range discussed in this article their flux is isotropic to good accuracy. The differential spectrum is thus averaged over the angular acceptance of the experiments. 

\section{A compact history of cosmic ray experiments in space}\label{history}

\begin{figure}[h]
\begin{center}
\includegraphics[width=0.5\textwidth,angle=0.3]{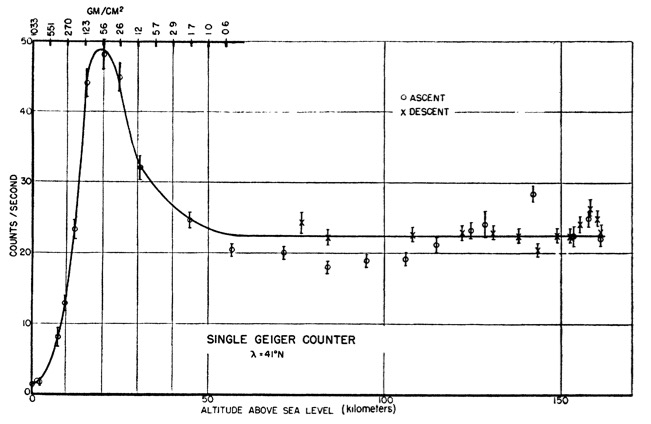}
\caption{\normalsize Count rate of a single unshielded Geiger-M\"uller counter installed at the tip of a V2 ballistic rocket as a function of altitude above sea level~\cite[Fig.4]{VanAllen_1948}.\label{fig_V2}}
\end{center}
\end{figure}

The exploration of primary cosmic radiation in Low Earth Orbit started together with the space age itself. Pioneering measurements were made by James A. Van Allen in 1947~\cite{VanAllen_1948} using a V2 rocket, spoils of World War II recuperated from Germany together with its inventor Wernher von Braun. The rate of a single Geiger-M\"uller counter installed at the tip of the ballistic rocket is shown in Figure~\ref{fig_V2}. It shows how the count rate falls beyond the Regener-Pfotzer maximum to reach a plateau above about $60\,\mathrm{km}$ altitude. Although these data are not corrected for back-splash from the atmosphere, this is arguably the first time that the direct observation of genuine primary cosmic rays has been reported. 

In the 1960s and 1970s, the Soviet Union PROTON and SOKOL satellite detectors made first attempts of calorimetric energy measurements combined with rudimentary particle identifications~\cite{Grigorov_1970,Grigorov_1971a,Grigorov_1971b,Grigorov_1971c,Grigorov_2004}. The devices used in the PROTON space flights were sampling calorimeters of gradually increasing thickness preceded by a carbon target~\cite{Grigorov_1966}. The particle identification by proportional chambers and a Cherenkov counter was compromised by back-splash from the carbon target. However, the energy measurements stay a reference for low-energy all-particle spectra~\cite{Grigorov_2004}. The SOKOL mission also used a sampling calorimeter with scintillators as active elements, but preceded it by a dual set of Cherenkov counters to fight backsplash~\cite{Ivanenko_1989}.  The NUCLEON detector~\cite{Atkin_2015}, in orbit from 2014 to 2017, was a modern implementation of this basic design. 

\begin{figure}[htb]
\begin{center}
\includegraphics[width=0.5\textwidth] {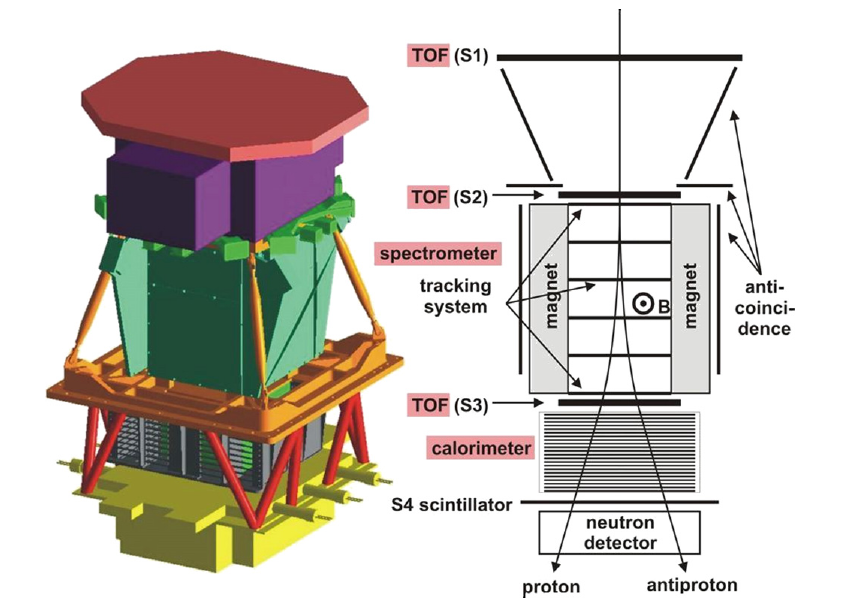}
\end{center}
\caption{\normalsize Schematic of the PAMELA satellite spectrometer and its components~\cite{Bonechi_2007}. A time-of-flight (TOF) system with three layers determines the particle flight direction and triggers the data taking. A spectrometer with a permanent magnet and six layers of solid state tracking detectors measures the magnetic rigidity. A calorimeter with tail catcher and neutron detector completes the set-up.} \label{fig_pamela}
\end{figure}

The pioneer of cosmic rays spectrometers in space was the PAMELA detector, launched in 2006 as an attached payload on a Russian Earth observation satellite~\cite{Spillantini_2009}, with the specific task of studying cosmic antimatter. A sketch of the set-up is shown in Figure~\ref{fig_pamela}. Its heart was a small magnetic spectrometer of $13 \times 16\,\mathrm{cm}^2$ active surface inside a permanent magnet providing a uniform field of $\sim 0.48\,\mathrm{T}$. Ferromagnetic shielding outside the magnet reduced field leakage. Six layers of double-sided silicon micro-strip detectors tracked particles with high precision and measured their specific energy loss $dE/dx \propto Z^2$, for a total geometrical acceptance of $21.5\,\mathrm{cm}^2 \mathrm{sr}$. The set-up was completed by scintillation detectors and an electromagnetic calorimeter, with tail catcher and neutron detector to help distinguish hadronic showers.  PAMELA's orbit was first a quasi-polar ($70^\circ$) elliptical one at altitudes between $355$ and $584\,\mathrm{km}$, changed in 2010 to a circular orbit at $550\,\mathrm{km}$. The duration of the mission, originally planned to be three years, turned out to be about ten years, during which the efficiency of the device changed from about 90\% in 2006 to 20\% after 2014, due to radiation damage to the tracking detector front-end chips~\cite{Adriani_2017}. A rich physics output resulted from the mission, a summary is presented in Reference~\cite{Adriani_2017}. In particular, results for the positron spectrum showed for the first time an excess with respect to pure secondary production~\cite{Adriani_2009b,Adriani_2013b}. 

With the PAMELA project, not only experience in modern particle detection but also the strict analysis techniques of modern particle physics migrated into the field of cosmic ray physics, which had suffered from little control of experimental systematics, as can be concluded from the large spread of flux results even for the most abundant components~\cite[Fig. 2.18]{Bindi_2023}. This dramatically changed with the advent of modern methodology in recent detectors, starting with PAMELA. Their sophisticated and redundant hardware, large acceptance and thorough calibration in accelerator particle beams yields better controlled systematics, together with excellent statistical accuracy of the data. The next such detector was the Alpha Magnetic Spectrometer, built by a collaboration of accelerator particle physicists led by Samuel C.C. Ting of MIT.  NASA required the collaboration to first build a demonstrator prototype, called AMS-01, to travel on the Space Shuttle. The detector had about one half of the spectrometer equipped inside a permanent magnet. Simplified particle identification equipment surrounded the spectrometer. 

\begin{figure}[!htb]
\begin{minipage}[c]{0.48\textwidth}
\includegraphics[width=\textwidth] {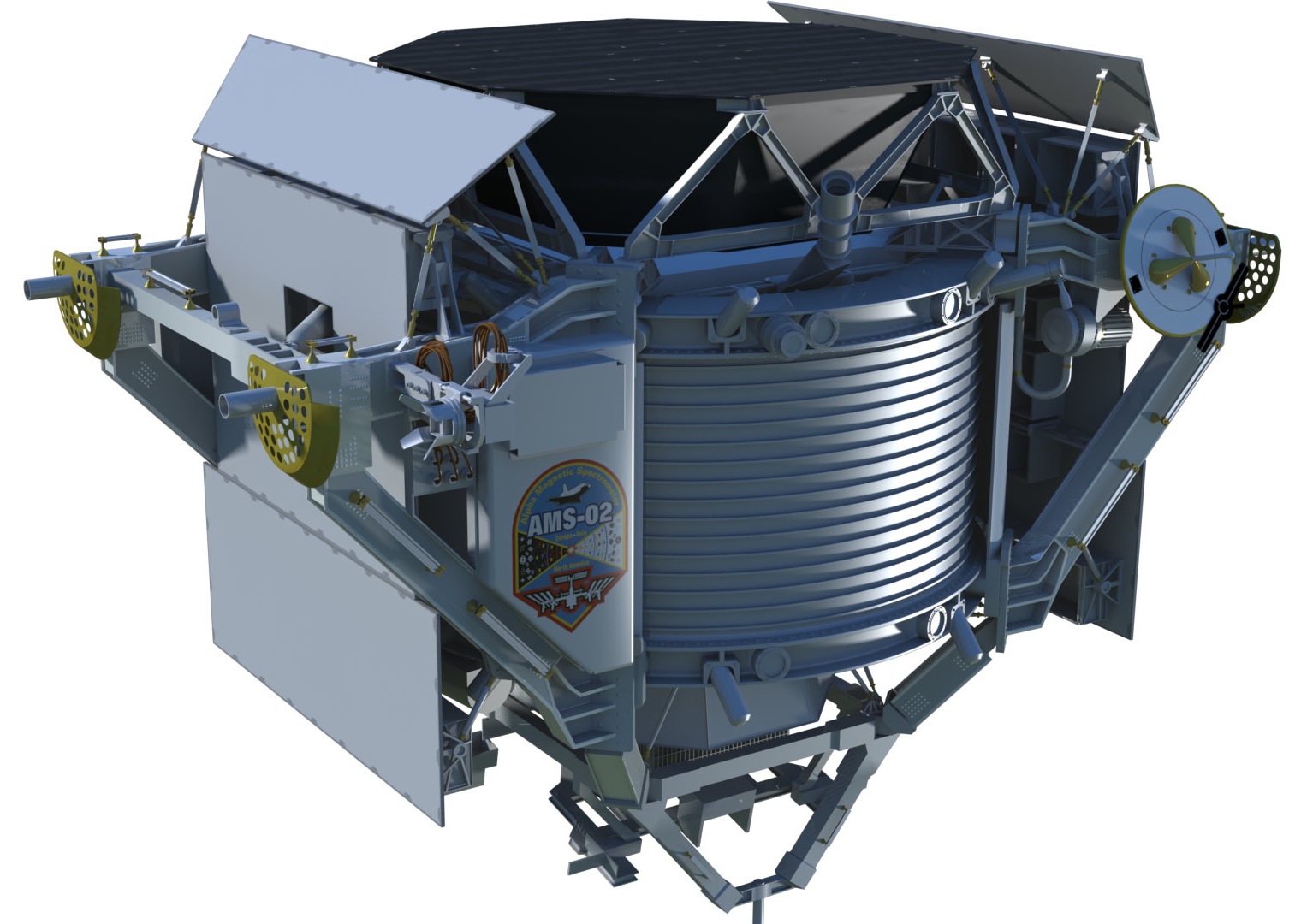}
\end{minipage} \hfill.
\begin{minipage}[c]{0.49\textwidth}
\includegraphics[width=\textwidth] {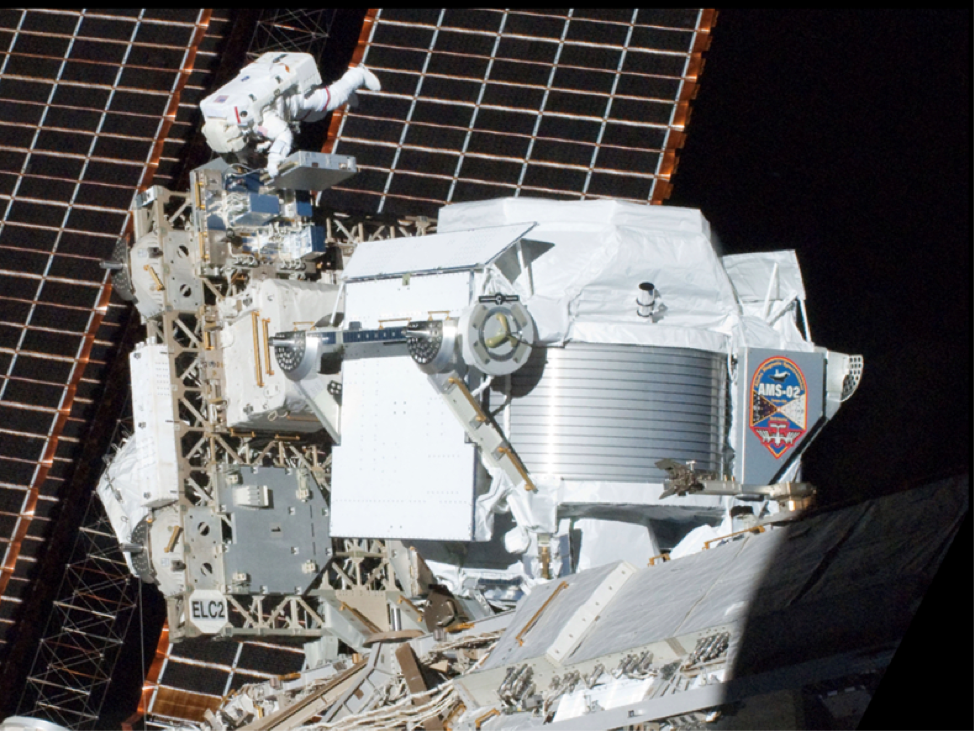}
\end{minipage} 
\caption{\normalsize Left: Artists impression of the AMS-02 detector (Credit: NASA). Right: The AMS-02 instrument installed on the main truss of the ISS with an astronaut working in a nearby site (Credit: NASA)\label{fig_ams02}.}
\end{figure}

The prototype detector flew on Space Shuttle Discovery's flight STS-91 in June 1998, the last Space Shuttle mission to the Russian Mir station. It successfully collected 70 million cosmic rays in ten days. Details of the set-up, performance and physics result are summarised in a  comprehensive report~\cite{Aguilar_2002}. Based on the success of this precursor, the collaboration decided to construct a more ambitious detector for a long-term installation on the ISS, AMS-02. It was intended to exchange the permanent magnet for a superconducting one of 0.87\,T~\cite{Blau_2002}, including the complex supplies by a cooling system with superfluid helium. From 2000 to 2008 the sub-detectors were constructed, mostly in Europe and financed by the international partners. The collaboration decided in 2010  to abandon the short-lived superconducting magnet in favour of the weaker permanent one from AMS-01~\cite{Overbye_2010}. In a crash program involving the whole collaboration~\cite{Lubelsmeyer_2011}, the permanent magnet was fitted into the vacuum case of the superconducting one. To rescue the spectrometer rigidity resolution, one tracking plane outside the magnet bore was moved to the very top of the set-up. An additional plane was equipped with spare parts and put in front on the calorimeter, lengthening the lever arm for bend-angle measurement to 3\,m. An important figure of merit is the so-called maximum detectable rigidity (MDR), defined as the magnetic rigidity where the relative measurement error reaches 100\%. AMS-02 achieves an MDR of 2\,TV for singly charged particles.\footnote{For higher charges the tracker resolution improves and the MDR reaches 3.7\,TV for iron nuclei.}  At the expense of rigidity resolution at low rigidities, the set-up thus gained enormously in life expectancy and was fit to see rare and higher energy phenomena. The new detector, shown in Figure~\ref{fig_ams02}, was assembled in a short time and exposed to beam calibration at CERN.   

On August 26, 2010, AMS-02 was delivered from CERN to the Kennedy Space Center by a Lockheed C-5 Galaxy cargo plane. It was launched to the ISS on May 16, 2011 on shuttle Endeavour's flight STS-134, commanded by Mark Kelly. Two robot arms transported it from the shuttle cargo bay to its anchor point on the zenith side of the starboard truss of the ISS, where it was installed and connected to power and data lines on May 19. A few hours later, AMS was up and running and recorded its first helium nucleus. 

The tracking system releases a constant power of about 140\,W through its front-end electronics inside the magnet bore. This heat must be removed to keep the temperature constant. The tracker is thus equipped with a thermal control system~\cite{Vanes_2013} using bi-phase CO$_2$ as a coolant, with two redundant cooling circuits each equipped with two redundant pumps. By 2015, these were showing wear and it was clear that a replacement was required for long-term operation. In 2019, an improved replacement of the pumping system was sent up to the ISS including a set of special tools to replace the system in orbit and replenish the liquid CO$_2$. In a complex operation, this was done by the ISS crew in four space-walks.\footnote{See \url{https://ams02.space/upgraded-tracker-thermal-pump-system}.} It involved cutting and welding a pressurised system in orbit for the first time and was described as the ``most challenging since Hubble repairs''~\cite{ESA_2019}. 

\section{Cosmic ray detectors in space}\label{generic} 

Space experiments are a privileged way of pursuing cosmic ray research since they allow to study them \emph{in situ} before they interact in the Earth's atmosphere. The ISS provides an ideal platform for the long-term exposure of sophisticated particle detectors. The year 2020 marked the twentieth anniversary of constant manning for the ISS, 2021 the tenth year of data taking by the Alpha Magnetic Spectrometer. New cosmic ray calorimeters have been deployed in the meantime. 

The power law shape of the cosmic ray all-particle spectrum has driven a two-fold experimental approach to the detection of cosmic rays.  Space-borne and balloon detectors are employed to measure cosmic ray spectra and elemental composition at energies approaching $10^6\,\mbox{GeV}$. Above this energy, the particle flux summed over a whole hemisphere is less than a few particles per square meter per year and rapidly decreases to about 100 particles per square kilometre per year at $10^9\,\mbox{GeV}$, thus requiring kilometre-size observatories which can only be deployed on the Earth's surface. Space-born detectors have direct access to cosmic ray particles before they enter the Earth's atmosphere and cause air showers. The accurate identification of the incoming cosmic ray particle is thus experimentally feasible. 

Both spectrometry-based  and calorimetry-based cosmic ray observatories are currently active in space missions. The primary observable of spectro\-meters is the magnetic rigidity, i.e.~the signed ratio of particle momentum and electric charge. For all astrophysical processes involving magnetic fields, like in acceleration and transport of cosmic rays, rigidity is indeed the relevant quantity. Spectrometers are generally equipped with additional detectors for particle identification, to enhance their capability to assess particle charge and distinguish light from heavy particles. Cosmic ray antiparticles can only be distinguished from particles by direct detection with a magnetic spectrometer, since there is no way to measure the sign of the electric charge by calorimetric means. Pioneers in the direct measurement of the spectra for individual cosmic ray species have been the balloon-borne magnetic spectrometer BESS~\cite{Yamamoto_2013,Abe_2017} and the satellite experiment PAMELA shown in Figure~\ref{fig_pamela}.

Magnetic spectrometers  equipped with velocity measuring devices also have the ability of measuring the isotopic composition of light nuclei. However, with current state-of-the-art technology, isotopic composition can be measured only at energies below about $10 \,\mathrm{GeV/n}$. Thus nuclei spectra are usually summed over isotopes.

Calorimetric detectors, on the other hand, are more compact and can cover a large solid angle. Current experiments can thus make meaningful flux measurements up to PeV energies, but do not distinguish between particles and antiparticles. Calorimeters are robust and can take data unattended over long periods of time. Their primary observable is the kinetic energy of the incident particle. The energy resolution of a calorimeter generally improves with energy until reaching a plateau, while the resolution of magnetic spectrometers worsens with rigidity. When combined with devices for the measurement of particle charge, calorimeters deliver the energy spectrum of individual cosmic ray species, summing over particles and antiparticles as well as isotopes. 

Spectra are thus measured as a function of magnetic rigidity by spectrometers, as a function of kinetic energy by calorimeters. Conversion from one quantity to the other requires an assumption about the isotopic composition of the observed elements based on available measurements or theory.

\subsection{The Alpha Magnetic Spectrometer AMS-02}

\begin{figure}[htbp]
\begin{center}
\includegraphics[width=\textwidth] {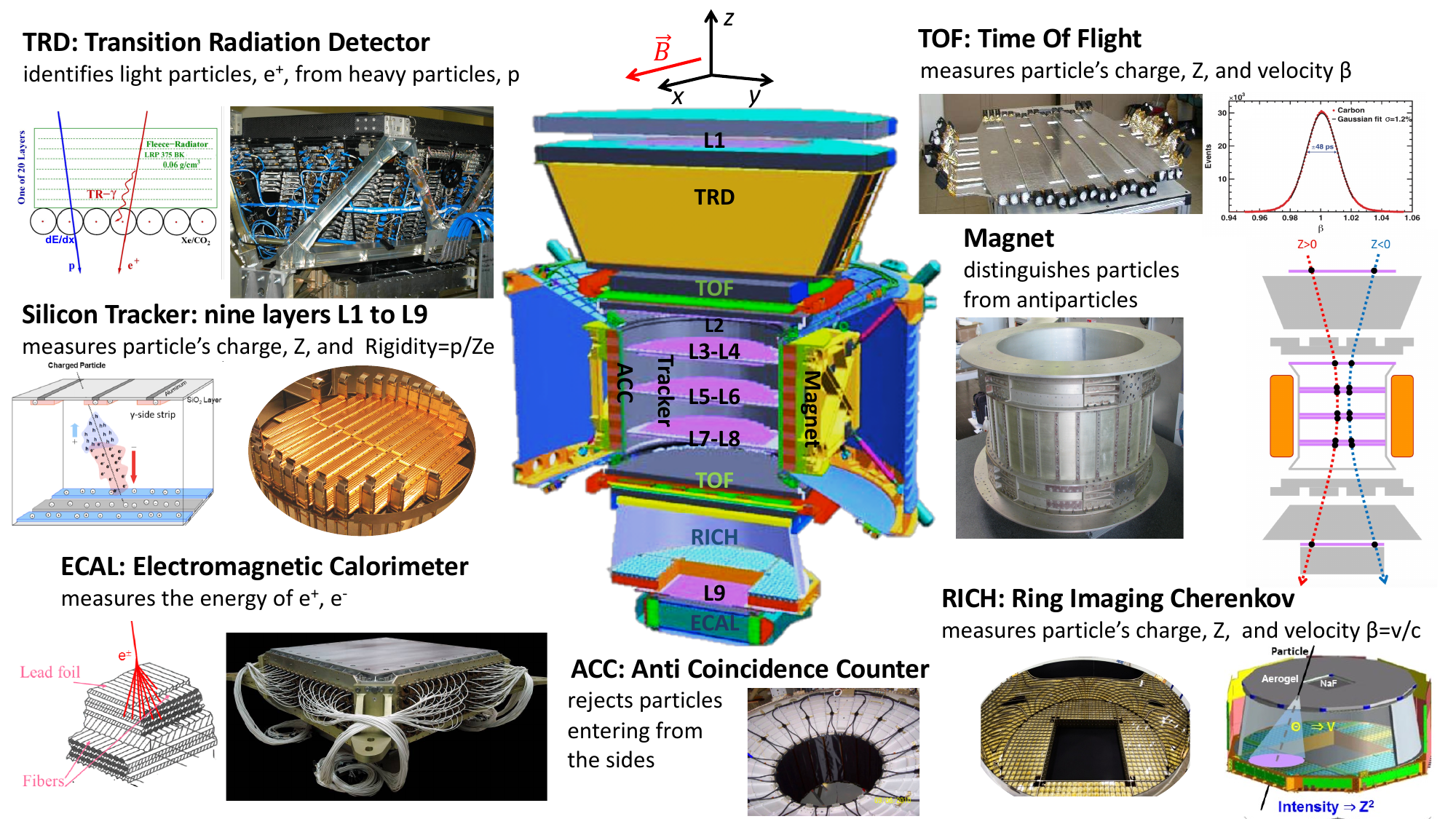}
\end{center}
\caption{\normalsize Schematic cut through the AMS-02 detector (centre) showing its components, TRD, TOF, Tracker, Magnet, ECAL, ACC, and RICH, with their main functions (adapted from~\cite{Aguilar_2021}). Cosmic rays are detected entering the instrument from the top.\label{fig_ams02_descr}}
\end{figure}

The only magnetic spectrometer currently acquiring cosmic ray data is the Alpha Magnetic Spectrometer AMS-02 shown in Figure~\ref{fig_ams02}. We briefly describe its principle components and their performance; more details are available in Reference~\cite{Aguilar_2021}. The AMS-02 components, shown in Figure \ref{fig_ams02_descr}, have been designed to perform simultaneous and accurate measurements of the individual spectra of positrons, electrons, antiprotons, protons and nuclei up to the nickel region ($Z \sim 30$). The spectrometer has a reach in magnetic rigidity from a fraction of a GV up to a few TV. 

Particles enter the detector from the top, in the negative $z$-direction in the coordinate system indicated in Figure~\ref{fig_ams02_descr}. The heart of the detector is a spectrometer comprising a cylindrical permanent magnet and a series of silicon sensors tracking the particle trajectory. The dipolar field is generated by 64 NdFeB blocks arranged in a Halbach array, reaching $0.15\, \mathrm{T}$ in the centre. It points in the $x$-direction, bending the trajectory in $y$. The trajectory is located by nine layers of silicon strip detectors. The tracking layers are arranged with six layers inside the magnet bore and three outside. The rigidity measurement thus results from a mixture of sagitta and bend-angle measurements. The relative rigidity resolution is about $\Delta R/R \simeq 0.1$ for $R < 20\, \mathrm{GV}$ and increases with rigidity. The maximum detectable rigidity varies between $2.0\, \mathrm{TV}$ for protons and $3.7\, \mathrm{TV}$ for iron nuclei. The spectrometer is completed by a large set of partially redundant devices identifying particles and measuring their kinetic energy, as described in Figure~\ref{fig_ams02_descr}. AMS-02 is installed on the ISS since May 2011. Until the end of 2024, it has collected over 245 billion cosmic ray particles,\footnote{The current count rate is constantly updated at \url{https://ams02.space}.} arguably the largest sample since the discovery of cosmic rays. The collaboration plans to continue operating the observatory until the end of the ISS lifetime. 

The AMS-02 detector components are listed in Figure~\ref{fig_ams02_descr}. Particles enter the detector from the top, opposite the $z$-direction in the coordinate system indicated. We describe the components in the order in which they are encountered:
\begin{itemize}
\item The {\bf Transition Radiation Detector} (TRD) has twenty layers.  Each layer is composed of a $20\, \mathrm{mm}$ thick fleece radiator and a layer of proportional tubes for X-ray detection. The TRD distinguishes light from heavy particles, by measuring the transition radiation emitted by highly relativistic particles. It identifies one positron in a background of more than 1000 protons at $90\%$ positron detection efficiency.
\item The {\bf Time-Of-Flight} system (TOF) consists of two double layers of scintillation detectors placed above and below the magnet, covering about $1.6\,\mathrm{m}^2$ each. They measures the time a particle takes to traverse from the top to the bottom layer with a resolution ranging from $160\, \mathrm{ps}$ for singly charged ($Z=1$) particles down to $50\,\mathrm{ps}$ for heavier ($Z>6$) nuclei. The TOF measures the cosmic ray arrival direction and velocity, and gives the {\bf trigger} for charged particles to the overall data acquisition system. 
\item Sixteen {\bf Anticoincidence Counters} (ICC) surround the inner bore of the magnet to reject particles entering the detector from the sides with an efficiency of 0.99999.
\item The heart of the detector is a {\bf magnetic spectrometer} composed of a cylindrical permanent magnet, and a tracker. The magnet generates a dipolar field with the main component directed along the $x$-axis. The tracker has nine layers of double-sided silicon micro-strip detectors. Layers L3 to L8 are inside the magnet bore, L2 above the magnet, L1 on top of the TRD and L9 just above the electromagnetic calorimeter (ECAL). The total lever arm from L1 to L9 is $3\, \mathrm{m}$. At each tracker layer the $x$- and $y$- coordinates of the particle impact point are measured with accuracies of $13$ to $20\, \mu\mathrm{m}$ and $5$ to $10\, \mu\mathrm{m}$  respectively. The bending of the particle trajectory inside the magnetic field gives the rigidity, $R=p/(Ze)$, and its direction allows to distinguish positively charged particles from negatively charged particles. 
\item The {\bf Ring Imaging Cherenkov counter} (RICH) is composed of a radiator plane made of NaF and silica aerogel, a conical mirror on the sides, and a photo-detection plane at the bottom. The particle's velocity is obtained from the aperture angle of the Cherenkov light cone with a relative resolution better than $0.1\%$, allowing to measure isotopic composition of light nuclei up to $\sim 10\, \mathrm{GeV}/\mathrm{n}$. The {\bf particle charge} squared, $Z^2$, is measured independently from the intensity of the emitted light in the RICH, and from the energy deposited in the active detector materials of TRD, TOF, Tracker and ECAL.
\item The 3D sampling {\bf Electromagnetic Calorimeter} (ECAL) is made of nine multi-layered lead and scintillating fibre sandwiches for a total of $17\,X_0$. The nine sandwiches are stacked such that the fibres run alternatively along the $x$-coordinate (five layers) and the $y$-coordinate (four layers). ECAL measures the energy of electrons and positrons with few percent accuracy. It also gives the {\bf trigger} to the overall data acquisition for photons and measures their energy. The 3D reconstruction of the particle shower in ECAL, and the matching of the energy measured in ECAL with the momentum measured in the tracker give an additional discrimination power to separate leptons ($\mathrm{e}^\pm$) from hadrons ($\mathrm{p}$ and $\bar{\mathrm{p}}$) better than 1 positron over 10'000 protons at $90\%$ positron detection efficiency, independent of the TRD. The ECAL also allows to calibrate the rigidity scale of the spectrometer \emph{in situ} using cosmic rays electrons and positrons. 
\end{itemize}
The complete detector in its flight configuration has been carefully calibrated in particle beams at CERN just before launch. Calibration and alignment of components are constantly updated and refined during data taking on the ISS~\cite[Sec. 1.]{Aguilar_2021}.

\subsection{The calorimetric cosmic ray detectors CALET and DAMPE}

Since 2015, the energy region beyond a few TeV is being explored with large-size calorimeters, such as DAMPE~\cite{Chang_2017} on a free-flying Chinese satellite, and CALET~\cite{Torii_2019} on the ISS. This pushes the frontier of current direct cosmic ray measurements to $10^6\,\mbox{GeV}$ and potentially towards knee energies as more data are collected.  The DAMPE and CALET missions both started in 2015 and are currently operational. They have collected several billion cosmic rays until the end of 2023.

\begin{figure}[htb]
\begin{minipage}[c]{0.49\textwidth}
\includegraphics[width=\textwidth] {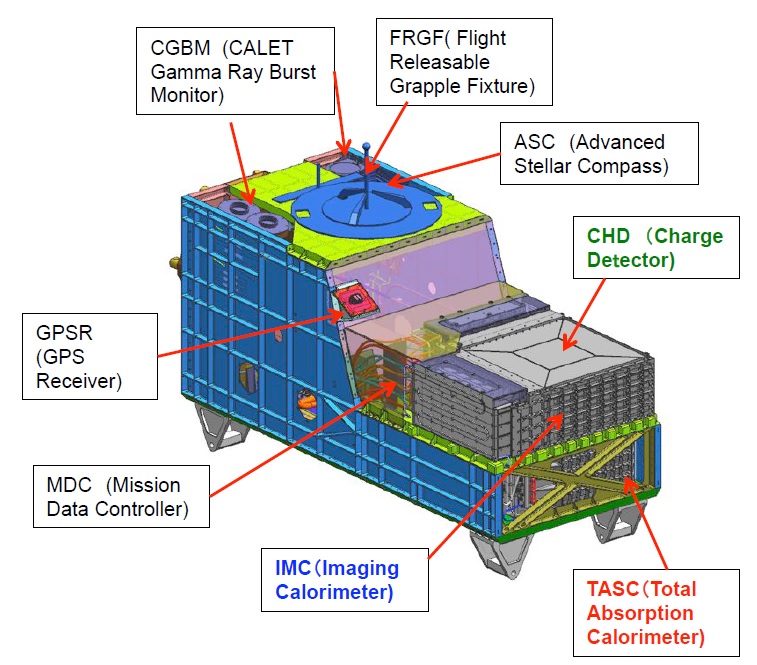}
\end{minipage} \hfill
\begin{minipage}[c]{0.49\textwidth}
\includegraphics[width=\textwidth] {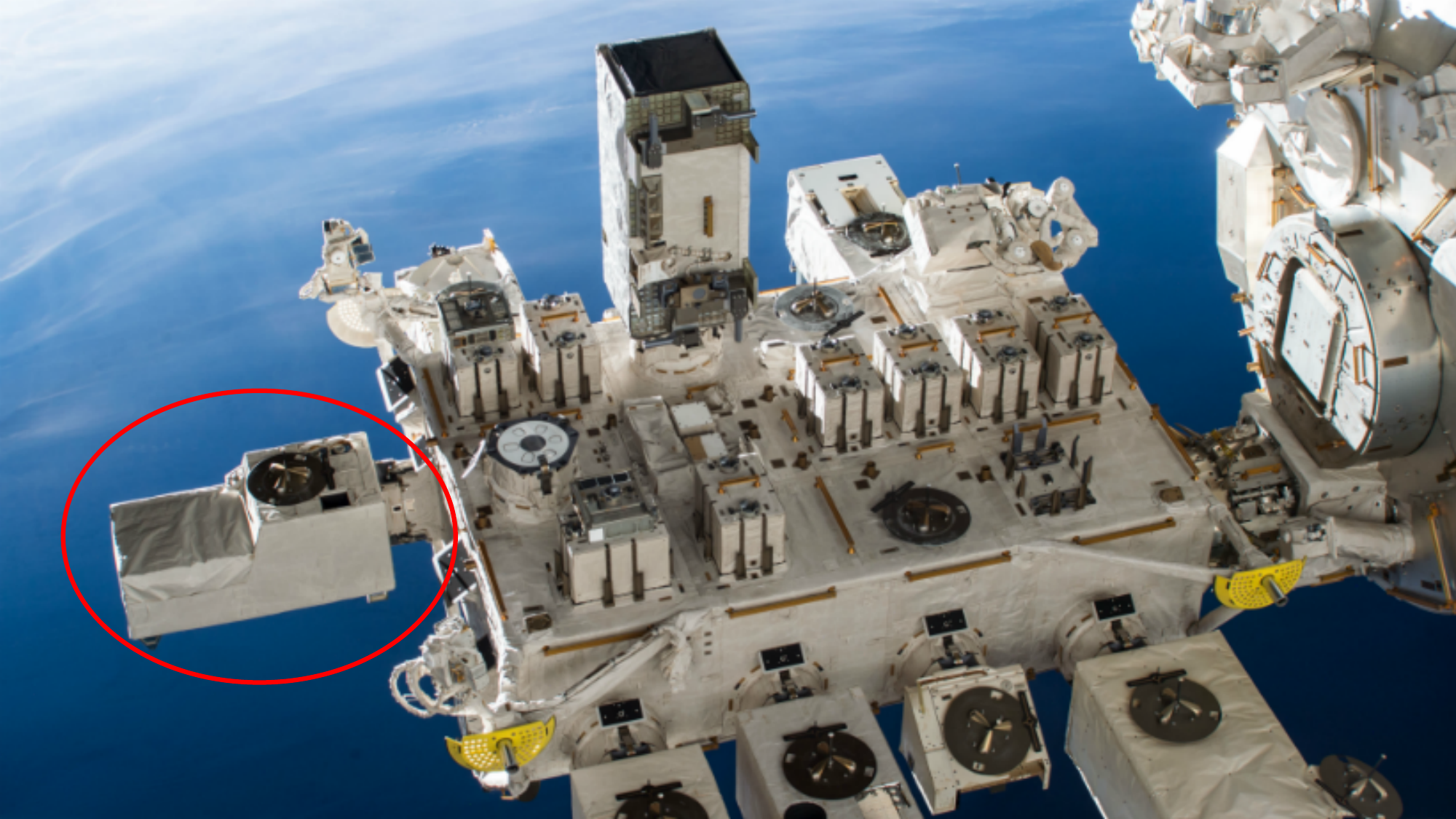}
\end{minipage} 
\caption{\normalsize Left: Principle components of the CALET calorimetric detector~\cite{Asaoka_2019} with the charge measurement device (CHD) followed by an imaging calorimeter (IMC) and a total absorption calorimeter (TASC) for electromagnetic showers. The CALET payload is also equipped with a gamma-ray burst monitor (CGBM).
Right: The CALET instrument (ellipse) installed on the Exposed Facility of the Japanese Experiment Module on the ISS (Credit: JAXA/NASA).}\label{fig_spacecal_calet} 
\end{figure}

The Calorimetric Electron Telescope (CALET) has been developed to measure the cosmic ray $\mathrm{e}^\pm$ spectrum in the kinetic energy range from $1\,\mathrm{GeV}$ to $20\,\mathrm{TeV}$, the gamma-ray spectrum up to $10\,\mathrm{TeV}$ and the proton spectrum from 50 GeV to 1000 TeV. The individual spectra of nuclei can be measured in the kinetic energy range from $10\, \mathrm{GeV}$ to $1000\, \mathrm{TeV}$, for nuclei up to iron as well as trans-iron nuclei up to $Z \sim 40$~\cite{Torii_2019}. CALET has been launched by a Japanese mission to the ISS and takes data since October 2015. The CALET set-up and its position on the ISS are shown in Figure~\ref{fig_spacecal_calet}. Cosmic rays entering from the top are measured in the following three detectors:
\begin{itemize}
\item The {\bf Charge Detector} (CHD) is made of a double layer of scintillators read by photomultiplier tubes (PMT). The signal of each CHD layer is used as input to trigger the overall data acquisition. The CHD measures the absolute value of the particle charge from the energy deposited in the scintillator. Its large dynamic range allows to identify nuclei up to $|Z|\sim 40$. 
\item The {\bf Imaging Calorimeter} (IMC) is a 3D sampling calorimeter with a total depth equivalent to $3\,X_0$, covering a surface of $45 \times 45\,\mathrm{cm}^2$. It is made of seven layers of tungsten alternated with two layers of scintillating fibres (SciFi) arranged orthogonally,  and capped by additional SciFi double layers for a total of 16 active layers. The fibres are read individually by multi-anode photomultiplier tubes (MAPMT). The IMC provides a 3D reconstruction of the early phase of the incoming particle shower allowing to determine the shower starting point and the cosmic ray arrival direction with angular resolutions of $0.14^{\circ}$ for electrons and $0.24^{\circ}$ for photons. The IMC also measures the absolute value of the charge for nuclei up to silicon ($Z=14$) from the energy deposited in the SciFi fibres. For each IMC double layer the signals from the two SciFi layers are combined to generate input to the trigger system.
\item The {\bf Total Absorption Calorimeter} (TASC) is a homogenous calorimeter equivalent to  $27\,X_0$. It consists of twelve layers of lead tungstate (PWO), each composed of 16 PWO logs. The layers are arranged alternatively  with the logs running along orthogonal directions to allow 3D reconstruction of the particle shower. The PWO logs of the first layer are read individually by a photomultiplier tube (PMT) to provide additional input to the trigger system. The remaining eleven layers are read by silicon photodiodes and silicon avalanche photodiodes (Dual PD/APD). The read-out system is configured to provide sufficient dynamic range to measure energy depositions from MIP to showers induced by a $1\,\mathrm{PeV}$ proton.The TASC measures the energy of electrons with a resolution better than $2\%$ above $20\, \mathrm{GeV}$. IMC and TASC together represent a thickness of $\sim 1.3 \lambda_I$ for hadrons.
\item The {\bf trigger} for the data acquisition is obtained combining the signals from the CHD layers, the IMC layers, and the first TASC layer~\cite{Asaoka_2017}. Leptons ($\mathrm{e}^\pm$) are distinguished from protons and nuclei by comparing the 3D reconstruction of their shower in the IMC and TASC. The discrimination power is of the order of 1 electron over a background of $10^5$ protons with less than $10\%$ proton contamination up to $7.5\,\mathrm{TeV}$~\cite{Adriani_2023}, using an innovative boosted decision tree analysis.
\end{itemize}
 CALET components have been calibrated before launch in particle beams at CERN, the calibration is constantly updated using flight data~\cite{Asaoka_2017}. Until the end of 2023, CALET has registered about 2 billion cosmic rays with an energy exceeding $10\,\mathrm{GeV}$.
 
Based on the successful balloon detectors CREAM, an evolved version ISS-CREAM~\cite{Seo_2014,Seo_2020} was installed on the ISS on the same platform as CALET in 2017. The detector components are similar to the balloon version described in the encyclopedia article about balloon cosmic ray detectors. It took data for 18 months from August 2017 to February 2019, before it was stopped prematurely by NASA management, after an estimated life-time of only 228 days~\cite{Choi_2022}.

\begin{figure}[hbt]
\begin{center} \includegraphics[width=0.59\textwidth] {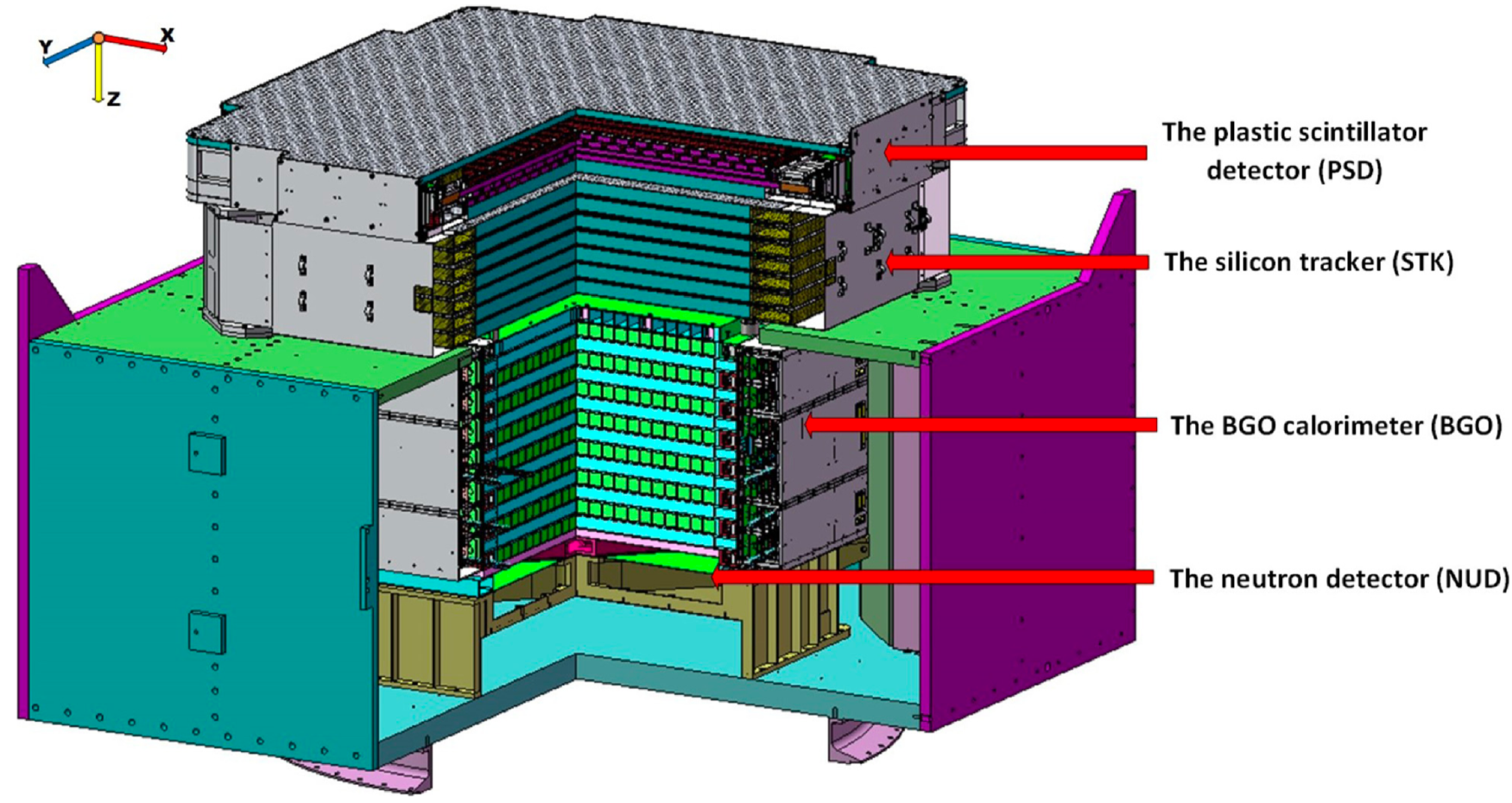}
\end{center}
\caption{\normalsize Principle components of the  DAMPE~\cite{Chang_2017} free-flying calorimetric cosmic ray detector, composed of a Plastic Scintillator Detector (PSD) and a Silicon-Tungsten tracker-converter (STK) for charge measurement, a BGO imaging calorimeter of $32\,X_0$ depth for energy measurement, and a Neutron Detector (NUD), which together with the calorimeter provides lepton/hadron identification.} \label{fig_spacecal_dampe}
\end{figure}

The Dark Matter Particle Explorer (DAMPE) measures the spectra of photons and $\mathrm{e}^\pm$  in the energy range  $5\, \mathrm{GeV}$ to $20\, \mathrm{TeV}$, and the spectra of protons and nuclei up to iron ($Z\leq 26$) in the kinetic energy range from $50\, \mathrm{GeV}$ to  $1\, \mathrm{PeV}$. The DAMPE satellite was launched in December 2015 on a Chinese carrier. It follows a sun-synchronous orbit of about $500\,\mathrm{km}$ radius. The principle components of the detector are shown in Figure~\ref{fig_spacecal_dampe}:
\begin{itemize}
\item The {\bf Plastic Scintillator Detector} (PSD) measures the charge of incident particles and provides charged-particle background rejection for gamma rays. The PSD has an active area of $82.5 \times 82.5\,\mathrm{cm}^2$. It consists of 82 plastic scintillator bars arranged in two orthogonal planes, each equipped with a double layer. Each bar is read out at both ends with a wide dynamic range extending up to 1400 times the energy deposition of a minimum ionising particle (MIP), so that it can identify nuclei up to $Z\simeq40$~\cite{Sun_2023}.  
\item The role of the {\bf Silicon-Tungsten Tracker Converter} (STK)  is to provide precise particle track reconstruction, measure the electrical charge of incoming cosmic rays and convert incoming photons to electron-positron pairs. It consists of six position-sensitive orthogonally oriented double planes of silicon detectors with a total area of about $7\,\mathrm{m}^2$, with a single hit resolution of about $50\,\mu\mathrm{m}$. Multiple thin tungsten layers enhance the photon conversion rate while keeping multiple scattering of electron-positron pairs at a negligible level for energies above $50\,\mathrm{GeV}$. The total thickness of STK is about $1\, X_0$.
\item The {\bf BGO calorimeter} (BGO) measures the energy deposition of incident particles. It images the 3D profile (both longitudinal and transverse) of the shower development, and thus provides electron/hadron discrimination. And it provides the main trigger for the DAMPE data acquisition. It consists of bismuth germanate (BGO) bars, arranged in 14 layers of alternating orthogonal layers with 22 bars each. Its total area coverage is $60\times 60\,\mathrm{cm}^2$, the total depth amounts to $32\, X_0$ at normal incidence. For electromagnetic showers, it has the percent-level energy resolution typical for this material. For hadronic showers, the resolution is about 30\%~\cite{Alemanno_2024c}.
\item The {\bf Neutron Detector} (NUD) acts as a tail catcher for hadronic showers. Its blocks of boron-loaded plastic scintillator are sensitive to neutron capture and give a signal after a fixed delay of $2\,\mu\mathrm{s}$ with respect to the trigger. The NUD thus contributes to the distinction between hadronic and electromagnetic showers. It enhances the rejection of hadronic showers by more than a factor of ten.  
\end{itemize}
Components have been calibrated before launch in particle beams at CERN, the calibration is constantly updated using flight data. Until the end of 2023, DAMPE has registered about 15 billion cosmic rays and photons with an energy exceeding $20\,\mathrm{GeV}$.

\section{From data to physics}\label{data} 

Cosmic ray experiments identify and count cosmic ray particles as a function of particle type, magnetic rigidity or energy,  direction and time of arrival. These observables are used to measure the differential cosmic ray flux, for single species, groups of particles or all particles. 

The differential flux $d\Phi/dE$ is given by the number of particles $\Delta N(E)$ counted in a time window $\Delta t$ and an energy window between $E$ and $E + \Delta E$. When it is measured by a detector covering an acceptance $A$, defined by its active surface, solid angle coverage and total detection efficiency, it is given by:
\begin{equation}
\dtod{\Phi(E)}{E} = \frac{\Delta N(E)-B}{A\ {\Delta t} \ {\Delta E}}
\end{equation} 
The raw number of particle counts $\Delta N$ needs to be corrected for the background counts $B$. The acceptance $A$ of a detector includes three main factors:
\begin{itemize}
\item A geometrical factor $G$ ($[G] = [\mbox{m}^2\mbox{sr}]$) defined by the active surface and solid angle coverage. It may depend on particle type, since different sub-systems may be required for their identification. 
\item The efficiency $\epsilon_T$ to trigger the data acquisition when a particle of the given species passes through $G$.
\item The efficiency $\epsilon_S$ of the event reconstruction and selection criteria used to assign particles to species and energy bin.
\end{itemize}
The geometrical factor depends on details of the analysis. Especially for the elongated geometry of AMS-02, the inclusion of the ECAL in the required components reduces it significantly. For the compact geometry of calorimetric detectors, this is less important.  For high energy electrons and positrons, the geometrical acceptance is of the order of $550\,\mathrm{cm}^2\mathrm{sr}$ for AMS-02~\cite{Aguilar_2013}, $1'200\,\mathrm{cm}^2\mathrm{sr}$ for CALET~\cite{Adriani_2023} and $3'000\,\mathrm{cm}^2\mathrm{sr}$ for DAMPE~\cite{Fan_2014}. 

For high-precision results, care must be taken when placing the data points inside their respective bins, especially energy-type variables like kinetic energy (total or per nucleon) or magnetic rigidity (momentum per unit charge). This is done here wherever possible using the prescription of reference~\cite{Lafferty_1995}. For high-precision measurements, it is also necessary to account for the bin-to-bin migration of events caused by energy or rigidity resolution. This is usually done by an unfolding procedure.  
 
Three features of generic space detectors are particularly crucial. The first is their ability to establish and monitor the rigidity or energy scale of the measurement. This is important, since an error on the absolute energy or rigidity scale results in an incorrect flux normalisation. We will multiply differential fluxes by a power of the rigidity or energy to make spectral features visible. In this article, the appropriate scale factor is $(d\Phi/dR) \cdot R^{2.7}$ for nuclei and $(d\Phi/dR) \cdot R^3$ for antiprotons, electrons and positrons. A scale error in rigidity or energy thus translates into a shift in the scaled flux. 

All modern detectors undergo tests in particle beams at accelerators before flight and the responses of the subdetectors measuring the energy or rigidity are calibrated based on these data. However, test beam particles are only available up to few hundred GeV, well below the TeV range. Calorimeters are also calibrated in-flight to correct for time-dependent variation of the energy responses using minimum ionising particles (MIP), which also have low energies. Therefore the check of the energy scale up to TeV requires assumptions on the functional behaviour of the energy response for increasing energy of the incoming particles. Moreover the vibrations and shocks during the launch and the thermal environment in space might change the alignment among the detector layers or the response of the detector. The stability of the rigidity or energy scale is checked in light by different means:
\begin{itemize}
\item In {\bf AMS-02}, the tracker in-flight rigidity scale shift and its uncertainty is obtained comparing the absolute rigidity values measured for electrons and positrons with the energy measured by the electromagnetic calorimeter in 72 energy points from 2\,GeV to 300\,GeV \cite{Berdugo_2017}. The energy of electrons  and positrons is determined with excellent linearity and a resolution of 1.4\% at 1 TeV~\cite{Adloff_2013, Kounine_2017}. The stability of the rigidity scale as function of time is evaluated analysing the data taken in different time periods and time-dependent corrections are applied. The alignment of the tracker layers is constantly monitored in-flight and corrections are applied. The rigidity scale is thus determined with an accuracy of 3\% at $1\,\mathrm{TV}$~\cite{Aguilar_2021}.
\item In {\bf CALET}, the detector energy calibration is performed in-flight by checking the consistency of the energy response for proton and helium minimum ionising particles at equivalent rigidity cutoffs. The helium minimum ionising particle peak as a function of the rigidity cutoff, derived from flight data, is compared to Monte Carlo simulations. In addition, the absolute energy scale for nuclei is determined with an accuracy of 3\% for protons~\cite{Adriani_2019} and 2\% of carbon, oxygen, and iron~\cite{Adriani_2020, Adriani_2021}, based on the precision of the beam test calibration.
\item In {\bf DAMPE} the absolute energy scale uncertainty was estimated in-flight at 13 GeV using the geomagnetic cutoff for electrons and positrons and resulted in an about 1.3\% accuracy~\cite{Zang_2017, Alemanno_2021}.
\end{itemize}

The second important feature concerns the identification of rare species, like positrons or antiprotons. Since their matter counterparts, protons and electrons, are orders of magnitude more abundant, experiments need strong rejection power on the dominant species to determine their flux. So in experiments, light particles like $\mathrm{e}^\pm$ must be well separated from heavy particles of like absolute charge, such as protons. In spectrometers, this especially applies to like-sign signal and background, where a rare species must be distinguished from an abundant one. Examples are positrons in a background of protons, antiprotons in a background of electrons. In AMS-02, redundant sub-detectors are used to increase the background rejection power.  

There is a third recurring problem not only for the measurement of elemental spectra but also for their interpretation in terms of cosmic ray propagation. It comes from uncertainties in Monte Carlo simulation. These uncertainties concern the interaction properties of certain cosmic ray species, in particular nuclei. Examples are the energy response of calorimeters or backsplash due to early interactions. The dominant uncertainty, however, comes from our poor knowledge of cross sections for nuclear reactions. Such reactions can transform a heavier element into a lighter one by inelastic interaction. The target can be interstellar matter, mostly protons and helium nuclei, met before the cosmic ray reaches us. The transformation probability must then be taken into account in modelling spectra and composition of cosmic rays. The extraction of propagation parameters strongly relies on the knowledge of spallation cross sections. The identification of signals from dark matter or primordial antimatter requires a precise determination of the background from secondary production of positrons, antiprotons and anti-nuclei. If one could firmly establish the probability for secondary production, this would serve as benchmark to identify excesses or deficits in observed  cosmic ray spectra and pin down their origin, acceleration mechanism and propagation history. 

The target for nuclear reactions can also be met inside the detector itself, the loss and gain of the initial and final species must then be corrected for. The poor knowledge of nuclear fragmentation cross sections  thus affects also cosmic ray measurements themselves and contributes a sizeable systematic error. In particular in direct detection, the knowledge of the amount of incoming cosmic ray nuclei fragmenting in the detector material is of paramount importance to correctly assess the overall flux normalisation. There are two principle approaches to this problem. One can estimate the survival probability of a given nucleus using simulation. The systematic error is then estimated comparing different nuclear interaction models or comparing Monte Carlo simulations to test beam data~\cite{Alemanno_2021}.  An alternative and more robust approach is to compare the simulation to data collected with the detector itself~\cite{Aguilar_2015b, Aguilar_2021, Adriani_2020}. At  energies below a few hundred GeV/n,  this can be done using test beam data. At high energy, cosmic ray data can be used to measure survival probabilities in the detector, provided the detector design allows to clearly identify incoming cosmic ray nuclei before they start to fragment, to use a portion of the detector as target, and to measure the amount of fragmentation products on the downstream side. With this approach, the AMS-02 detector has measured charge-changing nuclear fragmentation cross sections on carbon target for the most abundant primary cosmic ray nuclei from helium to iron of rigidities from few GV up to TV ~\cite{Yan_2021, Aguilar_2021a}. DAMPE has made a similar study for protons and helium nuclei impinging on heavy target material~\cite{Alemanno_2024a}.

\section{Physics of galactic cosmic rays}\label{physics} 

In this section, we discuss experimental results on the differential fluxes of cosmic ray nuclei, antinuclei, electrons and positrons. We start with the lightest and most abundant nuclear species, hydrogen and helium, and then continue upwards in nuclear mass. We end this section with a discussion of electron and positron spectra. 

\subsection{Protons and helium}\label{sec:hhe} 

The most abundant species in cosmic rays are hydrogen nuclei, i.e.~$Z=+1$ nuclei including protons and deuterons, often simply called protons in the literature. The tritium life time is too short to contribute significantly. Indeed protons make up the bulk of hydrogen nuclei, the ratio $^2\mathrm{H}/^1\mathrm{H}$ being around $2\%$ at $1\,\mathrm{GeV}/\mathrm{n}$. Likewise, the spectra for $Z=+2$ include $\phantom{}^3\mathrm{He}$ and $\phantom{}^4\mathrm{He}$, with predominance of $^4\mathrm{He}$. Spectra are usually plotted multiplied by $R^{2.7}$, for measurements from magnetic spectrometers, or by $E^{2.7}$, for calorimetric instruments, to put in evidence deviations from the featureless power law predicted by traditional cosmic ray models. 

The first instrument to connect the GV region to the TV region with precision measurements has been the PAMELA magnetic spectrometer, which measured the proton and helium spectra in the rigidity range from $1\, \mathrm{GV}$ to $1.2\, \mathrm{TV}$~\cite{Adriani_2011,Adriani_2017}. PAMELA demonstrated for the first time a spectral hardening in both the proton and helium spectra at rigidities in the range $230\,\mathrm{GV}$ to $240\,\mathrm{GV}$ for protons and helium. PAMELA observed that the proton-to-helium flux ratio was a smooth decreasing function of the rigidity over the entire rigidity range of their measurements, finally assessing that hydrogen and helium nuclei have different spectral shapes at modest energies. And neither spectrum follows a featureless power law with constant spectral index.

\begin{figure}[!hbt]
\begin{center}
\includegraphics[width=0.49\textwidth] {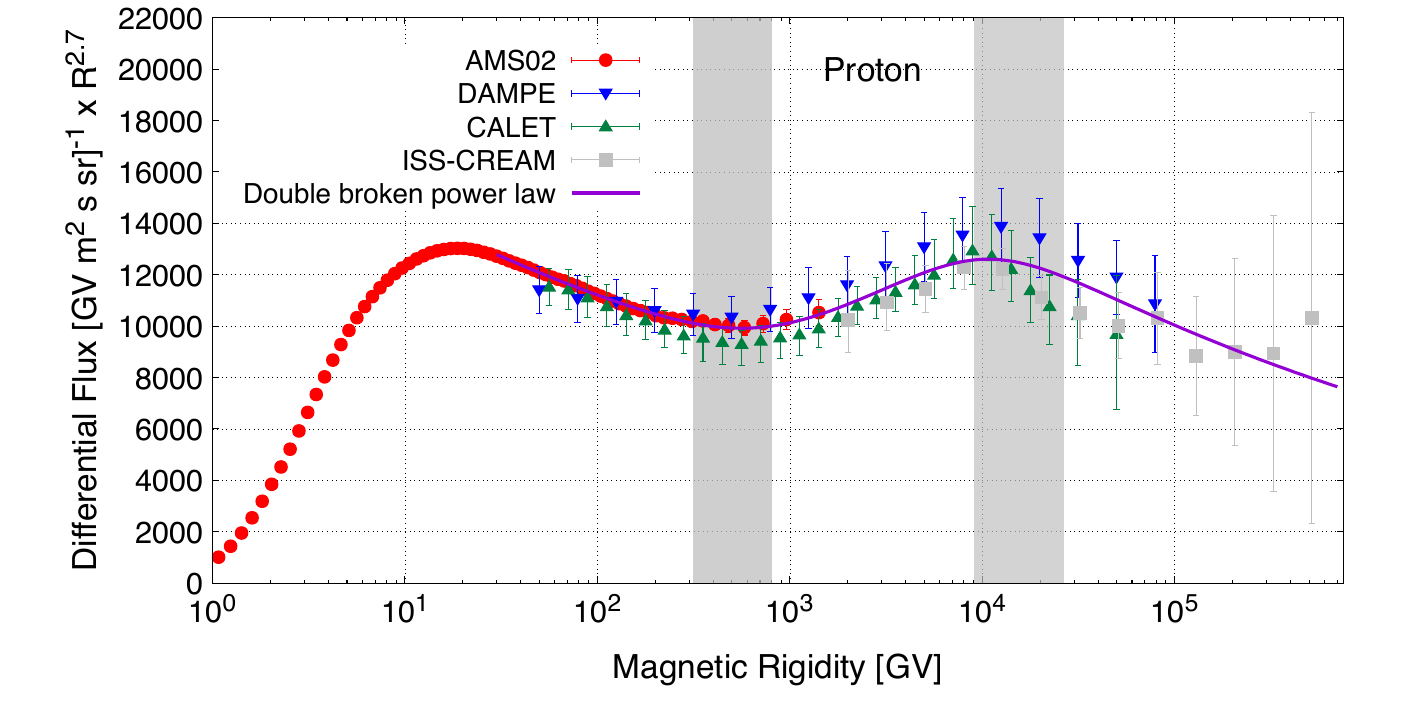} \hfill
\includegraphics[width=0.49\textwidth] {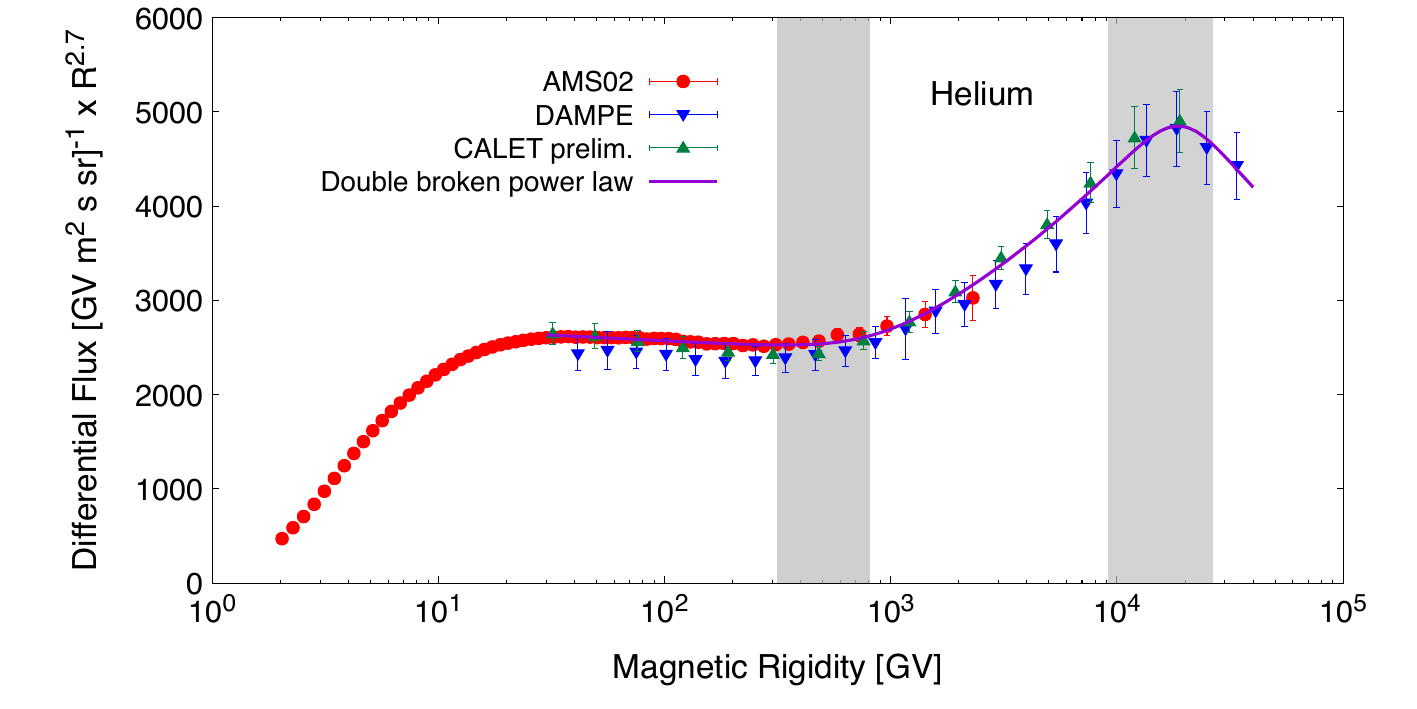}
\end{center}
\caption{\normalsize  The proton and helium differential spectra multiplied by $R^{2.7}$ as measured by the recent space experiments AMS-02 (full dots)~\cite{Aguilar_2021}, CALET (upward triangles)~\cite{Adriani_2022b,Brogi_2021},  DAMPE (downward triangles)~\cite{An_2019,Alemanno_2021} and ISS-CREAM (squares)~\cite{Choi_2022}. Conversion of kinetic energy to magnetic rigidity assumes dominance of the $^1\mathrm{H}$ and $^4\mathrm{He}$ isotopes. The error bars correspond to the quadratic sum of statistical and systematic errors. The full lines are the result of the fit to a double broken power law, Equation~\ref{equ_phe} and Table~\ref{tab_2break}. The vertical bands indicate the position of the two transition regions.}\label{fig_p_he_flux2}
\end{figure}

Few years later the AMS-02 magnetic spectrometer~\cite{Aguilar_2015a,Aguilar_2015b} measured the proton and helium spectra in the rigidity ranges $1\,\mathrm{GV}$ to $1.8\,\mathrm{TV}$ and $1.9\,\mathrm{GV}$ to $3\, \mathrm{TV}$, respectively, with unprecedented statistics and systematic accuracy. The calorimeters CALET~\cite{Adriani_2019,Brogi_2021} and DAMPE~\cite{An_2019,Alemanno_2021} extended the energy range, although with larger uncertainties. The results of modern spectrometric and calorimetric measurements of the differential spectra for these dominating elements are shown in Figure~\ref{fig_p_he_flux2}. The spectra are well described by a double broken power law~\cite{Lipari_2020,Kobayashi_2021,Vecchi_2021,Vecchi_2022}:
\begin{equation}\label{equ_phe}
\dtod{\Phi}{R} = c R^{-\gamma} \left[ 1+ \left( \frac{R}{R_l} \right)^{\Delta \gamma_l/s} \right]^s  \left[ 1+ \left( \frac{R}{R_h} \right)^{\Delta \gamma_h/s} \right]^{-s} 
\end{equation}
where $c$ is a normalisation constant. The first term with spectral index $\gamma$ describes a general power law. It is broken at two rigidities, $R_l$ and $R_h$, by the amounts $\Delta \gamma_l$ and $\Delta \gamma_h$, respectively, described by the two remaining terms. The parameter $s$ describes the smoothness of the two transitions; it is taken to be the same at both breaks, since the high rigidity data have little sensitivity to it. Parameters obtained from a fit of this function to all data in Figure~\ref{fig_p_he_flux2} are given in Table~\ref{tab_2break}, the result is shown in the figure. The fit starts at a rigidity of $30\, \mathrm{GV}$, chosen such that effects of solar modulation are negligible. The rigidity ranges where the two transitions occur are indicated in the figure by the vertical bands, a few hundred GV for the first and a few tens of TV for the second. CALET proton data with an extended energy range~\cite{Adriani_2022b} confirm the turn-over of the spectrum at high energies seen by DAMPE. The general spectral index for protons is slightly larger than the one for helium. This is due to an additional soft component in the proton spectrum which is clearly visible in Figure~\ref{fig_p_he_flux2}. The transitions in spectral index occur at about the same rigidities for protons and helium, and the indices change by about the same amount. Similar studies of the spectral shapes (see e.g.~\cite{Lipari_2020,Akaike_2024}) give similar results. 

\begin{table}
\begin{center}
\begin{tabular}{|c|rccccc|}
\hline
Element
& $c$  & $\gamma$ & $\Delta \gamma_l$ & $R_l\, [\mathrm{GV}]$ & $\Delta \gamma_h$ & $R_h\, [\mathrm{TV}]$\\
\hline
H
& $12777 \pm 29$  & $2.711 \pm 0.005$ & $0.263 \pm 0.072$ & $654 \pm 207$ & $-0.29 \pm 0.07$ & $10 \pm 5$ \\
He
& $  2634 \pm   8$  & $2.702 \pm 0.003$ & $0.252 \pm 0.020$ & $799 \pm   96$ & $-0.51 \pm 0.14$ & $19 \pm 3$ \\
\hline
\end{tabular}
\end{center}
\caption{Results of a fit of the proton and helium spectra with a power law and two breaks. The fits start at a rigidity of $30\,\mathrm{GV}$ to eliminate solar effects. The fits show little sensitivity to the smoothness parameter $s$, which is obtained as $s= 0.122 \pm 0.042$ for both proton and helium spectra.}\label{tab_2break}
\end{table} 

\begin{figure}[hbtp]
\begin{center}
\includegraphics[width=0.49\textwidth] {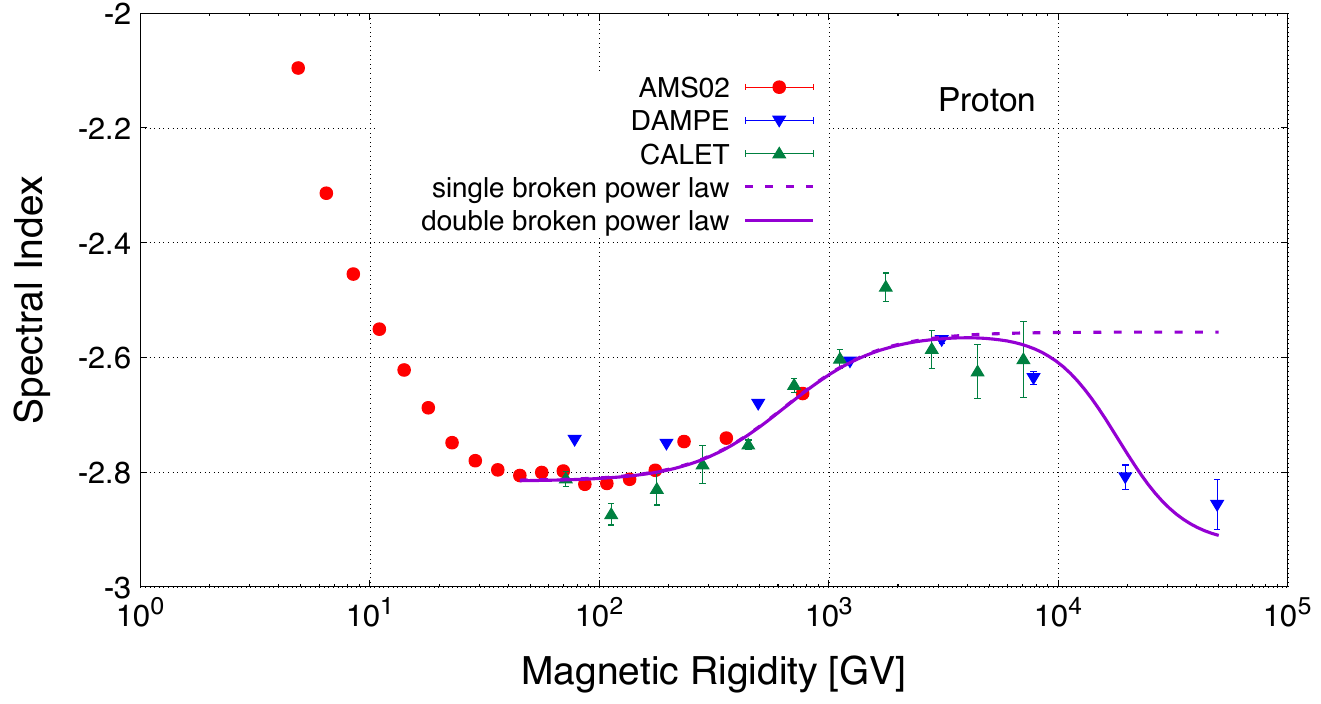} \hfill
\includegraphics[width=0.49\textwidth] {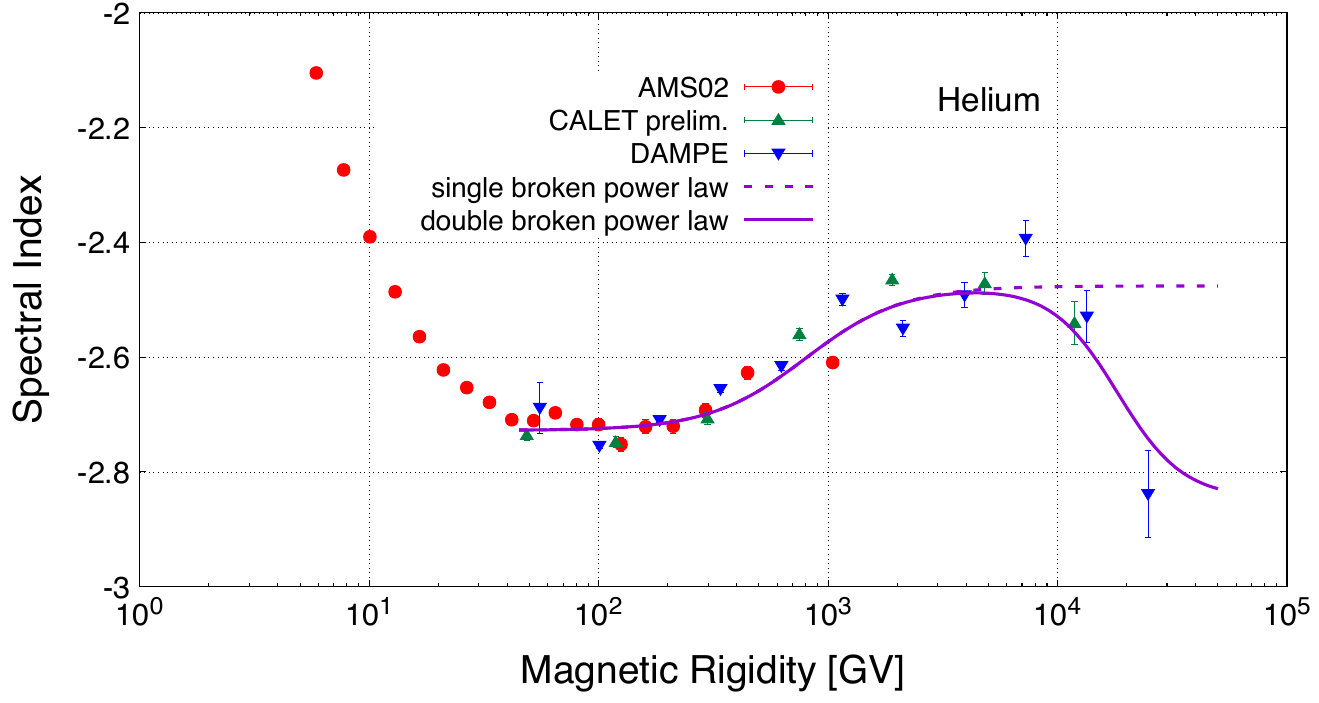}
\end{center}
\caption{\normalsize The spectral indices for proton and helium as a function of magnetic rigidity, derived from the data of the recent experiments AMS-02 (full dots)~\cite{Aguilar_2021}, CALET (upward triangles)~\cite{Adriani_2019,Brogi_2021} and DAMPE (downward triangles)~\cite{An_2019,Alemanno_2021}. The curves correspond to the indices $\gamma = d(\log \Phi)/d(\log R)$ for the broken power laws of Equation~\ref{equ_phe} with parameters of  Table~\ref{tab_2break}. Since the behaviour in the TV range is similar, we use the average of helium and proton parameters for the high-energy break. Conversion of kinetic energy to magnetic rigidity assumes dominance of the $^1\mathrm{H}$ and $^4\mathrm{He}$ isotopes.}\label{fig_p_he_index2}
\end{figure}

The AMS-02 data allowed the first detailed characterisation of the proton and helium spectral shapes with the rigidity dependence of their spectral indices, $\gamma = d(\log \Phi)/d(\log R)$. Calorimetric experiments allow to extend this approach to higher energies. The result is shown in Figure~\ref{fig_p_he_index2}. Rather than an abrupt spectral break, the proton spectrum exhibits a smooth progressive hardening as rigidity increases above $200\, \mathrm{GV}$. The proton spectral index progressively increases from $\gamma=-2.8$ below $200\, \mathrm{GV}$ to $\gamma = -2.6$ at about $1\, \mathrm{TV}$ and then back to about $-2.8$ around $20\, \mathrm{TV}$, in agreement with the fit results quoted above. The spectrum of helium nuclei also exhibits a smooth and progressive hardening above $200\, \mathrm{GV}$ but with an offset in spectral index of about $+0.1$ units. It starts out with a significantly harder spectrum at rigidities below $200\, \mathrm{GV}$ and reaches spectral indices similar to the proton values only above about $1\, \mathrm{TV}$. This convergence has first been observed by AMS through the proton-to-helium flux ratio $\Phi_{\mathrm{H}}/\Phi_{\mathrm{He}}$. The tendency for the two spectral indices to converge is also supported by higher energy data from DAMPE and CALET as shown in Figure~\ref{fig_p_he_index2}.

There are several astrophysical phenomena which may be invoked to explain the unexpected deviations from the simple consensus model~\cite{Serpico_2018,Gabici_2019}.  The source of the break can be searched in the injection or propagation phase of their life cycle.  An obvious way is to postulate multiple accelerating sites inside our galaxy with different characteristics and/or different distances to the Solar system~\cite{Vladimirov_2012}. A source origin of the break is described in models by introducing spectral breaks in the injected spectra or adding a high-energy population to the propagated primary spectra. An alternative is to consider a time dependence of the accelerating mechanism in supernovae remnants~\cite{Zhang_2020}. On the propagation side of the problem, it has been considered that there might be two different spacial zones, like e.g~the galactic disk and the galactic halo, where cosmic rays propagate differently~\cite{Tomassetti_2015}. They might lose different amounts of energy in these regions due to the density of the respective interstellar medium~\cite{Jaupart_2018}.  Alternatively one might consider the transition between different regimes in the diffusion process~\cite{Aloisio_2015}. In a phenomenological approach, a propagation effect is described introducing changes in the diffusion coefficient at appropriate rigidities~\cite{Genolini_2017,Vecchi_2022}. 

\subsection{Primary and secondary nuclei}\label{sec:primsec}

More information about the astrophysical origin of the breaks can be gained comparing the spectra of primary and secondary nuclei. Primary nuclei like helium, carbon or oxygen come to us from their sources mostly unaltered. Secondary nuclei like lithium, beryllium or boron, are dominantly generated by nuclear reaction of their primary progenitors with interstellar matter.   

\begin{figure}[h]
\begin{center}
\includegraphics[width=0.5\textwidth] {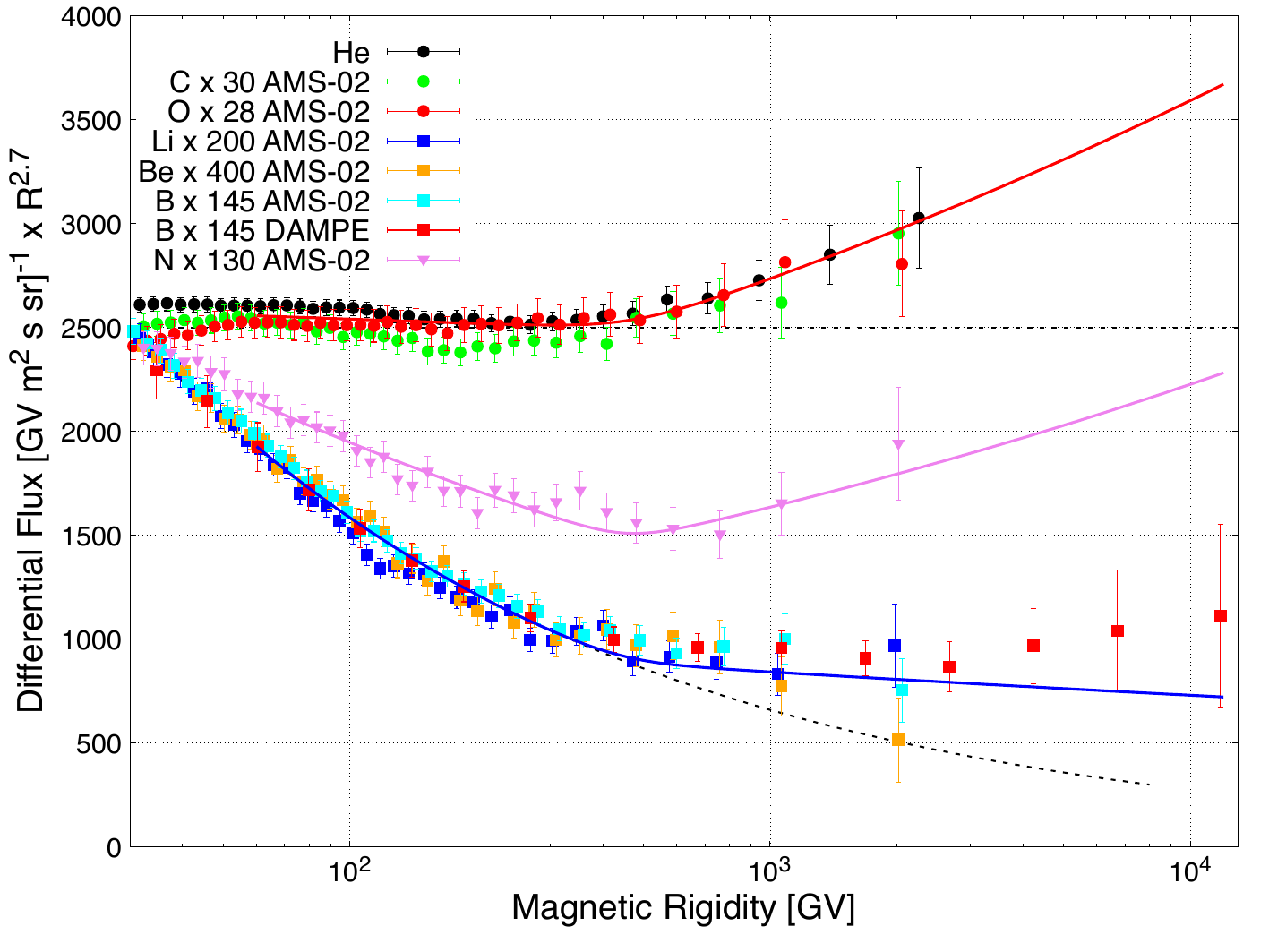}
\end{center} 
\caption{\normalsize Differential flux (times $R^{2.7}$) as a function of magnetic rigidity for primary nuclei (He, C, O) and secondary products of interactions with interstellar matter (Li, Be, B) as measured by AMS-02~\cite{Aguilar_2021,Aguilar_2023} and DAMPE~\cite{Alemanno_2024b}. Nitrogen (N) consists of a roughly 50:50 mixture of secondaries and primaries, so its spectral shape lies between the two. The fluxes are scaled so that they roughly match the helium flux at $30\,\mathrm{GV}$. The dashed lines indicate the dependency expected with a constant spectral index. The solid lines result from a fit to the spectra of the three groups with a broken power law and parameters from Table~\ref{tab_power}. The dashed line shows what the spectra would look like if the spectral indices were constant.}\label{fig_primsec}
\end{figure}

Helium nuclei are abundantly produced both in primordial and in stellar nucleosynthesis. Helium is also a very tightly bound nucleus and appears to receive no additional soft contribution in contrast to protons. Carbon and oxygen nuclei show roughly the same power spectrum as helium nuclei above $60\,\mathrm{GV}$~\cite{Aguilar_2017}, although their abundance is one order of magnitude lower. This fact is qualitatively demonstrated in Figure~\ref{fig_primsec} which compares the differential spectra of helium, carbon and oxygen, all measured by AMS-02, and roughly scaled to the helium flux at modest rigidities. A quantitative characterisation of their spectral shape is given in~\cite{Aguilar_2017}.  The CALET measurements of the carbon and oxygen spectra~\cite{Adriani_2020} confirm the spectral hardening observed by AMS-02, but do not agree with the absolute flux normalisation. 

Carbon and oxygen are the main progenitors of the light secondary nuclei lithium, beryllium and boron. The latter serve as raw materials and intermediate products in the breeding of heavier nuclei inside stars, and are therefore strongly underrepresented in stellar matter. They are found three orders of magnitude more often in cosmic rays, thanks to the splitting of heavier nuclei by spallation on the way from their sources to us. As secondary spallation products they have a significantly softer spectrum as shown in Figure~\ref{fig_primsec} with measurements from AMS-02. The light secondary Li, Be and B have very similar spectra above 30 GV, and exhibit a spectral break at a few hundred $\mathrm{GV}$, like their primary progenitors C and O, but with a spectral index change which is twice as large as for primaries.  Parameters are reported in Table~\ref{tab_power}, where a single spectral break $\Delta \gamma_l$ in the diffusion coefficient at rigidity $R_l$ is introduced, since the data do not cover the high energy break observed for protons and helium. 

One observes that all the three lightest primary nuclei -- He, C and O -- show compatible spectral indices and break parameters, as already obvious from Figure~\ref{fig_primsec}. The function corresponding to the average functional form of the three primary species is shown as the solid line overlapping He, C and O spectra in Figure~\ref{fig_primsec}. The dashed line shows what the spectra would look like if the spectral indices were constant. 

\begin{table}
\begin{center}
\begin{tabular}{|c|rcccc|}
\hline
Element
& $c$ $[\mathrm{m}^{-2}\mathrm{sr}^{-1}\mathrm{s}^{-1}\mathrm{GV}^{-1}]$ & $\gamma$ & $\Delta \gamma_l$ & $R_l\, [\mathrm{GV}]$ & $s$ \\
\hline
He & $(907.7 \pm 2.2) \times 10^{-4}$   & $2.728 \pm 0.003$ & $0.126 \pm 0.013$ & $364 \pm 28$  & $0.026 \pm 0.011$ \\
C   & $(29.61 \pm 0.26) \times 10^{-4}$ & $2.758 \pm 0.015$ & $0.150 \pm 0.028$ & $234 \pm 27$  & $0.035 \pm 0.027$ \\
O   & $(31.07 \pm 0.18) \times 10^{-4}$ & $2.712 \pm 0.014$ & $0.096 \pm 0.042$ & $354 \pm 118$ & $0.040 \pm 0.050$ \\
\hline
$\langle\mathrm{He,C,O}\rangle $ &     & $2.712 \pm 0.002$ & $0.131 \pm 0.011$ & $459 \pm 26$ & $0.023 \pm 0.009$ \\
\hline
Li  & $(3.63 \pm 0.06) \times 10^{-4}$   & $3.09 \pm 0.02$     & $0.43 \pm 0.09$  & & \\
Be & $(1.88 \pm 0.03) \times 10^{-4}$   & $3.09 \pm 0.02$     & $0.23 \pm 0.11$  & & \\
B   & $(5.06 \pm 0.05) \times 10^{-4}$   & $3.06 \pm 0.01$     & $0.35 \pm 0.05$  & & \\
\hline
$\langle\mathrm{Li,Be,B}\rangle $ &      & $3.08 \pm 0.01$     & $0.31 \pm 0.05$ & & \\
\hline
N    & $(5.96 \pm 0.06) \times 10^{-4}$  & $2.88 \pm 0.01$     & $0.32 \pm 0.05$  & & \\
\hline
Ne  & $(5.20 \pm 0.06) \times 10^{-4}$   & $2.75 \pm 0.01$     & $0.21 \pm 0.03$  & & \\
Si   & $(5.29 \pm 2.41) \times 10^{-4}$   & $2.71 \pm 0.20$     & $0.01 \pm 0.41$  & & \\
Mg & $(6.23 \pm 0.04) \times 10^{-4}$   & $2.74 \pm 0.01$     & $0.12 \pm 0.03$  & & \\
\hline
$\langle\mathrm{Ne,Si,Mg}\rangle$ &     & $2.74 \pm 0.01$     & $0.17 \pm 0.02$ & & \\
\hline
\end{tabular}
\end{center}
\caption{Parameters of a power law with a single break, fit  to the AMS-02 spectra of primary light nuclei  (He, C, O) and mostly secondary nuclei ( Li, Be, B) for $45\, \mathrm{GV} < R < 3 \, \mathrm{TV}$~\cite{Aguilar_2021,Aguilar_2023}. Note that the spectra of secondaries have little sensitivity to $R_l$ and $s$, we therefore fixed them to the values obtained for $\langle\mathrm{He,C,O}\rangle$. The results of DAMPE for B, $R_l = (391 \pm 53)\,\mathrm{GV}$, $\gamma= 3.02 \pm 0.01$, $\Delta \gamma_l = 0.31 \pm 0.05$ agree with the averages quoted. Among the heavier nuclei, N is of mixed primary and secondary origin. Ne, Si and Mg are mostly of primary origin.}\label{tab_power}
\end{table}

The three lightest secondary species -- Li, Be and B -- have also similar spectral shapes, which are however softer than the light primaries He, C and O as shown in Figure~\ref{fig_primsec}. This is expected since secondaries of a reaction naturally have a lower energy than primaries. Light secondaries also exhibit a spectral hardening at few hundreds of GV. The spectral hardening of light secondaries is about twice as important as for light primaries, as seen from the results of the individual spectra fits in Table~\ref{tab_power}. This is expected when the spectral break for all nuclei is due to a propagation effect where the diffusion coefficient changes at a given rigidity $R_l$. This change of spectral index will first apply to the primary spectra, then again to the secondary spectra. One would thus expect the spectral index change to be twice as large for secondaries  compared to primaries. Indeed this is what is observed.

Nitrogen is a good example of a species where neither primary nor secondary origin dominates. It is an important player in stellar nucleosynthesis. As an odd-odd nucleus, it is about a factor three to four less abundant than its neighbours carbon and oxygen, but equally so in stellar matter and in cosmic rays. One can thus conjecture that it represents a mixture of primary and secondary origins. And indeed, its spectral shape lies in between those of our prototype primary and secondary nuclei -- oxygen and boron -- as seen in Figure~\ref{fig_primsec} and Table~\ref{tab_power}. 

The measured cosmic ray spectra are used to tune and check analytical and numerical codes connecting spectra at the sources, where cosmic rays are injected into the interstellar medium of the Milk Way, to the spectra observed inside the heliosphere. An example is the GALPROP code~\cite{Strong_2007}, initiated by Igor Moskalenko and Andrew Strong in the 1990s~\cite{Moskalenko_1997,Strong_1998} and ever since continuously updated~\cite{Hanasz_2021, Boschini_2020, Boschini_2020a} (for the current status and results see \url{https://galprop.stanford.edu}). In order to directly compare the local interstellar spectra to the measured spectra near Earth, one needs to take into account the effect of solar modulation, as in the GALPROP-HelMod framework~\cite{Boschini_2017}. The solar modulation is obtained comparing cosmic ray spectra measured near-Earth with measurements in the local interstellar medium by the Voyager spacecraft~\cite{Cummings_2016}.

In particular, the GALPROP-HelMod authors have tested the two hypotheses mentioned before, the injection and the propagation scenario~\cite{Boschini_2020,Boschini_2020a}. They have randomly sampled the parameter space of injection and propagation to come up with an optimum solution. It turns out that both scenarios -- injection with variable spectral index and propagation with variable diffusion coefficients -- can fit the data. Other authors have invoked an injection scenario which explains the spectral indices change for all nuclei as transition from many distant sources to nearby sources with harder spectra~\cite{Lagutin_2021}. In this particular scenario, secondaries are accelerated together with primaries at the sources, and the classical diffusion coefficient  gets extra power law terms to describe the high inhomogeneities of the interstellar medium.

The fact that the spectral index change at few hundred GV is twice as large for secondaries compared to primaries strongly favours a transport effect as the origin of this spectral break~\cite{Serpico_2018}. The diffusion through the galactic disk could, for example, run differently than in the halo of the Milky Way~\cite{Tomassetti_2012}. Other spatial inhomogeneities in cosmic ray diffusion can also be invoked~\cite{Boschini_2017,Boschini_2018}, since heavier particles propagate shorter distances~\cite{Johannesson_2016}. Work is being done in the direction of performing numerical simulations where the cosmic ray equations of motion are solved in a realisation of the turbulent magnetic field, see reference~\cite{Mertsch_2020} for a review on test particle simulations of cosmic rays including practical examples.

Recently, a local reacceleration-propagation model has been proposed by Malkov and Moskalenko~\cite{Malkov_2021, Malkov_2022} to explain the spectral hardening at few hundred GV observed so far for all nuclei up to silicon and iron, and the TV spectral softening observed on protons and helium nuclei, which would together make a broad bump between $500\,\mathrm{GV}$ and $50\,\mathrm{TV}$. In this model, injection spectra are still featureless power laws, while a local shock generates the bump accelerating cosmic rays as it passes through the local interstellar medium. Particles below $500\,\mathrm{GV}$ are swept away with the interstellar medium, thus they could not reach the solar system. This creates the spectral index change for all nuclei, both primaries and secondaries. The shock parameters are derived using exclusively the proton data from AMS-02, CALET and DAMPE, and reproduce the spectra of helium and carbon and the B/C ratio~\cite{Malkov_2022}. The possible origin of the local shock could be a passing star, such as Epsilon Indi or Epsilon Eridani. In this scenario, the position of the spectral break at few hundred GV is expected to change on a time-scale of years as the shock gets farther away from the Sun~\cite{Malkov_2021}. 

Nuclei heavier than oxygen also contribute information on cosmic ray origin, acceleration and propagation, complementary to the one coming from light nuclei.  In the mass range from oxygen to iron several mostly secondary species are found, such as fluorine and the sub-iron elements, like scandium and vanadium, as well as the radioactive isotopes $ \phantom{}^{26}\mathrm{Al}$, $ \phantom{}^{36}\mathrm{Cl}$, and $ \phantom{}^{54}\mathrm{Mn}$. While silicon and iron nuclei are almost pure primary nuclei, other species, such as neon, magnesium, sulphur or calcium, have also a sizeable contribution from spallation of heavier nuclei with different fractions. Very high-mass nuclei, like iron, are expected to travel smaller distances in the galaxy because of their increased fragmentation cross section. They are expected to probe the local interstellar medium, rather than the entire galaxy, as do lighter nuclei~\cite{Johannesson_2016}. 

The abundances of cosmic ray nuclei heavier than oxygen are strongly suppressed. The three mostly primary nuclei nearest in mass -- Ne, Mg and Si -- are about a factor of seven rarer than oxygen. A further drop in abundance is observed beyond silicon nuclei. Sulphur, the most abundant nucleus in the mass range between silicon and iron, is a factor of five less abundant than silicon. Other nuclei in this mass range either have similar abundances, like argon and calcium, or are even more suppressed because they are mostly secondaries, like the nuclei from scandium to manganese. Iron, the most abundant primary nucleus heavier than silicon, is about three times less abundant than silicon. The reason is in the nucleosynthesis processes in stars for which the production of progressively heavier elements becomes more energy-intensive or requires the realisation of special temperature and pressure conditions in the core of the star. Accurate measurements thus require a large acceptance detector with a long exposure time to overcome their low abundance. In addition, the ability to estimate the nuclear fragmentation occurring inside the detector is required. 

\begin{figure}[!htb]
\begin{center}
\includegraphics[width=0.5\textwidth] {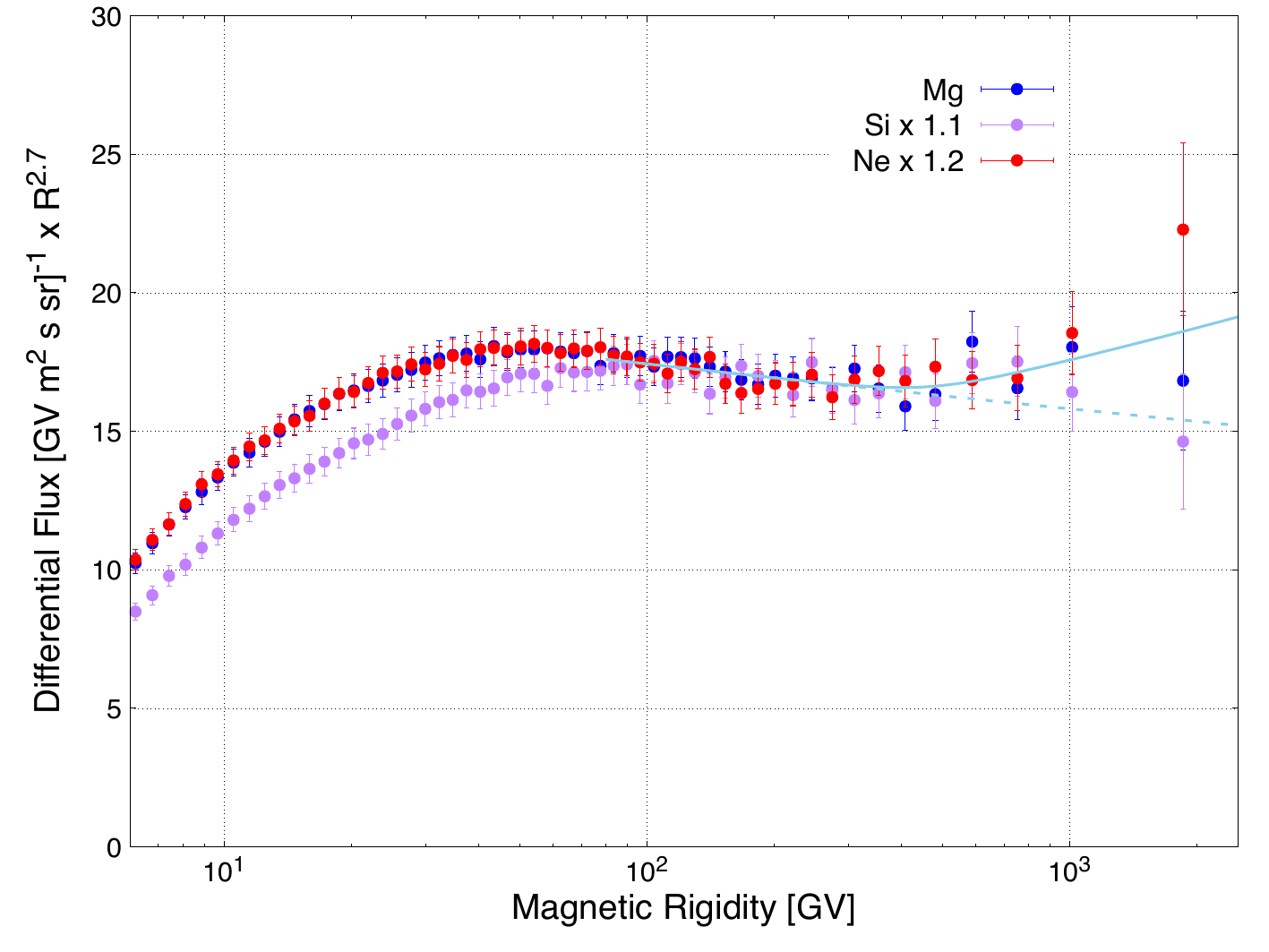}
\end{center} 
\caption{\normalsize  Differential fluxes of medium-mass primary nuclei, neon, magnesium and silicon as a function of rigidity measured by AMS-02~\cite{Aguilar_2020}, and multiplied by $R^{2.7}$. The three spectra follow the same rigidity dependence above $\sim 90\,\mathrm{GV}$. A spectral hardening is observed above $200\,\mathrm{GV}$. The solid lines result from a fit to the spectra of the three groups with a broken power law as used in~\cite{Aguilar_2020}, with parameters given in Table~\ref{tab_power}. The dashed line show the spectral shape expected for constant spectral indices.}
\label{fig_nemgsi}
\end{figure}

Neon, magnesium and silicon are mostly primary nuclei and share the same spectral shape at rigidities above $86.5\,\mathrm{GV}$ as shown in Figure~\ref{fig_nemgsi}. The neon and magnesium spectra are already identical from $3.65\,\mathrm{GV}$ onward, most likely because they have a sizeable secondary contribution from spallation while silicon nuclei are predominantly primary. The spectral indices of Ne, Mg and Si follow the same rigidity dependence with the same hardening above $\sim 200\,\mathrm{GV}$. However, comparison with the oxygen spectrum reveals differences. Heavy primary nuclei -- Ne, Mg and Si -- have a different spectral shape than light primary nuclei -- He, C and O -- exhibiting a slightly weaker spectral hardening with respect to the light primaries as shown in Table~\ref{tab_power}. The observed difference between the spectra of Ne-Mg-Si and He-C-O has been interpreted in the framework of the injection scenario, invoking either a local source or the superposition of several sources~\cite{Yuan_2020,Lagutin_2021,Niu_2022}.

\begin{figure}[!htb]
\begin{center}
\includegraphics[width=0.5\textwidth] {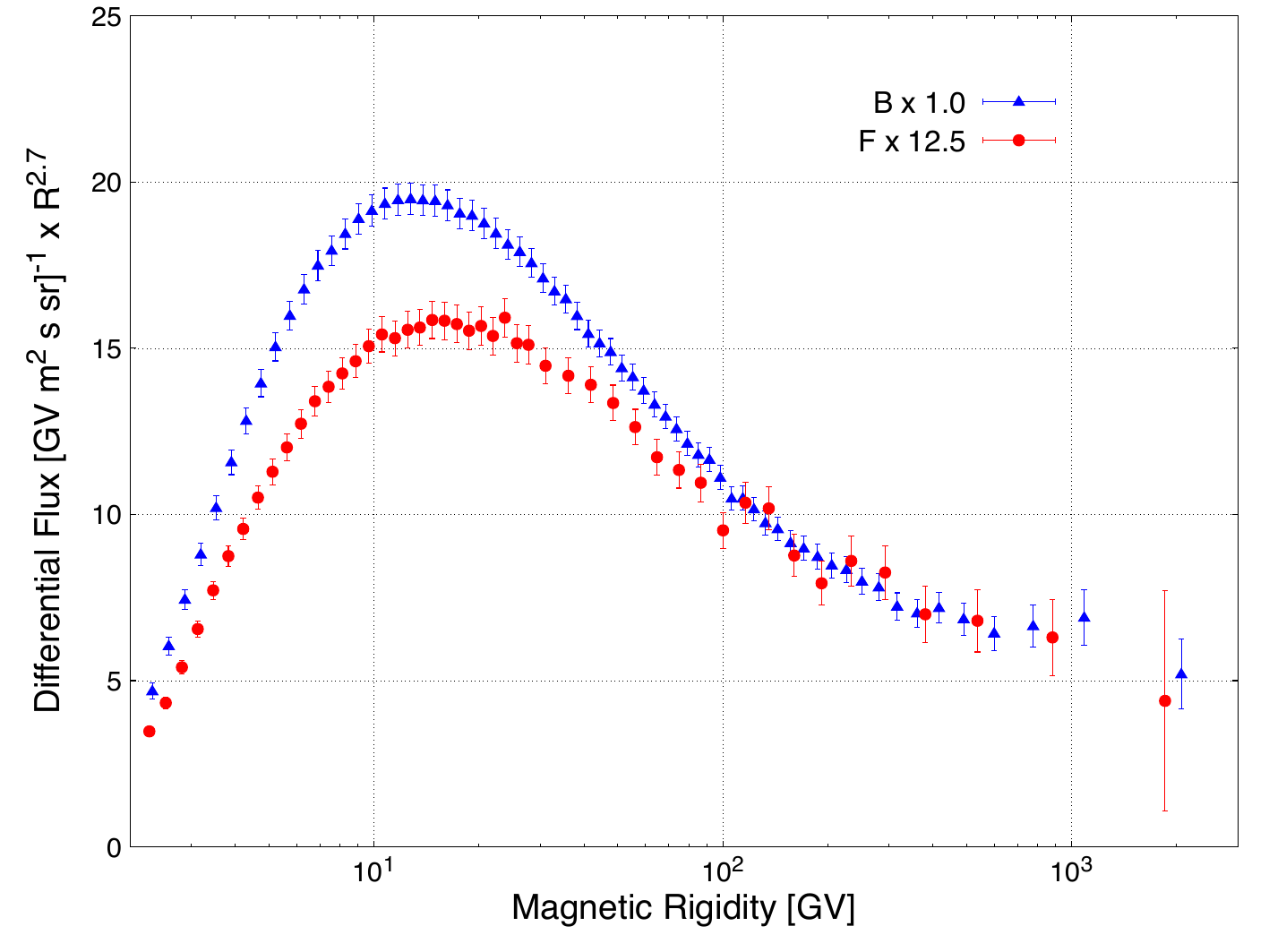}
\end{center} 
\caption{\normalsize  Differential fluxes of medium-mass secondary fluorine nuclei (F) compared to the flux of the light secondary boron nuclei (B) as measured by AMS-02~\cite{Aguilar_2023}. The fluxes have been multiplied by $R^{2.7}$ and rescaled by the factors indicated in the picture. The two spectral shapes are the same only at rigidities above $\sim 150\,\mathrm{GV}$.}
\label{fig_f_b}
\end{figure}

Neon, magnesium and silicon are the progenitors of the medium-mass secondary fluorine nuclei. In the mass range between oxygen and silicon, fluorine is less abundant than silicon by a factor of seven and it is the only one considered of pure secondary origin. Recent results from the AMS-02 magnetic spectrometer~\cite{Aguilar_2021d} over the rigidity range from $2.15\,\mathrm{GV}$ to $2.9\,\mathrm{TV}$ with a total uncertainty of 5.9\% at $100\,\mathrm{GV}$ have characterised the fluorine spectrum in detail.. The measurement of its spectral index as a function of rigidity shows that it also hardens above $200\,\mathrm{GV}$ as observed for all the other nuclei. However, when compared to the light secondary nuclei, for which we use the boron spectrum as template, the fluorine spectrum does not follow the same rigidity dependence below $150\,\mathrm{GV}$ as shown in Figure~\ref{fig_f_b}.
 
Differences in spectral shape indicate that particles traverse different distances. The GALPROP-HelMod team estimates, that the effective propagation distance  for silicon is about 10\% smaller than that of oxygen and neon~\cite{Boschini_2022}, which are expected to probe similar galactic volumes. A different distribution of sources for light and medium mass cosmic rays is thus not excluded. However, a similar combined study of the light and medium-mass primaries and secondaries using the USINE propagation tool~\cite{Vecchi_2021} comes to a different conclusion~\cite{Bueno_2022}. This underlines the importance of understanding cosmic ray propagation inside the Milky Way.

\begin{figure}[!htb]
\begin{center}
\includegraphics[width=0.5\textwidth] {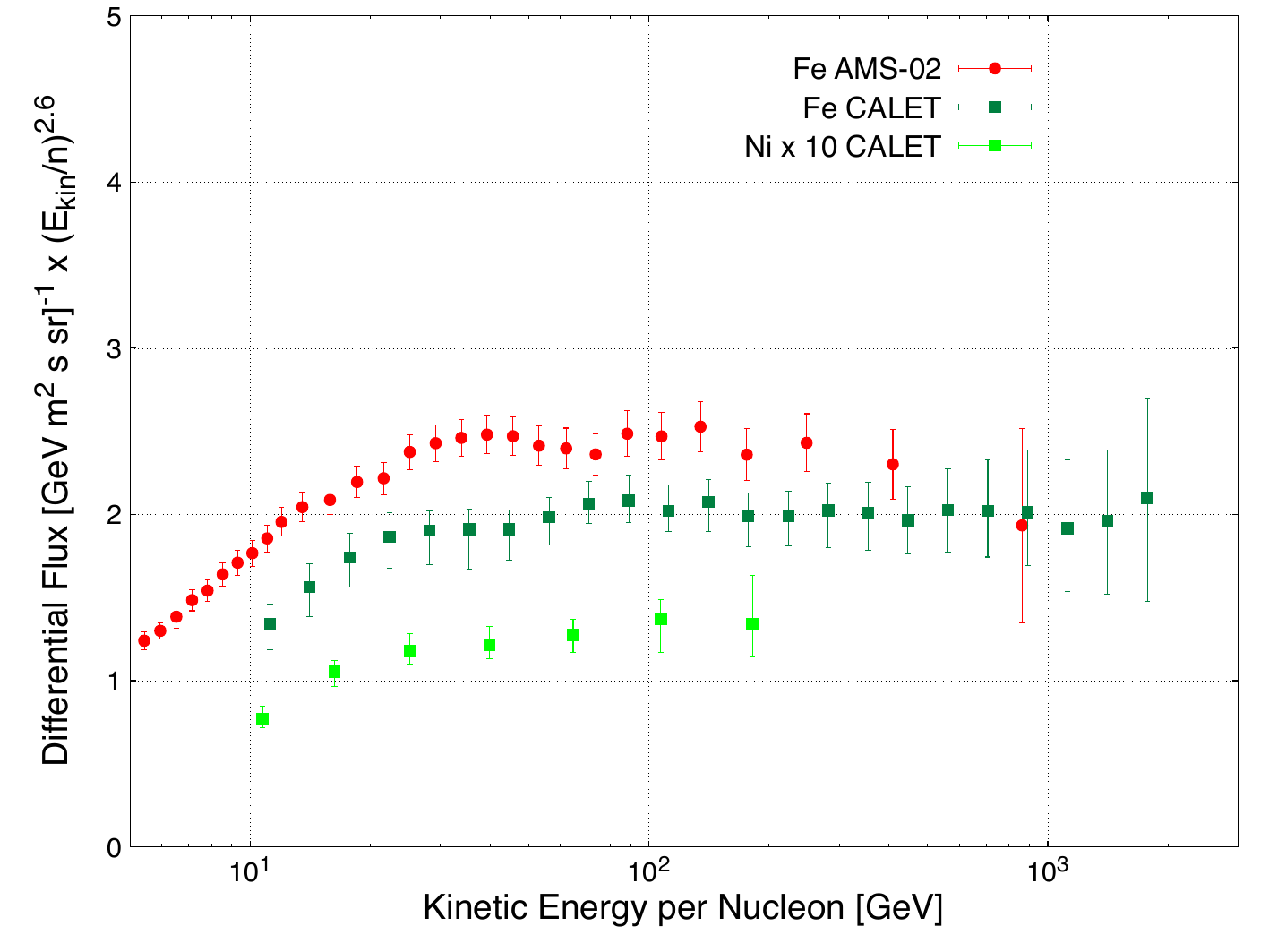}
\end{center} 
\caption{\normalsize Iron differential flux as measured by AMS-02~\cite{Aguilar_2021a} and CALET~\cite{Adriani_2021} as a function of kinetic energy per nucleon and multiplied by $(E_{kin}/\mathrm{n})^{2.6}$. The AMS-02 measurement has been converted from rigidity to kinetic energy per nucleon assuming iron is all $\phantom{}^{56}\mathrm{Fe}$. The CALET and AMS-02 measurements agree in shape but have a 20\% difference in absolute normalisation~\cite{Adriani_2021}. Also shown is the measurement of the nickel differential flux by CALET~\cite{Adriani_2022a}, multiplied by a factor of 10.}
\label{fig_feni_ams_calet}
\end{figure}

Cosmic ray iron nuclei are the only mostly primary nuclei beyond silicon and are found in a ratio of 1:3 with respect to silicon. Because of their high mass ($A=56$), their effective propagation distance is about 30\% smaller than that of oxygen. Therefore iron nuclei probe the local interstellar medium. The AMS-02 spectrometer~\cite{Aguilar_2021a} and the CALET calorimeter~\cite{Adriani_2021} have provided accurate measurements of the iron differential flux shown in Figure~\ref{fig_feni_ams_calet}. The measurements agree in shape but not in normalisation: CALET finds a 20\% lower flux than AMS-02. A possible source of this discrepancy can be the hadronic interaction model, systematics in the energy scale determination or the trigger efficiency.

The spectral shape of the Fe flux, when compared to other primary nuclei, follows the shape of light primaries~\cite{Aguilar_2021a}. The data have been analysed together with results from Voyager-1 and ACE-CRIS by the GALPROP-HelMod team to update their calculation of the local interstellar spectrum of iron in the range $1\,\mathrm{MeV/n}$ to $10\,\mathrm{TeV/n}$. They find an excess of iron at low energy hinting to a past supernova activity in the solar neighbourhood~\cite{Boschini_2021}. 

For nuclei beyond iron, the production mechanism changes radically from fusion reactions to neutron bombardment in explosive events. The spectra of these very heavy nuclei are beginning to be assessed by current high precision experiments~\cite{Aguilar_2021,Adriani_2022a,Akaike_2024}.  

\subsection{Nuclear antimatter}\label{sec:anti}
Antiparticles are excellent cosmic ray species to look for unconventional sources. This is because their production by conventional means is strongly suppressed, since antimatter does not appear to be present in stars. Thus antiparticles are extraordinarily rare in cosmic rays. The antiproton flux is suppressed with respect to the proton flux by four orders of magnitude. No observation of heavier antinuclei has so far been claimed by experiments.  Thus contributions from unconventional sources can be visible in antiparticle spectra, even when they are drowned in the spectra of dominant matter species by conventional contributions. 

Prominent among the potential non-conventional sources of antinuclei (and positrons, as we will see in Section~\ref{sec:elpos}) is the annihilation or decay of dark matter particles. The existence of dark matter has been put in firm evidence as discussed in the section of this encyclopedia covering phenomena beyond the standard model. Despite the fact that dark matter makes up about 85\% of the total matter in the Universe, the nature of this non-luminous form of matter still remains unknown. Among possible dark matter candidates are weakly interacting massive particles, which we will concentrate on here. If they exist, their masses may range from GeV to TeV.

The matter-antimatter asymmetry is also a long-standing issue in the understanding of our Universe. It is thought that at the time of the Big Bang matter particles (quarks and leptons) and antimatter particles (anti-quarks and anti-leptons) were produced in equal amounts. However, the world we observe today is made by matter particles, i.e.~atoms with positively charged nuclei surrounded by electrons. Anti-hydrogen atoms were synthesised at CERN for the first time more than 30 years ago~\cite{Baur_1997}. Today, they are used to systematically study possible differences between matter and antimatter~\cite{Charlton_2020}. Another experimental approach to address the issue of the apparent matter-antimatter asymmetry in nature is to search for leftovers of primordial anti-nuclei -- relics from the Big Bang -- in cosmic rays.

Accurate measurements of antiproton spectra require a magnetic spectrometer which reliably determines the sign of the particle charge and has a high rejection power against the more abundant protons and electrons. The AMS-02 detector indeed has the required capabilities. Confusion between positive and negative charges is at the level of $10^{-4}$ for rigidities up to $100\, \mathrm{GV}$ and less than 8\% up to $1\, \mathrm{TV}$~\cite[Sec. 1.2.4]{Aguilar_2021}. And antiprotons are distinguished from the like-sign, but much lighter electrons at a level of one in $10^4$. At lower energies, the BESS Polar II balloon mission~\cite{Abe_2008} and the PAMELA space spectrometer~\cite{Adriani_2009a,Adriani_2010} pioneered the study of cosmic ray antiprotons. 

\begin{figure}[t]
\begin{center}
\includegraphics[width=0.5\textwidth] {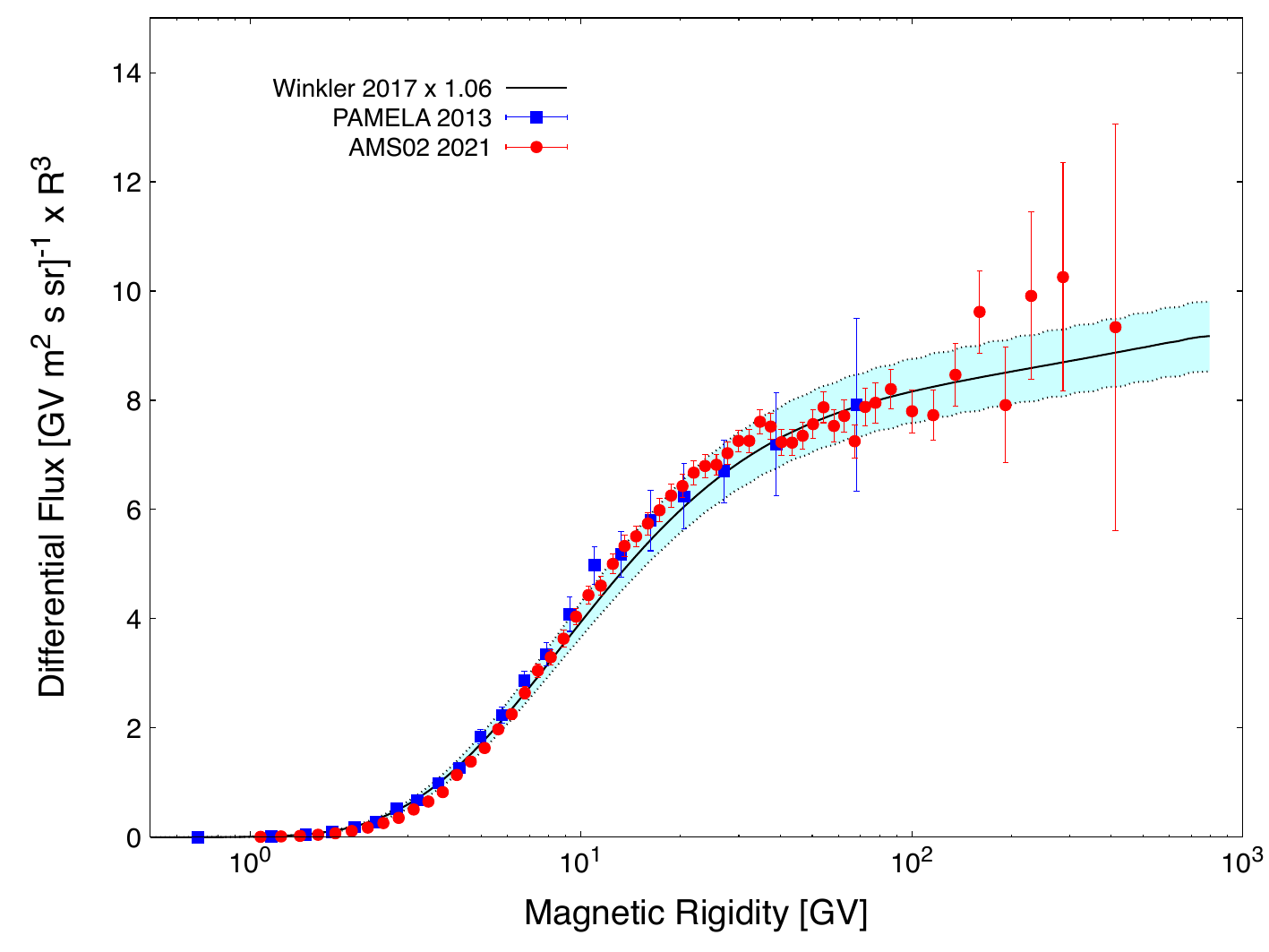}
\end{center} 
\caption{\normalsize Differential flux of antiprotons ($\times R^3$) measured by PAMELA~\cite{Adriani_2013c} and AMS-02~\cite{Aguilar_2021}, compared to a calculation based on secondary production only~\cite{Winkler_2017}. The shaded band corresponds to the uncertainty from the cosmic ray propagation model and the antiproton production cross sections.}\label{fig_pbar}
\end{figure}

Figure~\ref{fig_pbar} shows the antiproton spectra measured by the space observatories PAMELA and AMS-02, compared to a recent calculation~\cite{Winkler_2017} assuming a purely secondary origin of antiprotons. The agreement is satisfactory, given the uncertainties from propagation parameters and antiproton production cross sections represented by the error band of the calculation result. It is also clear that the data leave little room for contributions to the flux other than conventional processes. Interaction of matter particles with the interstellar medium appears to produce the observed antiprotons in the energy range currently studied. The conclusion that suspected dark matter signals are insignificant in this channel is also reached in other studies~\cite{Kahlhoefer_2021,DiMauro_2021,Calore_2022} and~\cite[Sec. 5]{Aguilar_2021}. Rather stringent upper limits on the dark matter annihilation rate $\langle \sigma v \rangle$ are obtained for masses of dark matter particles between ten and a few hundred GeV.  

The search for heavier antinuclei, like antideuterium or antihelium, is particularly fascinating. They are extremely difficult to produce through interactions between matter particles. The probability to form them decreases by orders of magnitude with each additional anti\-nucleon~\cite{Kachelriess_2020}. There are, however, a good handful of antihelium candidates in the AMS-02 data among 200 billion cosmic particles. Sam Ting presented some of them in a 2018 CERN seminar.\footnote{See \texttt{https://cds.cern.ch/record/2320166}} The rate is roughly equivalent to one $\overline{\mathrm{He}}$ per year, or one per 100 million He nuclei. On the one hand, this is a very small rate; the systematic significance is difficult to assess because, in particular, the expected background has to be quantified. The fact that this has not happened so far, and that there is no publication, is probably also due to the tiny rate. To ensure that there is no detector malfunction at the level of 1 in 100 million, one needs to know more about the properties of these rare events. Thus the AMS-02 collaboration does not (yet?) claim to have discovered complex cosmic antinuclei.

On the other hand, if at least some of these candidates are taken seriously, there are too many to explain them by conventional nuclear astrophysics. Thus there should be a small, ready-made supply of antimatter somewhere in the galaxy which might have been left over from the Big Bang. That motivated a new search for anti-stars in the Milky Way, i.e.~stars made of antimatter. In the catalog of the Fermi satellite there are 14 objects whose photon spectra are compatible with antimatter~\cite{Dupourque_2021}, about 2.5 candidates per million normal stars at a distance between a few $10^{14}$ and $10^{16}$\,km. They could be considered as sources of $\overline{\mathrm{He}}$ if one knew the release and acceleration mechanism. According to the authors, however, it is more likely that they are normal $\gamma$-ray emitters such as pulsars or black holes.

\subsection{Electrons and positrons}\label{sec:elpos} 
Electrons and positrons are the lightest cosmic ray species. Electrons are components of ordinary matter and easily liberated by ionisation. The pair production of electrons and positrons has a low ``cost'' in energy, thus electromagnetic processes can more easily generate these than other matter-antimatter pairs. In fact, electrons and especially positrons 
have long been suspected to partially come from unidentified sources, already since the results of the PAMELA space spectrometer~\cite{Adriani_2009b,Adriani_2013b}. On the other hand, electrons and positrons suffer stronger energy losses during propagation through interstellar matter. At high energy they thus do not travel very far, astrophysical and unconventional sources -- if any -- must thus be close to the solar system, no farther than a kpc or so.

\begin{figure}[htb]
\begin{center}
\includegraphics[width=0.5\textwidth] {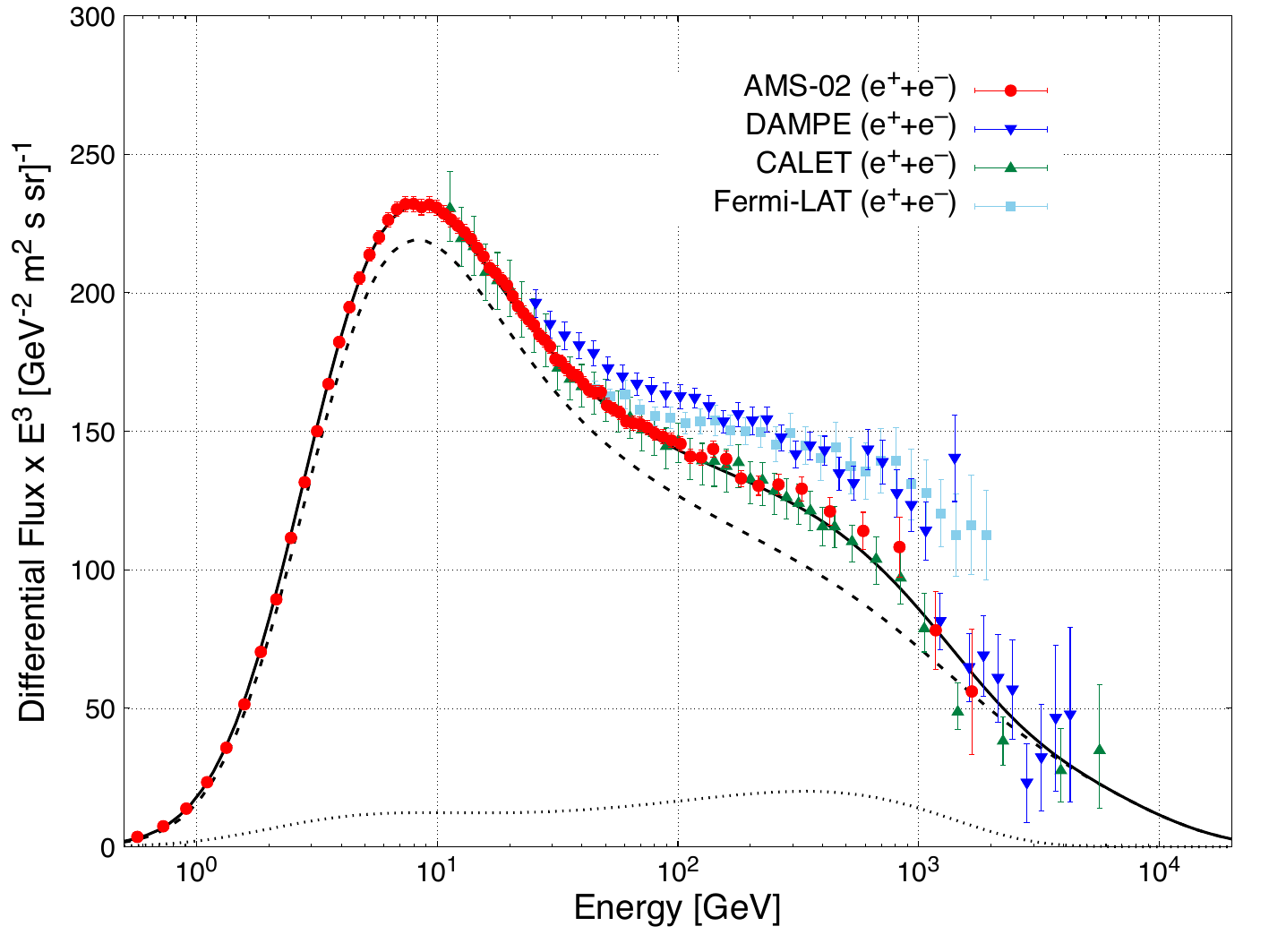}
\end{center} 
\caption{\normalsize The differential spectrum of the sum of electrons and positrons, $d\Phi_{\mathrm{e}^\pm}/dE$, scaled by $E^3$. Results of the AMS-02 spectrometer~\cite{Aguilar_2021} are shown with those of the calorimeters DAMPE~\cite{Ambrosi_2017}, CALET~\cite{Adriani_2023} and Fermi-LAT~\cite{Abdollahi_2017}. The solid curve is the sum of fits to the separate electron (dashed curve) and positron spectra (dotted curve) from AMS, imposing an overall cut-off at 8\,TeV.\label{fig_eplusminus}}
\end{figure}

The flux of the sum of electrons and positrons, $d\Phi_{\mathrm{e}^\pm}/dE$, can be measured by spectro\-meters and calorimeters alike. Figure~\ref{fig_eplusminus} shows recent results from the AMS-02 spectro\-meter~\cite{Aguilar_2021} and the DAMPE~\cite{Ambrosi_2017} and CALET calorimeters~\cite{Adriani_2023}. Results from the Fermi-LAT gamma ray detector are shown for comparison~\cite{Abdollahi_2017}. The differential flux is multiplied by the kinetic energy to the third power, $E^3$, since the spectrum is supposed to be softer than for nuclei due to the higher energy loss. The results of AMS-02 and CALET are in excellent agreement over the full range of energies, where the two measurements overlap. The results of DAMPE and Fermi-LAT deviate from these by more than the quoted systematics. The reason could in principle be sought in unidentified background from more abundant species or in uncertainties concerning the energy scale. Background is certainly not an issue for AMS-02, since the TRD signal, the comparison between calorimeter energy and spectrometer rigidity as well as the shower shape in the calorimeter give three handles for a very high background rejection power. The deeper calorimeters of CALET and DAMPE likewise provide excellent background rejection by shower shape alone. The energy scale of the space experiments, on the other hand, has an uncertainty of two to three percent~\cite{Aguilar_2021,Asaoka_2017,Zhang_2016,Abdollahi_2017,Zang_2017}. Using this margin, one can improve the agreement~\cite[Fig. 9.22]{Bindi_2023}.

The spectrum clearly has more than one component, as evident from the shoulder for energies beyond about thirty GeV. Before going into details about the properties of this joint spectrum, we look into the separate contributions of electrons and positron. Obviously only spectrometers are equipped to measure these separately.\footnote{An exception is an effort by the Fermi-LAT collaboration to use the Earth's magnetic field as a spectrometer magnet~\cite{Abdollahi_2017}.} 

\begin{figure}[htb]
\begin{center}
\includegraphics[width=0.5\textwidth]{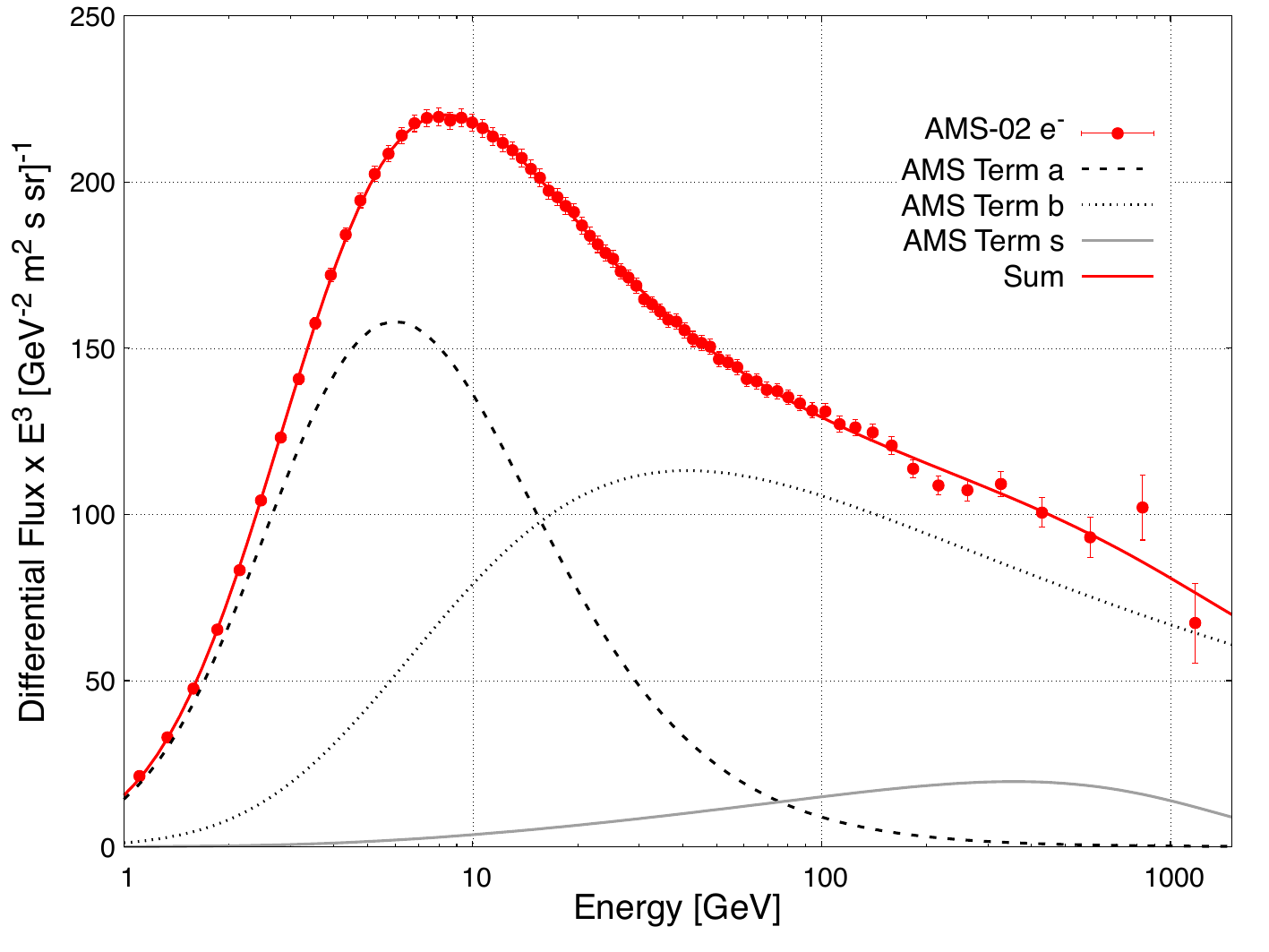}
\end{center}
\caption{\normalsize AMS-02 measurement of the differential flux of electrons as a function of the energy $E$, scaled with $E^3$~\cite{Aguilar_2021}. Error bars represent the total statistical and systematic errors. The solid curve is the result of a fit with three flux components, a soft (dashed) and a harder one (dotted). The third component (grey line) is fixed to be the same in shape and normalisation as the ``source'' term determined by the positron spectrum.}\label{fig_electron}
\end{figure}

Cosmic ray fluxes often suggest that more than one component contributes.\footnote{The proton and helium spectra can indeed also be described with a two-component model~\cite[Sec. 9.3.1]{Bindi_2023}. One can view the broken power law as an approximation of the two-component function for modest variations in spectral index.} Multiple contributions can be due to different particle sources -- e.g.~near or far ones -- or different astrophysical conditions at the sources. They can also arise from  different acceleration mechanisms at work or different diffusive conditions met during propagation. In the simplest case, such contributions just add. One can then describe the joint differential flux by the superposition of the components, e.g.~for electrons:
\begin{equation}
\dtod{\Phi_{\mathrm{e}^-}}{E}  =   c_a \dtod{\Phi_a}{E} + c_b \dtod{\Phi_b}{E} + \cdots
\end{equation}
The estimation of solar modulation, relevant at low energies, is done here in the force field approximation. This introduces an estimator $\hat{E} = E + \phi$ of the energy before particles enter the heliosphere, with the so-called solar potential $\phi$ dependent on the solar cycle and particle type. A model of the total power-law spectrum used by AMS reads~\cite{Kounine_2023}:
\begin{equation}\label{equ_electron}
\dtod{\Phi_{\mathrm{e}^-}}{E}  =  \left( \frac{E}{\hat{E}} \right)^2 \left[ c_a \left( \frac{\hat{E}}{E_a} \right)^{\gamma_1} + c_b \left( \frac{\hat{E}}{E_b} \right)^{\gamma_2} + c_s \left( \frac{\hat{E}}{E_s} \right)^{\gamma_s} e^{-\hat{E}/E_c}\right]
\end{equation}
This function is fitted to the observed electron spectrum by AMS-02 during its first seven years of exposure on the ISS~\cite{Aguilar_2021}. The constants $E_i$ have no physical significance, but can be chosen to minimise the correlation between the normalisations $c_i$ and the spectral indices $\gamma_i$. The differential flux is shown in Figure~\ref{fig_electron}, the fitted parameters are given in Table~\ref{tab_elpos}. There is a clear change in spectral index beyond an energy of about $30\, \mathrm{GeV}$. The total spectrum in the GeV to TeV energy range is well descried by such a multi-component fit. It consists of two power law components, a soft one dominant at low energies and a harder one apparent at high energies. Again, neither spectral index is equal to $-3$, as one would naively expect from  Fermi acceleration and diffusion alone. And numerical codes like GALPROP~\cite{Trotta_2011,Vladimirov_2012} do not match the spectral behaviour either. The origin can again be sought in different astrophysical mechanisms. One can e.g.~invoke separate sources for the two components, a break in the injection spectrum, a break in the diffusion coefficient, or a re-acceleration phenomenon. The third term with an exponential cut-off is inherited from the positron spectrum (see below). It is required by the shape of the positron spectrum, but the electron spectrum also admits such a contribution.

\begin{table}
\begin{center}
\begin{tabular}{|ccccc|}
\hline
$c_a$ 						& $\gamma_a$ 				& $c_b$  						& $\gamma_b$ & 	$\phi\, [\mathrm{GeV}]$ \\
\hline
$(1.80 \pm 0.03) \times 10^{-2}$	& $-4.65 \pm 0.03$ 				& $(3.33 \pm 0.02) \times 10^{-6}$	& $-3.23 \pm 0.10$ 	& $1.97 \pm 0.02$ \\
\hline
\hline
$c_d$ 						& $\gamma_d$ 				& $c_s$  						& $\gamma_s$ & 	$\phi\, [\mathrm{GeV}]$ 	\\
\hline
$(6.51 \pm 0.14) \times 10^{-2}$	& $-4.07 \pm 0.06$ 				& $(6.80 \pm 0.15) \times 10^{-5}$	& $-2.58 \pm 0.05$ 	& $1.10 \pm 0.03$ 		\\
\hline
\end{tabular}
\end{center}
\caption{Parameters of a multi-component fit which describes the electron and positron spectra measured by AMS-02~\cite{Bindi_2023}. Normalisation constants $c_i$ are in units of  $[\mathrm{m}^{-2}\mathrm{sr}^{-1}\mathrm{s}^{-1}\mathrm{GeV}^{-1}]$ and for $E_a = 20\, \mathrm{GeV}$, $E_b= 300\, \mathrm{GeV}$, $E_d = 7\, \mathrm{GeV}$ and $E_s = 60\, \mathrm{GeV}$. Note that the fit to the electron spectrum contains the source term with parameters fixed to the positron fit result.}\label{tab_elpos}
\end{table}

\begin{figure}[htb]
\begin{center}
\includegraphics[width=0.5\textwidth]{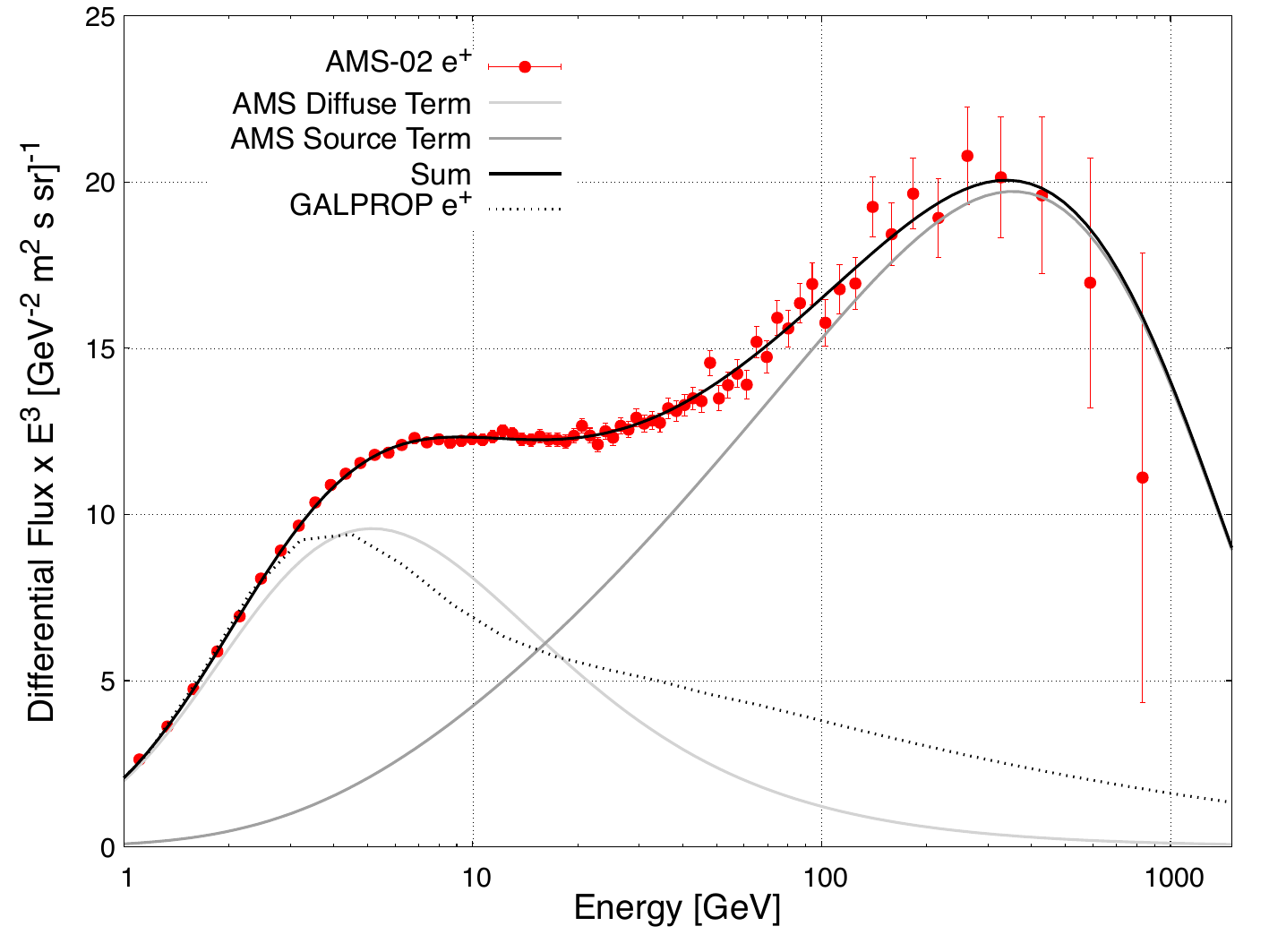}
\end{center}
\caption{\normalsize AMS-02 measurement of the differential flux of positrons as a function of the energy $E$, scaled with $E^3$~\cite{Aguilar_2021}. Error bars represent the total statistical and systematic errors. The black curve is the result of a fit with two contributions: the diffuse term (light grey) represents a secondary production of positrons through the interaction of cosmic rays with interstellar matter; the source term (dark grey) corresponds to primary production by a new, not yet identified source. The dotted curve is the result of GALPROP~\cite{Trotta_2011} for secondary positrons, similar to the fitted diffuse term.}\label{fig_positron}
\end{figure}

The spectrum of positrons measured by AMS in Figure~\ref{fig_positron} likewise shows a deviations from prejudice. A similar decomposition of the flux as for electrons applies:
\begin{equation}\label{equ_positron}
\dtod{\Phi_{\mathrm{e}^+}}{E}  =   \left( \frac{E}{\hat{E}} \right)^2 \left[ c_d \left( \frac{\hat{E}}{E_d} \right)^{\gamma_d} +  c_s \left( \frac{\hat{E}}{E_s} \right)^{\gamma_s} e^{-\hat{E}/E_c} \right]
\end{equation}
with an additional exponential cut-off at energy $E_s$ which applies to the second component. The indices chosen by AMS suggestively indicate a ``diffuse'' and a ``source'' term. The contribution of the ``diffuse'' term indeed comes out qualitatively similar to the GALPROP result~\cite{Trotta_2011} for secondary positrons shown in Figure~\ref{fig_positron}.  It dominates the spectrum at low energies. At energies beyond about $20\, \mathrm{GeV}$, a new component with a much harder spectrum becomes apparent. The curve in Figure~\ref{fig_positron} corresponds to Equation~\ref{equ_positron}, with parameters given in Table~\ref{tab_elpos}. The cut-off energy of the ``source'' term comes out as $1/E_c = (1.23 \pm 0.34)/\mathrm{TeV}$ or $E_c = 810^{+310}_{-180}\, \mathrm{GeV}$~\cite{Aguilar_2021}. The AMS electron spectrum does not require such a contribution, but allows to add one~\cite{Kounine_2023}, as has been done in Equation~\ref{equ_electron}. Thus, the ``source'' term can contribute symmetrically to electron and positron spectra. The sum of electron and positron spectra of Figure~\ref{fig_eplusminus} is well described by the sum of the fitted electron and positron spectra quoted above, but it does appear to require an overall cut-off at $(8 \pm 2)\, \mathrm{TeV}$. 

Since positrons are not present in normal matter, one would expect them to be of secondary nature, produced by electromagnetic phenomena. Not only interactions between ordinary cosmic rays and interstellar matter, but also environments with exceptionally high electromagnetic fields can be a source of positrons without invoking any unconventional physics. An astrophysical example for electromagnetic production of electron-positron pairs is the rapidly changing magnetic fields of pulsars. These are rotating neutron stars whose rotational axis does not coincide with the direction of the magnetic axis. This leads to strong fields in some regions, which can be suitable for generating high-energy electron-positron pairs. Single pulsars tend to produce a spectrum which is more peaked than the AMS measurement and the fit results. There is also no reason why a single pulsar should be the source of additional positrons. Since positrons easily disappear in interaction with interstellar matter, point sources should not be located too far from the solar system. Positrons of 100\,GeV can e.g.\ only come from sources some ten thousand years old and located no farther than about one kiloparsec ($\simeq 3\times 10^{16}\,\mbox{km} $) from us~\cite{Gabici_2019}. This limits the choice of pulsar source candidates to a few known ones and potentially also some undetected ones~\cite{Bitter_2022}.  

The CALET collaboration has worked out a complex astrophysical model of the $\mathrm{e}^+$ spectrum from AMS and their own recent measurement of the  $\mathrm{e}^\pm$ flux~\cite{Adriani_2023}. It includes diffuse secondary electrons and positrons, near-by pulsars and supernova remnants as additional point sources, all cut-off at $1\,\mathrm{TeV}$. The sum of all these contributions describes the spectra in a satisfactory manner.   

If point sources were the origin of the observed positron excess, one could expect a slight anisotropy in their directions of incidence. This is, however, not observed. AMS finds the distribution of positron arrival directions for $E>16\, \mathrm{GeV}$ compatible with isotropy and sets an upper limit on the amplitude of a possible dipole anisotropy of $\delta < 0.019$ at 95\% confidence level~\cite{Aguilar_2021}.  For the sum of electrons and positrons and lower limits on energy between $60\, \mathrm{GeV}$ and $480\, \mathrm{GeV}$, the Fermi-LAT collaboration finds limits on a dipole amplitude between $\delta < 0.005$ and $\delta < 0.1$~\cite{Ackermann_2010}.  

\begin{figure}[htb]
\begin{center}
\includegraphics[width=0.5\textwidth]{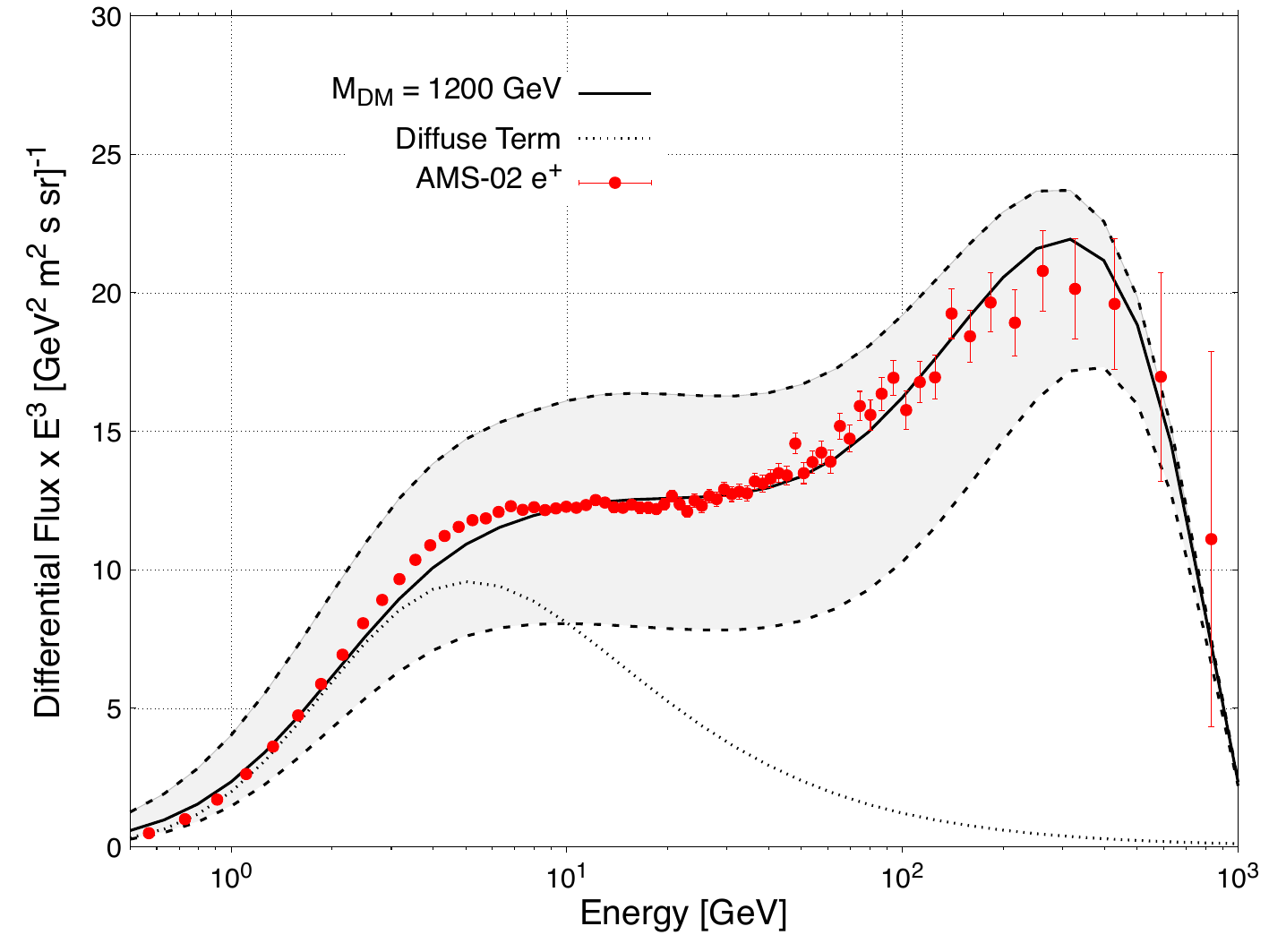}
\end{center}
\caption{\normalsize AMS-02 measurement of the differential flux of positrons as a function of the energy $E$, scaled by $E^3$~\cite{Aguilar_2021}, compared to the sum of the diffuse term and the expectation from dark matter annihilation calculated using \texttt{PPPC 4 DM ID} results~\cite{Cirelli_2011}. Dark matter is assumed to have a Navarro-Frank-White profile~\cite{Navarro_1996} in the Milky Way. The mass is $m_\chi = 1200\,\mathrm{GeV}$ and annihilation is supposed to produce heavy leptons and quarks. The assumed annihilation rate is $\langle \sigma v \rangle = 7.1 \times 10^{-25} \mathrm{cm}^3/\mathrm{s}$. Uncertainties due to the diffusion through the galaxy are indicated by the shaded band.}\label{fig_dm1200}
\end{figure}

While astrophysical point sources are so far not clearly identified, the way is open to consider more diffuse unconventional sources.   The shape and energy scale of the source term invites interpretations also in terms of dark matter annihilation. There is a large body of literature dealing with a possible dark matter origin of the excess; without an attempt for completeness we quote~\cite{Bergstrom_2013,Cholis_2013,Feng_2014,Ibarra_2014,Lin_2015,DiMauro_2016,Chen_2016,Feng_2018,Bai_2018,Ghosh_2021}.  If annihilation between two dark matter particles $\chi$ or their decays cause the high-energy feature in the positron spectrum, their origin follows the density profile of dark matter itself. There are several models for this, including  the Navarro-Frank-White profile~\cite{Navarro_1996,Navarro_2010} used here and others listed in~\cite{Cirelli_2011}.  For dark matter annihilation, the reaction rate is given by the product of cross section $\sigma$ and velocity $v$, $\langle \sigma v \rangle$. For cross sections of typical weak interactions and a thermal velocity distribution, the rates are of the order of $\langle \sigma v \rangle \simeq 10^{-26} \mathrm{cm}^3/\mathrm{s}$. To saturate the source term observed by AMS-02, much higher rates are usually required. An example is shown in Figure~\ref{fig_dm1200}, where we have overlayed the expected spectrum of positrons from the reactions $\chi \chi \rightarrow \mu^+ \mu^-$, $\tau^+ \tau^-$, $\mathrm{b} \bar{\mathrm{b}}$ and  $\mathrm{t} \bar{\mathrm{t}}$, using \texttt{PPPC 4 DM ID} results~\cite{Cirelli_2011}, for an annihilation rate about 70 times larger than the thermal one. Definite conclusions cannot be drawn, given the important uncertainties from cosmic ray propagation.  

\section{Complementarity with other experiments in the field}\label{sec:compl} 

Space experiments have obvious limitations in size and weight. The technology is also limited by the requirement that equipment survives harsh conditions during launch and the hostile environment in space~\cite{Pohl_2014}. The size transportable by rockets and the power supplied by satellites or the ISS limit the energy reach of current experiments to a little more than $100\,\mathrm{TeV}$. At higher energies, ground arrays take over the analysis of cosmic rays. These are covered under the heading \emph{Cosmic Rays Earth Telescopes} in this encyclopedia. 

Lower energies in the MeV to GeV range, not covered in this article, are relevant for cosmic ray influence on terrestrial life. Environmental conditions in the solar system have a large influence on charged particles in this energy domain. These conditions are often summarised under the heading of space weather. Cosmic ray experiments covered in this article observe these particles, and so do specialised satellite and balloon detectors. 

Transport cost and limited flight opportunities also limit the room for advanced technologies in space experiments. Balloon detectors, especially those on long duration flights around the poles, allow for more specialised hardware. On the other hand, their short exposure time limits their energy reach. Balloon cosmic ray experiments are covered in a separate article of this encyclopedia.

The astrophysical objects which are relevant in our galaxy for the production, acceleration and transport of cosmic rays are also observed by classical astronomy and by other messengers. Examples are supernovae and their remnants, pulsars and  active galactic nuclei. Thus the astrophysics at the beginning of the life cycle can be compared to the particle properties at the end of  the food chain. Messengers, which point back to their sources -- photons but also neutrinos and gravitational waves --, carry information on the astrophysics of cosmic ray generation and aceleration. Synergy by using more than one source of information in an astroparticle context is generically called multi-messenger astrophysics. Cosmic messengers other than charged particles are covered in separate articles of this encyclopedia.  

\section{Future developments}\label{sec:future}

Direct detection of cosmic rays in space will see future projects following the two basic approaches: calorimetric measurements and magnetic spectroscopy. Probably the next observatory in line will be the High Energy Radiation Detector (HERD)~\cite{Zhang_2014,Kyratzis_2022,Cagnoli_2024} of the Chinese Space Agency. The detector will be accommodated on the Chinese Space Station CSS, which is currently under construction. The core of the detector is a novel calorimeter made of small cubic elements with three-dimensional read-out, inspired by the CaloCube project~\cite{Adriani_2019a}. It will allow to not only accept cosmic rays entering through the zenith face but also through the lateral faces, thus greatly increasing the accepted solid angle. Despite its compact size of less than a $\mathrm{m}^3$ and a weight of less than $4\, \mathrm{t}$, the acceptance of HERD will reach several $\mathrm{m}^2\, \mathrm{sr}$. The fine segmentation, highly sensitive material and unprecedented depth of the calorimeter will allow to reliably separate $\mathrm{e}^\pm$ from nuclei up to $100\, \mathrm{TeV}$ and measure their energy with percent resolution. Large acceptance and long term exposure on the space station will enable identification and energy measurement of nuclei up to PeV energies. The calorimeter is embedded in a series of detectors for particle identification. It is currently foreseen to install the observatory on the CSS around 2027. It will orbit the Earth at an altitude of around $400\, \mathrm{km}$ with an orbit inclination angle of $42^\circ$.   

For the foreseeable future, AMS-02 will be the only magnetic spectrometer in space, i.e.~the only detector that can differentiate between matter and antimatter. The plan is to operate it until the end of the ISS operation, currently foreseen for January 2031. To improve the performance of AMS-02 further, in early 2026 the collaboration will add two large layers of tracking devices at the very top of the current set-up~\cite{Ambrosi_2025}. The two layers will cover the whole of the TRD surface with two layers of silicon microstrip detectors, about $4\, \mathrm{m}^2$ each, while in the existing configuration only about a third is covered. The detectors will be similar to the DAMPE sensors, with single-sided read-out. One layer will measure the bending direction, while the second will be rotated by $45^\circ$, tagging each particle entry with a space point. This addition will increase the detector acceptance by about a factor of three.

\begin{figure}[htb]
\begin{minipage}[b]{0.39\textwidth}
\caption{\normalsize Acceptance as a function of weight for operating (full dots) and proposed cosmic ray observatories (open dots)~\cite{Adriani_2022}. Note that the acceptance of a device depends on the analysis, thus the point positions represent only benchmark values. The diagonal lines correspond to a fixed acceptance per ton.}\label{fig_accton}
\end{minipage}\hfill
\begin{minipage}[b]{0.59\textwidth}
\includegraphics[width=\textwidth]{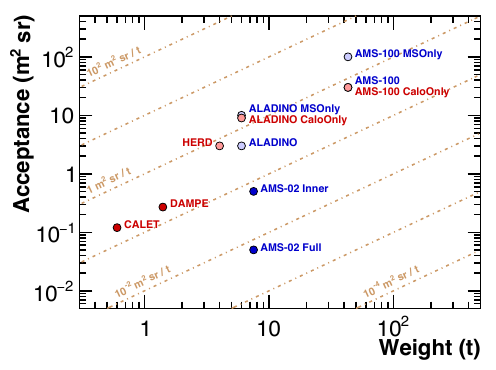}
\end{minipage}
\end{figure}

If one intends to extend the energy reach of a device beyond existing cosmic ray observatories, the aim must be to increase the detector acceptance by orders of magnitude. At the same time, one must take into account the transport capabilities of rockets as far as volume and weight is concerned. Figure~\ref{fig_accton} shows an interesting study~\cite{Adriani_2022} of the acceptance (in $\mathrm{m}^2\mathrm{sr}$) of calorimeters and spectrometers as a function of their weight, both for existing devices and for future projects. It shows that spectrometers can in fact compete with calorimeters for these basic figures of merit. 

Based on experience with AMS-02, colleagues from Germany~\cite{Schael_2019} and Italy~\cite{Battiston_2021} have thus submitted design studies for follow-up spectrometer projects in connection with the ESA program ``Voyage 2050''. These are called AMS-100~\cite{Chung_2022} and ALADInO~\cite{Adriani_2022}, respectively. Both intend to surround a 3D calorimeter inspired by the HERD design with a magnetic spectrometer. To avoid having to cool their magnet coils to superconducting temperatures, both designs foresee the use of warm superconductors, operating at a few tens of Kelvin, in thermal equilibrium with the local environment if sunshine is efficiently shielded. To ensure long term exposure of the observatory with minimum consumables,  both detectors are intended to operate at Lagrange point 2, where the combined gravitational fields of Sun and Earth are in equilibrium with the centrifugal force of a co-rotating object. This position, about $1.5\times 10^9\, \mathrm{km}$ away from Earth,  is also where the James Webb Space Telescope is orbiting. 

For AMS-100 the spectrometer magnet is a thin superconducting solenoid with tracking devices on the inside. With only a few mm thickness, the magnet will generate a homogeneous field of about $1\,\mathrm{T}$ inside its cylindrical volume of $75\, \mathrm{m}^3$. Since the solenoid cannot have a return yoke due to weight limitations, its stray field must be compensated on the outside. Otherwise the dipole magnetic field of the solenoid would interact with the ambient magnetic field of a few $\mathrm{nT}$ and create a torque aligning the two fields. Since the ambient field is not constant, the torque cannot be compensated by gyroscopes. 

For ALADInO, the magnetic field is configured as an open toroid filled with tracking layers. Since the field outside the toroid is negligible, no compensation is necessary. Ten coils will create a doughnut shape field of about $0.8\, \mathrm{T}$.

The basic design and depth of the calorimeter is similar for both studies, reaching an average thickness of the order of $70 \,X_0$ or $4\,\lambda_I$, reached by stacking individually read-out LYSO crystals into the required cylindrical shape.  Tracking devices are based on silicon microstrip detectors, as inherited from AMS-02. However, the required surfaces are about an order of magnitude larger. Both detectors are covered by scintillators for triggering, time-of-flight measurement and charge determination. On the time scale of these projects important technological advances may occur which will no doubt be incorporated into future design iterations. 

All in all, the ALADInO concept will provide an acceptance in excess of $10\, \mathrm{m}^2\mathrm{sr}$, roughly 20 times larger than the current AMS-02. Its spectrometer will have a maximum detectable rigidity of more than $20\, \mathrm{TV}$, commensurate with the increased acceptance. Its weight of less than $7\, \mathrm{t}$ and its dimensions allow transport to Lagrange point 2 by e.g.~the Ariane heavy lift launcher which already successfully deployed the James Webb Space Telescope. The more ambitious AMS-100 design aims for an acceptance yet another order of magnitude larger, of the order of  $100\, \mathrm{m}^2\mathrm{sr}$, and a spectrometer measuring rigidities up to $100\, \mathrm{TV}$. It will weigh of the order of  $40\, \mathrm{t}$ and its dimensions will require delivery by the next generation of heavy weight rockets, like the SLS Block 2 of NASA or the Long March 9 of CAST. 

\section{Conclusions}\label{sec:conclusions}

More than a hundred years after the discovery of the extraterrestrial origin of cosmic rays, a standard model of their life cycle is emerging. It covers nuclear physics inside stars, plasma physics for acceleration and transport of charged particles, as well as particle physics governing their interactions. The current description is far from complete, but the basic ingredients appear to be known. A simple picture is spoiled by the multitude of phenomena which contribute: multiple particle sources, multiple acceleration sites, multiple distance and energy scales. Thus basically all cosmic ray spectra show signs of multiple contributions, when they are measured with sufficient precision. Unconventional sources of particles, like the self-annihilation of dark matter or pockets of antimatter inside the Milky Way, are not yet excluded. 

Important challenges thus remain. The main experimental challenge is to extend the coverage of \emph{in situ} observation  to the region of the knees in the all-particle spectrum. This way, contact will be made with the coverage of ground-based observatories and the origin of these important spectral features will be clarified. Another challenge is to make more and better use of the synergy offered by the observation of multiple cosmic messengers, photons, charged particles, neutrinos and gravitational waves. 

\begin{ack}[Acknowledgments]%
I wish to thank Oscar Adriani, Mercedes Paniccia and Andrii Tykhonov for their careful reading of the draft of this article and important advice concerning contents and presentation. 
\end{ack}

\seealso{V. Bindi, M. Paniccia and M. Pohl, \emph{Cosmic Ray Physics: An Introduction to the Cosmic Laboratory}, CRC Press, Boca Raton, 2023}

\bibliographystyle{Numbered-Style} 
\bibliography{Section6_Cosmic_Ray_Space_Experiments.bib}

\begin{thebibliography*}{100}
\providecommand{\bibtype}[1]{}
\providecommand{\url}[1]{{\tt #1}}
\providecommand{\urlprefix}{URL }
\expandafter\ifx\csname urlstyle\endcsname\relax
  \providecommand{\doi}[1]{doi:\discretionary{}{}{}#1}\else
  \providecommand{\doi}{doi:\discretionary{}{}{}\begingroup
  \urlstyle{rm}\Url}\fi
\providecommand{\bibinfo}[2]{#2}
\providecommand{\eprint}[2][]{\url{#2}}
\makeatletter\def\@biblabel#1{\bibinfo{label}{[#1]}}\makeatother

\bibtype{Article}%
\bibitem{Greisen_1966}
\bibinfo{author}{K. Greisen}, \bibinfo{title}{{End to the cosmic-ray
  spectrum?}}, \bibinfo{journal}{Phys. Rev. Lett.} \bibinfo{volume}{16}
  (\bibinfo{year}{1966}) \bibinfo{pages}{748--750}.

\bibtype{Article}%
\bibitem{Zatsepin_1966}
\bibinfo{author}{G.T. Zatsepin}, \bibinfo{author}{V.A. Kuzmin},
  \bibinfo{title}{{Upper limit of the spectrum of cosmic rays}},
  \bibinfo{journal}{J. Exp. Theor. Phys. Lett.} \bibinfo{volume}{4}
  (\bibinfo{year}{1966}) \bibinfo{pages}{78--80}.

\bibtype{Article}%
\bibitem{Bird_1995}
\bibinfo{author}{D.J.~Bird \emph{et al.} (Fly's Eye~Coll.)},
  \bibinfo{title}{{Detection of a cosmic ray with measured energy well beyond
  the expected spectral cutoff due to cosmic microwave radiation}},
  \bibinfo{journal}{Astrophys. J.}  (\bibinfo{year}{1995}).

\bibtype{Book}%
\bibitem{Bindi_2023}
\bibinfo{author}{V. Bindi}, \bibinfo{author}{M. Paniccia}, \bibinfo{author}{M.
  Pohl}, \bibinfo{title}{Cosmic Ray Physics: An Introduction to the Cosmic
  Laboratory}, \bibinfo{publisher}{CRC Press, Boca Raton} \bibinfo{year}{2023}.

\bibtype{Article}%
\bibitem{Aguilar_2021}
\bibinfo{author}{M.~Aguilar \emph{et al.} (AMS~Coll.)}, \bibinfo{title}{The
  Alpha Magnetic Spectrometer (AMS) on the International Space Station: Part II
  -- Results from the first seven years}, \bibinfo{journal}{Physics Reports}
  \bibinfo{volume}{894} (\bibinfo{year}{2021}) \bibinfo{pages}{1--116}.

\bibtype{Article}%
\bibitem{VanAllen_1948}
\bibinfo{author}{J.A.~Van Allen}, \bibinfo{author}{H.E. Tatel},
  \bibinfo{title}{{The cosmic-ray counting rate of a single Geiger counter from
  ground level to 161 kilometers altitude}}, \bibinfo{journal}{Phys. Rev.}
  \bibinfo{volume}{73} (\bibinfo{year}{1948}) \bibinfo{pages}{245--251}.

\bibtype{Article}%
\bibitem{Grigorov_1970}
\bibinfo{author}{N.L.~Grigorov \emph{et al.}}, \bibinfo{title}{{Investigation
  of energy spectrum of primary cosmic particles with high and super-high
  energies of space station PROTON}}, \bibinfo{journal}{Yad. Fiz.}
  \bibinfo{volume}{11} (\bibinfo{year}{1970}) \bibinfo{pages}{1058--1069}.

\bibtype{Inproceedings}%
\bibitem{Grigorov_1971a}
\bibinfo{author}{N.L.~Grigorov \emph{et al.}}, \bibinfo{title}{{Energy spectrum
  of primary cosmic rays in the $10^{11}$--$10^{15}$eV energy range according
  to the data of Proton-4 measurements}}, in: \bibinfo{booktitle}{12th
  International Cosmic Ray Conference, Hobart, Australia, Volume 5}
  \bibinfo{year}{1971}, pp. \bibinfo{pages}{1746--1751}.

\bibtype{Inproceedings}%
\bibitem{Grigorov_1971b}
\bibinfo{author}{N.L.~Grigorov \emph{et al.}}, \bibinfo{title}{{On irregularity
  in the primary cosmic ray spectrum in the $10^{12}$ eV energy range}}, in:
  \bibinfo{booktitle}{12th International Cosmic Ray Conference, Hobart,
  Australia, Volume 5} \bibinfo{year}{1971}, pp. \bibinfo{pages}{1752--1759}.

\bibtype{Inproceedings}%
\bibitem{Grigorov_1971c}
\bibinfo{author}{N.L.~Grigorov \emph{et al.}}, \bibinfo{title}{{Energy spectrum
  of cosmic ray $\alpha$-particles in $5\rtimes 10^{10}$ -- $10^{12}$
  eV/nucleon energy range}}, in: \bibinfo{booktitle}{12th International Cosmic
  Ray Conference, Hobart, Australia, Volume 5} \bibinfo{year}{1971}, pp.
  \bibinfo{pages}{1760--1768}.

\bibtype{Article}%
\bibitem{Grigorov_2004}
\bibinfo{author}{N.L. Grigorov}, \bibinfo{author}{E.D. Tolstaya},
  \bibinfo{title}{{The spectrum of cosmic-ray particles and their prigin}},
  \bibinfo{journal}{JETP} \bibinfo{volume}{98} (\bibinfo{year}{2004})
  \bibinfo{pages}{643--650}.

\bibtype{Article}%
\bibitem{Grigorov_1966}
\bibinfo{author}{N.L.~Grigorov \emph{et al.}}, \bibinfo{title}{{Some problems
  and perspectives in cosmic-ray studies}}, \bibinfo{journal}{Space Sci. Rev.}
  \bibinfo{volume}{5} (\bibinfo{year}{1966}) \bibinfo{pages}{167--209}.

\bibtype{Article}%
\bibitem{Ivanenko_1989}
\bibinfo{author}{I.P.~Ivanenko \emph{et al.}}, \bibinfo{title}{Energy spectrum
  of primary cosmic-ray particles at 1--100 TeV from data from the Sokol
  package}, \bibinfo{journal}{JETP Lett.} \bibinfo{volume}{49}
  (\bibinfo{year}{1989}) \bibinfo{pages}{222--224}.

\bibtype{Article}%
\bibitem{Atkin_2015}
\bibinfo{author}{E.~Atkin \emph{et al.}}, \bibinfo{title}{{The NUCLEON space
  experiment for direct high energy cosmic rays investigation in
  TeV\textendash{}PeV energy range}}, \bibinfo{journal}{Nucl. Instrum. Meth. A}
  \bibinfo{volume}{770} (\bibinfo{year}{2015}) \bibinfo{pages}{189--196}.

\bibtype{Inproceedings}%
\bibitem{Spillantini_2009}
\bibinfo{author}{P. Spillantini}, \bibinfo{title}{{CR from space based
  observatories: History, results and perspectives of the PAMELA mission}}, in:
  \bibinfo{booktitle}{{9th Baikal Summer School on Physics of Elementary
  Particles and Astrophysics, Bol'shie Koty, Russia}} \bibinfo{year}{2009}, pp.
  \bibinfo{pages}{{213--234}}.

\bibtype{Article}%
\bibitem{Adriani_2017}
\bibinfo{author}{O.~Adriani \emph{et al.} (PAMELA~Coll.)}, \bibinfo{title}{{Ten
  years of PAMELA in space}}, \bibinfo{journal}{Riv. Nuovo Cim.}
  \bibinfo{volume}{40} (\bibinfo{year}{2017}) \bibinfo{pages}{473--522}.

\bibtype{Article}%
\bibitem{Adriani_2009b}
\bibinfo{author}{O.~Adriani \emph{et al.}}, \bibinfo{title}{{An anomalous
  positron abundance in cosmic rays with energies 1.5--100?GeV.}},
  \bibinfo{journal}{Nature} \bibinfo{volume}{458} (\bibinfo{year}{2009})
  \bibinfo{pages}{607--609}.

\bibtype{Article}%
\bibitem{Adriani_2013b}
\bibinfo{author}{O.~Adriani \emph{et al.} (PAMELA~Coll.)},
  \bibinfo{title}{Cosmic-ray positron energy spectrum measured by PAMELA},
  \bibinfo{journal}{Phys. Rev. Lett.} \bibinfo{volume}{111}
  (\bibinfo{year}{2013}) \bibinfo{pages}{081102}.

\bibtype{Article}%
\bibitem{Bonechi_2007}
\bibinfo{author}{L.~Bonechi \emph{et al.} (PAMELA~Coll.)},
  \bibinfo{title}{Status of the PAMELA silicon tracker},
  \bibinfo{journal}{Nucl. Instrum. Meth. A} \bibinfo{volume}{570}
  (\bibinfo{year}{2007}) \bibinfo{pages}{281--285}.

\bibtype{Article}%
\bibitem{Aguilar_2002}
\bibinfo{author}{M.~Aguilar \emph{et al.} (AMS~Coll.)}, \bibinfo{title}{The
  Alpha Magnetic Spectrometer (AMS) on the International Space Station: Part I
  -- results from the test flight on the space shuttle},
  \bibinfo{journal}{Physics Reports} \bibinfo{volume}{366}
  (\bibinfo{year}{2002}) \bibinfo{pages}{331--405}.

\bibtype{Article}%
\bibitem{Blau_2002}
\bibinfo{author}{B.~{Blau} \emph{et al.}}, \bibinfo{title}{The superconducting
  magnet system of AMS-02 -- a particle physics detector to be operated on the
  International Space Station}, \bibinfo{journal}{IEEE Transactions on Applied
  Superconductivity} \bibinfo{volume}{12} (\bibinfo{year}{2002})
  \bibinfo{pages}{349--352}.

\bibtype{Misc}%
\bibitem{Overbye_2010}
\bibinfo{author}{D. Overbye}, \bibinfo{title}{{A Costly Quest for the Dark
  Heart of the Cosmos}}, \bibinfo{howpublished}{The New York Times, November
  16} \bibinfo{year}{2010}.

\bibtype{Article}%
\bibitem{Lubelsmeyer_2011}
\bibinfo{author}{K.~L{\"u}belsmeyer \emph{et al.}}, \bibinfo{title}{Upgrade of
  the Alpha Magnetic Spectrometer (AMS-02) for long term operation on the
  International Space Station (ISS)}, \bibinfo{journal}{Nucl. Instrum. Meth. A}
  \bibinfo{volume}{654} (\bibinfo{year}{2011}) \bibinfo{pages}{639--648}.

\bibtype{Misc}%
\bibitem{Vanes_2013}
\bibinfo{author}{J. van Es~\emph{et al.}}, \bibinfo{title}{{AMS02 Tracker
  Thermal Control Cooling System commissioning and operational results}}
  \bibinfo{year}{2013},
  \bibinfo{url}{\urlprefix\url{https://core.ac.uk/download/pdf/53034866.pdf}}.

\bibtype{Misc}%
\bibitem{ESA_2019}
\bibinfo{author}{ESA}, \bibinfo{title}{{Luca to lead most challenging
  spacewalks since Hubble repairs}} \bibinfo{year}{2019},
  \bibinfo{url}{\urlprefix\url{http://www.esa.int/Science_Exploration/Human_and_Robotic_Exploration/}}.

\bibtype{Article}%
\bibitem{Yamamoto_2013}
\bibinfo{author}{{A. Yamamoto \emph{et al,}}}, \bibinfo{title}{Search for
  cosmic-ray antiproton origins and for cosmological antimatter with BESS},
  \bibinfo{journal}{Adv. Space Res.} \bibinfo{volume}{51}
  (\bibinfo{year}{2013}) \bibinfo{pages}{227--233}.

\bibtype{Article}%
\bibitem{Abe_2017}
\bibinfo{author}{{K. Abe \emph{et al.}}}, \bibinfo{title}{{The results from
  BESS-Polar experiment}}, \bibinfo{journal}{Adv. Space Res.}
  \bibinfo{volume}{60} (\bibinfo{year}{2017}) \bibinfo{pages}{806--814}.

\bibtype{Article}%
\bibitem{Chang_2017}
\bibinfo{author}{J.~Chang \emph{et al.} (DAMPE~Coll.)}, \bibinfo{title}{The
  DArk Matter Particle Explorer mission}, \bibinfo{journal}{Astrop. Phys.}
  \bibinfo{volume}{95} (\bibinfo{year}{2017}) \bibinfo{pages}{6--24}.

\bibtype{Article}%
\bibitem{Torii_2019}
\bibinfo{author}{S. Torii}, \bibinfo{author}{P.S. Marrocchesi~(CALET Coll.)},
  \bibinfo{title}{{The CALorimetric Electron Telescope (CALET) on the
  International Space Station}}, \bibinfo{journal}{Adv. Space Res.}
  \bibinfo{volume}{64} (\bibinfo{year}{2019}) \bibinfo{pages}{2531--2537}.

\bibtype{Article}%
\bibitem{Asaoka_2019}
\bibinfo{author}{{Y. Asaoka \emph{et al.} (CALET Coll.)}}, \bibinfo{title}{{The
  CALorimetric Electron Telescope (CALET) on the International Space Station:
  Results from the first two years on orbit}}, \bibinfo{journal}{J. Phys.:
  Conf. Ser.} \bibinfo{volume}{1181} (\bibinfo{year}{2019})
  \bibinfo{pages}{012003}.

\bibtype{Article}%
\bibitem{Asaoka_2017}
\bibinfo{author}{Y.~Asaoka \emph{et al.} (CALET~Coll.)}, \bibinfo{title}{Energy
  calibration of CALET onboard the International Space Station},
  \bibinfo{journal}{Astroparticle Physics} \bibinfo{volume}{91}
  (\bibinfo{year}{2017}) \bibinfo{pages}{1--10}.

\bibtype{Article}%
\bibitem{Adriani_2023}
\bibinfo{author}{O.~Adriani \emph{et al.} (CALET~Coll.)},
  \bibinfo{title}{Direct Measurement of the Spectral Structure of Cosmic-Ray
  $\text{Electrons}+\text{Positrons}$ in the TeV Region with CALET on the
  International Space Station}, \bibinfo{journal}{Phys. Rev. Lett.}
  \bibinfo{volume}{131} (\bibinfo{year}{2023}) \bibinfo{pages}{191001},
  \bibinfo{url}{\urlprefix\url{https://link.aps.org/doi/10.1103/PhysRevLett.131.191001}}.

\bibtype{Article}%
\bibitem{Seo_2014}
\bibinfo{author}{E.S.~Seo \emph{et al.} (ISS-CREAM~Coll.)},
  \bibinfo{title}{Cosmic Ray Energetics And Mass for the International Space
  Station (ISS-CREAM)}, \bibinfo{journal}{Advances in Space Research}
  \bibinfo{volume}{53} (\bibinfo{year}{2014}) \bibinfo{pages}{1451--1455}.

\bibtype{Article}%
\bibitem{Seo_2020}
\bibinfo{author}{E.S.~Seo \emph{et al.} (CREAM~Coll.)}, \bibinfo{title}{{Cosmic
  Ray Energetics And Mass for the International Space Station (ISS-CREAM)}},
  \bibinfo{journal}{PoS} \bibinfo{volume}{ICRC2019} (\bibinfo{year}{2020})
  \bibinfo{pages}{137}.

\bibtype{Article}%
\bibitem{Choi_2022}
\bibinfo{author}{G.H.~Choi \emph{et al.} (ISS-CREAM~Coll.)},
  \bibinfo{title}{{Measurement of High-energy Cosmic-Ray Proton Spectrum from
  the ISS-CREAM Experiment}}, \bibinfo{journal}{Astrophys. J.}
  \bibinfo{volume}{940} (\bibinfo{year}{2022}) \bibinfo{pages}{107}.

\bibtype{Article}%
\bibitem{Sun_2023}
\bibinfo{author}{{H.R. Sun \emph{et al.} (DAMPE Coll.)}},
  \bibinfo{title}{{Measurement of Heavy Nuclei beyond Iron in Cosmic Rays with
  the DAMPE experiment}}, \bibinfo{journal}{PoS} \bibinfo{volume}{ICRC2023}
  (\bibinfo{year}{2023}) \bibinfo{pages}{174}.

\bibtype{Article}%
\bibitem{Alemanno_2024c}
\bibinfo{author}{{F. Alemanno \emph{et al.}}} (\bibinfo{collaboration}{DAMPE
  Collaboration}), \bibinfo{title}{Measurement of the cosmic $p+\mathrm{He}$
  energy spectrum from 50 GeV to 0.5 PeV with the DAMPE space mission},
  \bibinfo{journal}{Phys. Rev. D} \bibinfo{volume}{109} (\bibinfo{year}{2024})
  \bibinfo{pages}{L121101}, \bibinfo{doi}{\doi{10.1103/PhysRevD.109.L121101}},
  \bibinfo{url}{\urlprefix\url{https://link.aps.org/doi/10.1103/PhysRevD.109.L121101}}.

\bibtype{Article}%
\bibitem{Aguilar_2013}
\bibinfo{author}{{M. Aguilar \emph{et al.}}} (\bibinfo{collaboration}{AMS
  Collaboration}), \bibinfo{title}{Electron and Positron Fluxes in Primary
  Cosmic Rays Measured with the Alpha Magnetic Spectrometer on the
  International Space Station}, \bibinfo{journal}{Phys. Rev. Lett.}
  \bibinfo{volume}{113} (\bibinfo{year}{2014}) \bibinfo{pages}{121102},
  \bibinfo{doi}{\doi{10.1103/PhysRevLett.113.121102}},
  \bibinfo{url}{\urlprefix\url{https://link.aps.org/doi/10.1103/PhysRevLett.113.121102}}.

\bibtype{Misc}%
\bibitem{Fan_2014}
\bibinfo{author}{{R. R. Fan \emph{et al,}}}, \bibinfo{title}{The silicon matrix
  for the prototype for the Dark Matter Particle Explorer}
  \bibinfo{year}{2014}, \eprint{1403.1679},
  \bibinfo{url}{\urlprefix\url{https://arxiv.org/abs/1403.1679}}.

\bibtype{Article}%
\bibitem{Lafferty_1995}
\bibinfo{author}{G.D. Lafferty}, \bibinfo{author}{T.R. Wyatt},
  \bibinfo{title}{{Where to stick your data points: The treatment of
  measurements within wide bins}}, \bibinfo{journal}{Nucl. Instrum. Meth. A}
  \bibinfo{volume}{355} (\bibinfo{year}{1995}) \bibinfo{pages}{541--547}.

\bibtype{Article}%
\bibitem{Berdugo_2017}
\bibinfo{author}{J. Berdugo}, \bibinfo{author}{V. Choutko}, \bibinfo{author}{C.
  Delgado}, \bibinfo{author}{Q. Yan}, \bibinfo{title}{Determination of the
  rigidity scale of the Alpha Magnetic Spectrometer}, \bibinfo{journal}{Nucl.
  Instrum. Meth. A} \bibinfo{volume}{869} (\bibinfo{year}{2017})
  \bibinfo{pages}{10--14}.

\bibtype{Article}%
\bibitem{Adloff_2013}
\bibinfo{author}{C.~Adloff \emph{et al.}}, \bibinfo{title}{The AMS-02
  lead-scintillating fibres electromagnetic calorimeter},
  \bibinfo{journal}{Nucl. Instrum. Meth. A} \bibinfo{volume}{714}
  (\bibinfo{year}{2013}) \bibinfo{pages}{147--154}.

\bibtype{Article}%
\bibitem{Kounine_2017}
\bibinfo{author}{A. Kounine}, \bibinfo{author}{Z. Weng}, \bibinfo{author}{W.
  Xu}, \bibinfo{author}{C. Zhang}, \bibinfo{title}{Precision measurement of
  $0.5~\mathrm{GeV}-3~\mathrm{TeV}$ electrons and positrons using the AMS
  electromagnetic calorimeter}, \bibinfo{journal}{Nucl. Instrum. Meth. A}
  \bibinfo{volume}{869} (\bibinfo{year}{2017}) \bibinfo{pages}{110--117}.

\bibtype{Article}%
\bibitem{Adriani_2019}
\bibinfo{author}{{O. Adriani \emph{et al.} (CALET Coll.)}},
  \bibinfo{title}{{Direct measurement of the cosmic-ray proton spectrum from 50
  GeV to 10 TeV with the Calorimetric Electron Telescope on the International
  Space Station}}, \bibinfo{journal}{Phys. Rev. Lett.} \bibinfo{volume}{122}
  (\bibinfo{year}{2019}) \bibinfo{pages}{181102}.

\bibtype{Article}%
\bibitem{Adriani_2020}
\bibinfo{author}{{O. Adriani \emph{et al.} (CALET Coll.)}},
  \bibinfo{title}{{Direct measurement of the cosmic-ray carbon and oxygen
  spectra from 10 GeV/n to 2.2 TeV/n with the Calorimetric Electron Telescope
  on the International Space Station}}, \bibinfo{journal}{Phys. Rev. Lett.}
  \bibinfo{volume}{125} (\bibinfo{year}{2020}) \bibinfo{pages}{251102}.

\bibtype{Article}%
\bibitem{Adriani_2021}
\bibinfo{author}{O.~Adriani \emph{et al.} (CALET~Coll.)},
  \bibinfo{title}{Measurement of the iron spectrum in cosmic rays from
  $10\,\mathrm{GeV}/n$ to $2.0\,\mathrm{TeV}/n$ with the Calorimetric Electron
  Telescope on the International Space Station}, \bibinfo{journal}{Phys. Rev.
  Lett.} \bibinfo{volume}{126} (\bibinfo{year}{2021}) \bibinfo{pages}{241101}.

\bibtype{Article}%
\bibitem{Zang_2017}
\bibinfo{author}{J.J.~Zang \emph{et al.} (DAMPE~Coll.)},
  \bibinfo{title}{{Measurement of absolute energy scale of ECAL of DAMPE with
  geomagnetic rigidity cutoff}}, \bibinfo{journal}{PoS}
  \bibinfo{volume}{ICRC2017} (\bibinfo{year}{2018}) \bibinfo{pages}{197}.

\bibtype{Article}%
\bibitem{Alemanno_2021}
\bibinfo{author}{F.~Alemanno \emph{et al.} (DAMPE~Coll.)},
  \bibinfo{title}{{Measurement of the cosmic ray helium spectrum from 70 GeV to
  80 TeV with the DAMPE space mission}}, \bibinfo{journal}{Phys. Rev. Lett.}
  \bibinfo{volume}{126} (\bibinfo{year}{2021}) \bibinfo{pages}{201102}.

\bibtype{Article}%
\bibitem{Aguilar_2015b}
\bibinfo{author}{M.~Aguilar \emph{et al.} (AMS~Coll.)},
  \bibinfo{title}{{Precision Measurement of the helium flux in primary cosmic
  rays of rigidities 1.9 GV to 3 TV with the Alpha Magnetic Spectrometer on the
  International Space Station}}, \bibinfo{journal}{Phys. Rev. Lett.}
  \bibinfo{volume}{115} (\bibinfo{year}{2015}) \bibinfo{pages}{211101}.

\bibtype{Article}%
\bibitem{Yan_2021}
\bibinfo{author}{Q. Yan}, \bibinfo{author}{V. Choutko}, \bibinfo{author}{A.
  Oliva}, \bibinfo{author}{M. Paniccia}, \bibinfo{title}{Measurements of
  nuclear interaction cross sections with the Alpha Magnetic Spectrometer on
  the International Space Station}, \bibinfo{journal}{Nucl. Phys. A}
  \bibinfo{volume}{996} (\bibinfo{year}{2020}) \bibinfo{pages}{121712}.

\bibtype{Article}%
\bibitem{Aguilar_2021a}
\bibinfo{author}{M.~Aguilar \emph{et al.} (AMS~Coll.)},
  \bibinfo{title}{Properties of Iron primary cosmic rays: Results from the
  Alpha Magnetic Spectrometer}, \bibinfo{journal}{Phys. Rev. Lett.}
  \bibinfo{volume}{126} (\bibinfo{year}{2021}) \bibinfo{pages}{0411104}.

\bibtype{Article}%
\bibitem{Alemanno_2024a}
\bibinfo{author}{F.~Alemanno \emph{et al.} (DAMPE~Coll.)},
  \bibinfo{title}{Hadronic cross section measurements with the DAMPE space
  mission using 20GeV--10TeV cosmic-ray protons and $^4$He},
  \bibinfo{journal}{Phys. Rev. D} \bibinfo{volume}{111} (\bibinfo{year}{2025})
  \bibinfo{pages}{012002}, \bibinfo{doi}{\doi{10.1103/PhysRevD.111.012002}},
  \bibinfo{url}{\urlprefix\url{https://link.aps.org/doi/10.1103/PhysRevD.111.012002}}.

\bibtype{Article}%
\bibitem{Adriani_2011}
\bibinfo{author}{O.~Adriani \emph{et al.}}, \bibinfo{title}{PAMELA measurements
  of cosmic-ray proton and helium spectra}, \bibinfo{journal}{Science}
  \bibinfo{volume}{332} (\bibinfo{year}{2011}) \bibinfo{pages}{69--72}.

\bibtype{Article}%
\bibitem{Adriani_2022b}
\bibinfo{author}{O.~Adriani \emph{et al.} (CALET~Coll.)},
  \bibinfo{title}{Observation of spectral structures in the flux of cosmic-ray
  protons from 50 GeV to 60 TeV with the Calorimetric Electron Telescope on the
  International Space Station}, \bibinfo{journal}{Phys. Rev. Lett.}
  \bibinfo{volume}{129} (\bibinfo{year}{2022}) \bibinfo{pages}{101102}.

\bibtype{Article}%
\bibitem{Brogi_2021}
\bibinfo{author}{P. Brogi}, \bibinfo{author}{K.~Kobayashi \emph{et al.}
  (CALET~Coll.)}, \bibinfo{title}{{Measurement of the energy spectrum of
  cosmic-ray helium with CALET on the International Space Station}},
  \bibinfo{journal}{PoS} \bibinfo{volume}{ICRC2021} (\bibinfo{year}{2021})
  \bibinfo{pages}{101}.

\bibtype{Article}%
\bibitem{An_2019}
\bibinfo{author}{Q.~An \emph{et al.} (DAMPE~Coll.)},
  \bibinfo{title}{{Measurement of the cosmic-ray proton spectrum from 40 GeV to
  100 TeV with the DAMPE satellite}}, \bibinfo{journal}{Sci. Adv.}
  \bibinfo{volume}{5} (\bibinfo{year}{2019}) \bibinfo{pages}{eaax3793}.

\bibtype{Article}%
\bibitem{Aguilar_2015a}
\bibinfo{author}{{M. Aguilar \emph{et al.} (AMS Coll.)}},
  \bibinfo{title}{{Precision measurement of the proton flux in primary cosmic
  rays from rigidity 1 GV to 1.8 TV with the Alpha Magnetic Spectrometer on the
  International Space Station}}, \bibinfo{journal}{Phys. Rev. Lett.}
  \bibinfo{volume}{114} (\bibinfo{year}{2015}) \bibinfo{pages}{171103}.

\bibtype{Article}%
\bibitem{Lipari_2020}
\bibinfo{author}{P. Lipari}, \bibinfo{author}{S. Vernetto},
  \bibinfo{title}{{The shape of the cosmic ray proton spectrum}},
  \bibinfo{journal}{Astropart. Phys.} \bibinfo{volume}{120}
  (\bibinfo{year}{2020}) \bibinfo{pages}{102441}.

\bibtype{Article}%
\bibitem{Kobayashi_2021}
\bibinfo{author}{K. Kobayasi}, \bibinfo{author}{P.S.~Marrochesi \emph{et al.}
  (CALET~Coll.)}, \bibinfo{title}{{Extended measurement of the proton spectrum
  with CALET on the International Space Station}}, \bibinfo{journal}{PoS}
  \bibinfo{volume}{ICRC2021} (\bibinfo{year}{2021}) \bibinfo{pages}{098}.

\bibtype{Article}%
\bibitem{Vecchi_2021}
\bibinfo{author}{M.~Vecchi \emph{et al.}}, \bibinfo{title}{{Combined analysis
  of AMS-02 secondary-to-primary ratios: Universality of cosmic-ray propagation
  and consistency of nuclear cross-sections}}, \bibinfo{journal}{PoS}
  \bibinfo{volume}{ICRC2021} (\bibinfo{year}{2021}) \bibinfo{pages}{174}.

\bibtype{Article}%
\bibitem{Vecchi_2022}
\bibinfo{author}{M.~Vecchi \emph{et al.}}, \bibinfo{title}{{The rigidity
  dependence of galactic cosmic-ray fluxes and its connection with the
  diffusion coefficient}}, \bibinfo{journal}{Frontiers in Physics}
  \bibinfo{volume}{10} (\bibinfo{year}{2022}),
  \bibinfo{url}{\urlprefix\url{https://www.frontiersin.org/journals/physics/articles/10.3389/fphy.2022.858841}}.

\bibtype{Article}%
\bibitem{Akaike_2024}
\bibinfo{author}{Y.~Akaike \emph{et al.} (CALET~Coll.)}, \bibinfo{title}{The
  Calorimetric Electron Telescope (CALET) on the International Space Station:
  Results from the first eight years on orbit}, \bibinfo{journal}{Advances in
  Space Research} \bibinfo{volume}{74} (\bibinfo{year}{2024})
  \bibinfo{pages}{4353--4367},
  \bibinfo{url}{\urlprefix\url{https://www.sciencedirect.com/science/article/pii/S027311772400382X}}.

\bibtype{Article}%
\bibitem{Serpico_2018}
\bibinfo{author}{P.D. Serpico}, \bibinfo{title}{Entering the cosmic ray
  precision era}, \bibinfo{journal}{J. Astrophys. Astr.} \bibinfo{volume}{39}
  (\bibinfo{year}{2018}) \bibinfo{pages}{41}.

\bibtype{Article}%
\bibitem{Gabici_2019}
\bibinfo{author}{S.~Gabici \emph{et al.}}, \bibinfo{title}{The origin of
  galactic cosmic rays: Challenges to the standard paradigm},
  \bibinfo{journal}{Int. J. Mod. Phys. D} \bibinfo{volume}{28}
  (\bibinfo{year}{2019}) \bibinfo{pages}{1930022}.

\bibtype{Article}%
\bibitem{Vladimirov_2012}
\bibinfo{author}{A.E.~Vladimirov \emph{et al.}}, \bibinfo{title}{{Testing the
  origin of high-energy cosmic rays}}, \bibinfo{journal}{Astrophys. J.}
  \bibinfo{volume}{752} (\bibinfo{year}{2012}) \bibinfo{pages}{68}.

\bibtype{Article}%
\bibitem{Zhang_2020}
\bibinfo{author}{H.G.~Zhang \emph{et al.}}, \bibinfo{title}{{Performance of the
  ISS-CREAM calorimeter in a calibration beam test}},
  \bibinfo{journal}{Astropart. Phys.} \bibinfo{volume}{130}
  (\bibinfo{year}{2021}) \bibinfo{pages}{102583}.

\bibtype{Article}%
\bibitem{Tomassetti_2015}
\bibinfo{author}{N. Tomassetti}, \bibinfo{title}{Cosmic-ray protons, nuclei,
  electrons, and antiparticles under a two-halo scenario of diffusive
  propagation}, \bibinfo{journal}{Phys. Rev. D} \bibinfo{volume}{92}
  (\bibinfo{year}{2015}) \bibinfo{pages}{081301}.

\bibtype{Article}%
\bibitem{Jaupart_2018}
\bibinfo{author}{E. Jaupart}, \bibinfo{author}{E. Parizot}, \bibinfo{author}{D.
  Allard}, \bibinfo{title}{Contribution of the galactic centre to the local
  cosmic-ray flux}, \bibinfo{journal}{Astron. Astrophys.} \bibinfo{volume}{619}
  (\bibinfo{year}{2018}) \bibinfo{pages}{A12}.

\bibtype{Article}%
\bibitem{Aloisio_2015}
\bibinfo{author}{R. Aloisio}, \bibinfo{author}{P. Blasi}, \bibinfo{author}{P.D.
  Serpico}, \bibinfo{title}{Nonlinear cosmic ray galactic transport in the
  light of AMS-02 and Voyager data}, \bibinfo{journal}{Astron. Astrophys.}
  \bibinfo{volume}{583} (\bibinfo{year}{2015}) \bibinfo{pages}{A95}.

\bibtype{Inproceedings}%
\bibitem{Genolini_2017}
\bibinfo{author}{Y.~G\'enolini \emph{et al.}}, \bibinfo{title}{{Indications for
  a high-rigidity break in the cosmic-ray diffusion coefficient}}, in:
  \bibinfo{booktitle}{35th International Cosmic Ray Conference, Busan, Korea}
  \bibinfo{year}{2017}, p. \bibinfo{pages}{268}.

\bibtype{Article}%
\bibitem{Aguilar_2017}
\bibinfo{author}{{M. Aguilar \emph{et al.} (AMS Coll.)}},
  \bibinfo{title}{{Observation of identical rigidity dependence of He, C and O
  cosmic rays at high rigidities by the Alpha Magnetic Spectrometer on the
  International Space Station}}, \bibinfo{journal}{Phys. Rev. Lett.}
  \bibinfo{volume}{119} (\bibinfo{year}{2017}) \bibinfo{pages}{251101}.

\bibtype{Article}%
\bibitem{Aguilar_2023}
\bibinfo{author}{M.~\emph{et al.} M.~Aguilar} (\bibinfo{collaboration}{AMS
  Collaboration}), \bibinfo{title}{Properties of Cosmic-Ray Sulfur and
  Determination of the Composition of Primary Cosmic-Ray Carbon, Neon,
  Magnesium, and Sulfur: Ten-Year Results from the Alpha Magnetic
  Spectrometer}, \bibinfo{journal}{Phys. Rev. Lett.} \bibinfo{volume}{130}
  (\bibinfo{year}{2023}) \bibinfo{pages}{211002},
  \bibinfo{doi}{\doi{10.1103/PhysRevLett.130.211002}},
  \bibinfo{url}{\urlprefix\url{https://link.aps.org/doi/10.1103/PhysRevLett.130.211002}}.

\bibtype{Misc}%
\bibitem{Alemanno_2024b}
\bibinfo{author}{F.~Alemanno \emph{et al.} (DAMPE~Coll.)},
  \bibinfo{title}{Observation of a spectral hardening in cosmic ray boron
  spectrum with the DAMPE space mission} \bibinfo{year}{2024},
  \eprint{2412.11460},
  \bibinfo{url}{\urlprefix\url{https://arxiv.org/abs/2411.11460}}.

\bibtype{Article}%
\bibitem{Strong_2007}
\bibinfo{author}{A.W. Strong}, \bibinfo{author}{I.V. Moskalenko},
  \bibinfo{author}{V.S. Ptuskin}, \bibinfo{title}{Cosmic-ray propagation and
  interactions in the galaxy}, \bibinfo{journal}{Ann. Rev. Nucl. Part. Sci.}
  \bibinfo{volume}{57} (\bibinfo{year}{2007}) \bibinfo{pages}{285--327}.

\bibtype{Article}%
\bibitem{Moskalenko_1997}
\bibinfo{author}{I.V. Moskalenko}, \bibinfo{author}{A.W. Strong},
  \bibinfo{title}{{Production and propagation of cosmic ray positrons and
  electrons}}, \bibinfo{journal}{Astrophys. J.} \bibinfo{volume}{493}
  (\bibinfo{year}{1998}) \bibinfo{pages}{694--707}.

\bibtype{Article}%
\bibitem{Strong_1998}
\bibinfo{author}{A.W. Strong}, \bibinfo{author}{I.V. Moskalenko},
  \bibinfo{title}{Propagation of cosmic-ray nucleons in the galaxy},
  \bibinfo{journal}{Astrophys. J.} \bibinfo{volume}{509} (\bibinfo{year}{1998})
  \bibinfo{pages}{212--228}.

\bibtype{Article}%
\bibitem{Hanasz_2021}
\bibinfo{author}{M. Hanasz}, \bibinfo{author}{A.W. Strong}, \bibinfo{author}{P.
  Girichidis}, \bibinfo{title}{{Simulations of cosmic ray propagation}},
  \bibinfo{journal}{Living Rev. Comp. Astrophys.} \bibinfo{volume}{7}
  (\bibinfo{year}{2021}) \bibinfo{pages}{2}.

\bibtype{Article}%
\bibitem{Boschini_2020}
\bibinfo{author}{M.J.~Boschini \emph{et al.}}, \bibinfo{title}{Deciphering the
  local interstellar spectra of secondary nuclei with the Galprop/HelMod
  framework and a hint for primary lithium in cosmic rays},
  \bibinfo{journal}{Astrophys. J.} \bibinfo{volume}{889} (\bibinfo{year}{2020})
  \bibinfo{pages}{167}.

\bibtype{Article}%
\bibitem{Boschini_2020a}
\bibinfo{author}{M.J.~Boschini \emph{et al.}}, \bibinfo{title}{Inference of the
  local interstellar spectra of cosmic-ray nuclei $Z \leq 28$ with the
  Galprop-HelMod framework}, \bibinfo{journal}{Astrophys. J. Suppl.}
  \bibinfo{volume}{250} (\bibinfo{year}{2020}) \bibinfo{pages}{27}.

\bibtype{Article}%
\bibitem{Boschini_2017}
\bibinfo{author}{M.J.~Boschini \emph{et al.}}, \bibinfo{title}{Energy
  calibration of CALET onboard the International Space Station},
  \bibinfo{journal}{Astrophys. J.} \bibinfo{volume}{840} (\bibinfo{year}{2017})
  \bibinfo{pages}{115}.

\bibtype{Article}%
\bibitem{Cummings_2016}
\bibinfo{author}{A.C.~Cummings \emph{et al.}}, \bibinfo{title}{Galactic cosmic
  rays in the local interstellar medium: Voyager-1 observations and model
  results}, \bibinfo{journal}{Astrophys. J.} \bibinfo{volume}{831}
  (\bibinfo{year}{2016}) \bibinfo{pages}{18}.

\bibtype{Article}%
\bibitem{Lagutin_2021}
\bibinfo{author}{A.A. Lagutin}, \bibinfo{author}{N.V. Volkov},
  \bibinfo{title}{Features of the energy spectra of primary and secondary
  nuclei of cosmic rays: A consistent astrophysical interpretation},
  \bibinfo{journal}{Bull. Russ. Acad. Sci.} \bibinfo{volume}{85}
  (\bibinfo{year}{2021}) \bibinfo{pages}{375--378}.

\bibtype{Article}%
\bibitem{Tomassetti_2012}
\bibinfo{author}{N. Tomassetti}, \bibinfo{title}{{Entering the cosmic ray
  precision era}}, \bibinfo{journal}{Astrophys. J. Lett.} \bibinfo{volume}{752}
  (\bibinfo{year}{2012}) \bibinfo{pages}{L13}.

\bibtype{Article}%
\bibitem{Boschini_2018}
\bibinfo{author}{M.J.~Boschini \emph{et al.}}, \bibinfo{title}{Deciphering the
  local interstellar spectra of primary cosmic ray species with HelMod},
  \bibinfo{journal}{Astrophys. J.} \bibinfo{volume}{858} (\bibinfo{year}{2018})
  \bibinfo{pages}{61}.

\bibtype{Article}%
\bibitem{Johannesson_2016}
\bibinfo{author}{G.~J\'ohannesson \emph{et al.}}, \bibinfo{title}{Bayesian
  analysis of cosmic-ray propagation: evidence against homogeneous diffusion},
  \bibinfo{journal}{Astrophys. J.} \bibinfo{volume}{824} (\bibinfo{year}{2016})
  \bibinfo{pages}{16}.

\bibtype{Article}%
\bibitem{Mertsch_2020}
\bibinfo{author}{P. Mertsch}, \bibinfo{title}{Test particle simulations of
  cosmic rays}, \bibinfo{journal}{Astrophys. Space Sci.} \bibinfo{volume}{365}
  (\bibinfo{year}{2020}) \bibinfo{pages}{135}.

\bibtype{Article}%
\bibitem{Malkov_2021}
\bibinfo{author}{M.~A. Malkov}, \bibinfo{author}{I.V. Moskalenko},
  \bibinfo{title}{The {TeV} cosmic-ray bump: A message from the Epsilon Indi or
  Epsilon Eridani Star?}, \bibinfo{journal}{Astrophys. J.}
  \bibinfo{volume}{911} (\bibinfo{year}{2021}) \bibinfo{pages}{151}.

\bibtype{Article}%
\bibitem{Malkov_2022}
\bibinfo{author}{M.A. Malkov}, \bibinfo{author}{I.V. Moskalenko},
  \bibinfo{title}{On the origin of observed cosmic-ray spectrum below 100
  {TV}}, \bibinfo{journal}{Astrophys. J.} \bibinfo{volume}{933}
  (\bibinfo{year}{2022}) \bibinfo{pages}{78}.

\bibtype{Article}%
\bibitem{Aguilar_2020}
\bibinfo{author}{M.~Aguilar \emph{et al.} (AMS~Coll.)},
  \bibinfo{title}{Properties of neon, magnesium, and silicon primary cosmic
  rays: Results from the Alpha Magnetic Spectrometer}, \bibinfo{journal}{Phys.
  Rev. Lett.} \bibinfo{volume}{124} (\bibinfo{year}{2020})
  \bibinfo{pages}{211102}.

\bibtype{Article}%
\bibitem{Yuan_2020}
\bibinfo{author}{Q.Yuan \emph{et al.}}, \bibinfo{title}{Nearby source
  interpretation of differences among light and medium composition spectra in
  cosmic rays}, \bibinfo{journal}{Frontiers of Physics} \bibinfo{volume}{16}
  (\bibinfo{year}{2020}) \bibinfo{pages}{24501}.

\bibtype{Article}%
\bibitem{Niu_2022}
\bibinfo{author}{J.-S. Niu}, \bibinfo{title}{Hybrid origins of the cosmic-ray
  nucleus spectral hardening at a few hundred {GV}},
  \bibinfo{journal}{Astrophys. J.} \bibinfo{volume}{932} (\bibinfo{year}{2022})
  \bibinfo{pages}{37}.

\bibtype{Article}%
\bibitem{Aguilar_2021d}
\bibinfo{author}{M.~Aguilar \emph{et al.} (AMS~Coll.)},
  \bibinfo{title}{Properties of heavy secondary fluorine cosmic rays: Results
  from the Alpha Magnetic Spectrometer}, \bibinfo{journal}{Phys. Rev. Lett.}
  \bibinfo{volume}{126} (\bibinfo{year}{2021}) \bibinfo{pages}{081102}.

\bibtype{Article}%
\bibitem{Boschini_2022}
\bibinfo{author}{M.J.~Boschini \emph{et al.}}, \bibinfo{title}{A hint of a
  low-energy excess in cosmic-ray fluorine}, \bibinfo{journal}{Astrophys. J.}
  \bibinfo{volume}{925} (\bibinfo{year}{2022}) \bibinfo{pages}{108}.

\bibtype{Article}%
\bibitem{Bueno_2022}
\bibinfo{author}{E.~Ferrunato~Bueno \emph{et al.}}, \bibinfo{title}{Transport
  parameters from AMS-02 F/Si data and fluorine source abundance},
  \bibinfo{journal}{A\&A} \bibinfo{volume}{688} (\bibinfo{year}{2024})
  \bibinfo{pages}{A17}, \bibinfo{doi}{\doi{10.1051/0004-6361/202244660}},
  \bibinfo{url}{\urlprefix\url{https://doi.org/10.1051/0004-6361/202244660}}.

\bibtype{Article}%
\bibitem{Adriani_2022a}
\bibinfo{author}{O.~Adriani \emph{et al.} (CALET~Coll.)},
  \bibinfo{title}{Direct measurement of the nickel spectrum in cosmic rays in
  the energy range from $8.8\,\mathrm{GeV}/n$ to $240\,\mathrm{GeV}/n$ with
  CALET on the International Space Station}, \bibinfo{journal}{Phys. Rev.
  Lett.} \bibinfo{volume}{128} (\bibinfo{year}{2022}) \bibinfo{pages}{131103}.

\bibtype{Article}%
\bibitem{Boschini_2021}
\bibinfo{author}{M.J.~Boschini \emph{et al.}}, \bibinfo{title}{The discovery of
  a low-energy excess in cosmic-ray iron: Evidence of the past supernova
  activity in the local bubble}, \bibinfo{journal}{Astrophys. J.}
  \bibinfo{volume}{913} (\bibinfo{year}{2021}) \bibinfo{pages}{5}.

\bibtype{Article}%
\bibitem{Baur_1997}
\bibinfo{author}{G.~Baur \emph{et al.}}, \bibinfo{title}{{Observation of
  antihydrogen production in flight at CERN}}, \bibinfo{journal}{Hyperfine
  Interact.} \bibinfo{volume}{109} (\bibinfo{year}{1997})
  \bibinfo{pages}{191--203}.

\bibtype{Book}%
\bibitem{Charlton_2020}
\bibinfo{author}{M. Charlton}, \bibinfo{author}{S. Eriksson},
  \bibinfo{author}{G.M. Shore}, \bibinfo{title}{{Antihydrogen and Fundamental
  Physics}}, \bibinfo{publisher}{Springer} \bibinfo{year}{2020}.

\bibtype{Article}%
\bibitem{Abe_2008}
\bibinfo{author}{K.~Abe \emph{et al.}}, \bibinfo{title}{{Measurement of
  cosmic-ray low-energy antiproton spectrum with the first BESS-Polar Antarctic
  flight}}, \bibinfo{journal}{Phys. Lett. B} \bibinfo{volume}{670}
  (\bibinfo{year}{2008}) \bibinfo{pages}{103--108}.

\bibtype{Article}%
\bibitem{Adriani_2009a}
\bibinfo{author}{O.~Adriani \emph{et al.}}, \bibinfo{title}{{A new measurement
  of the antiproton-to-proton flux ratio up to 100 GeV in the cosmic
  radiation}}, \bibinfo{journal}{Phys. Rev. Lett.} \bibinfo{volume}{102}
  (\bibinfo{year}{2009}) \bibinfo{pages}{051101}.

\bibtype{Article}%
\bibitem{Adriani_2010}
\bibinfo{author}{O.~Adriani \emph{et al.} (PAMELA~Coll)},
  \bibinfo{title}{PAMELA Results on the Cosmic-Ray Antiproton Flux from 60 MeV
  to 180 GeV in Kinetic Energy}, \bibinfo{journal}{Phys. Rev. Lett.}
  \bibinfo{volume}{105} (\bibinfo{year}{2010}) \bibinfo{pages}{121101},
  \bibinfo{url}{\urlprefix\url{https://link.aps.org/doi/10.1103/PhysRevLett.105.121101}}.

\bibtype{Article}%
\bibitem{Adriani_2013c}
\bibinfo{author}{O.~Adriani \emph{et al.} (PAMELA~Coll.)},
  \bibinfo{title}{Measurement of the flux of primary cosmic ray antiprotons
  with energies of 60 MeV to 350 GeV in the PAMELA experiment},
  \bibinfo{journal}{JETP Lett.} \bibinfo{volume}{96} (\bibinfo{year}{2013})
  \bibinfo{pages}{621--627}.

\bibtype{Article}%
\bibitem{Winkler_2017}
\bibinfo{author}{M.W. Winkler}, \bibinfo{title}{Cosmic ray antiprotons at high
  energies}, \bibinfo{journal}{J. Cosm. Astropart. Phys.}
  \bibinfo{volume}{2017} (\bibinfo{year}{2017}) \bibinfo{pages}{048}.

\bibtype{Article}%
\bibitem{Kahlhoefer_2021}
\bibinfo{author}{F.~Kahlhoefer \emph{et al.}}, \bibinfo{title}{{Constraining
  dark matter annihilation with cosmic ray antiprotons using neural networks}},
  \bibinfo{journal}{JCAP} \bibinfo{volume}{12} (\bibinfo{year}{2021})
  \bibinfo{pages}{037}.

\bibtype{Article}%
\bibitem{DiMauro_2021}
\bibinfo{author}{M.~Di Mauro}, \bibinfo{author}{M.W. Winkler},
  \bibinfo{title}{{Multimessenger constraints on the dark matter interpretation
  of the Fermi-LAT Galactic center excess}}, \bibinfo{journal}{Phys. Rev. D}
  \bibinfo{volume}{103} (\bibinfo{year}{2021}) \bibinfo{pages}{123005}.

\bibtype{Article}%
\bibitem{Calore_2022}
\bibinfo{author}{F.~Calore \emph{et al.}}, \bibinfo{title}{{AMS-02 antiprotons
  and dark matter: Trimmed hints and robust bounds}}, \bibinfo{journal}{SciPost
  Phys.} \bibinfo{volume}{12} (\bibinfo{year}{2022}) \bibinfo{pages}{163}.

\bibtype{Article}%
\bibitem{Kachelriess_2020}
\bibinfo{author}{M. Kachelriess}, \bibinfo{author}{S. Ostapchenko},
  \bibinfo{author}{J. Tjemsland}, \bibinfo{title}{Revisiting cosmic ray
  antinuclei fluxes with a new coalescence model}, \bibinfo{journal}{JCAP}
  \bibinfo{volume}{08} (\bibinfo{year}{2020}) \bibinfo{pages}{048}.

\bibtype{Article}%
\bibitem{Dupourque_2021}
\bibinfo{author}{S. Dupourqu\'e}, \bibinfo{author}{L. Tibaldo},
  \bibinfo{author}{P. von Ballmoos}, \bibinfo{title}{{Constraints on the
  antistar fraction in the solar system neighborhood from the 10-year Fermi
  Large Area Telescope gamma-ray source catalog}}, \bibinfo{journal}{Phys. Rev.
  D} \bibinfo{volume}{103} (\bibinfo{year}{2021}) \bibinfo{pages}{083016}.

\bibtype{Article}%
\bibitem{Ambrosi_2017}
\bibinfo{author}{G.~Ambrosi \emph{et al.} (DAMPE~Coll.)},
  \bibinfo{title}{{Direct detection of a break in the teraelectronvolt
  cosmic-ray spectrum of electrons and positrons}}, \bibinfo{journal}{Nature}
  \bibinfo{volume}{552} (\bibinfo{year}{2017}) \bibinfo{pages}{63--66}.

\bibtype{Article}%
\bibitem{Abdollahi_2017}
\bibinfo{author}{S.~Abdollahi \emph{et al.} (Fermi-LAT~Coll.)},
  \bibinfo{title}{Cosmic-ray electron-positron spectrum from 7 GeV to 2 TeV
  with the Fermi Large Area Telescope}, \bibinfo{journal}{Phys. Rev. D}
  \bibinfo{volume}{95} (\bibinfo{year}{2017}) \bibinfo{pages}{082007}.

\bibtype{Article}%
\bibitem{Zhang_2016}
\bibinfo{author}{Z.Y.~Zhang \emph{et al.} (DAMPE~Coll.)}, \bibinfo{title}{{The
  calibration and electron energy reconstruction of the BGO ECAL of the DAMPE
  detector}}, \bibinfo{journal}{Nucl. Instrum. Meth. A} \bibinfo{volume}{836}
  (\bibinfo{year}{2016}) \bibinfo{pages}{98--104}.

\bibtype{Article}%
\bibitem{Kounine_2023}
\bibinfo{author}{A.~Kounine \emph{et al.}}, \bibinfo{title}{{Understanding the
  Origin of Cosmic-Ray Electrons}}, \bibinfo{journal}{PoS}
  \bibinfo{volume}{ICRC2023} (\bibinfo{year}{2023}) \bibinfo{pages}{065}.

\bibtype{Article}%
\bibitem{Trotta_2011}
\bibinfo{author}{R.~Trotta \emph{et al.}}, \bibinfo{title}{Constraints on
  cosmic-ray propagation models from a global Bayesian analysis},
  \bibinfo{journal}{Astrophys. J.} \bibinfo{volume}{729} (\bibinfo{year}{2011})
  \bibinfo{pages}{106}.

\bibtype{Article}%
\bibitem{Bitter_2022}
\bibinfo{author}{O.M. Bitter}, \bibinfo{author}{D. Hooper},
  \bibinfo{title}{{Constraining the Milky Way's pulsar population with the
  cosmic-ray positron fraction}}, \bibinfo{journal}{J. Cosm. Astropart. Phys.}
  \bibinfo{volume}{2022} (\bibinfo{year}{2022}) \bibinfo{pages}{081},
  \bibinfo{doi}{\doi{10.1088/1475-7516/2022/10/081}},
  \bibinfo{url}{\urlprefix\url{https://dx.doi.org/10.1088/1475-7516/2022/10/081}}.

\bibtype{Article}%
\bibitem{Ackermann_2010}
\bibinfo{author}{M.~Ackermann \emph{et al.} (Fermi LAT~Coll.)},
  \bibinfo{title}{{Searches for cosmic-ray electron anisotropies with the Fermi
  Large Area Telescope}}, \bibinfo{journal}{Phys. Rev. D} \bibinfo{volume}{82}
  (\bibinfo{year}{2010}) \bibinfo{pages}{092003}.

\bibtype{Article}%
\bibitem{Bergstrom_2013}
\bibinfo{author}{L.~Bergstrom \emph{et al.}}, \bibinfo{title}{{New limits on
  dark matter annihilation from AMS cosmic ray positron data}},
  \bibinfo{journal}{Phys. Rev. Lett.} \bibinfo{volume}{111}
  (\bibinfo{year}{2013}) \bibinfo{pages}{171101}.

\bibtype{Article}%
\bibitem{Cholis_2013}
\bibinfo{author}{I. Cholis}, \bibinfo{author}{D. Hooper}, \bibinfo{title}{{Dark
  matter and pulsar origins of the rising cosmic ray positron fraction in light
  of new data from AMS}}, \bibinfo{journal}{Phys. Rev. D} \bibinfo{volume}{88}
  (\bibinfo{year}{2013}) \bibinfo{pages}{023013}.

\bibtype{Article}%
\bibitem{Feng_2014}
\bibinfo{author}{L.~Feng \emph{et al.}}, \bibinfo{title}{{AMS-02 positron
  excess: new bounds on dark matter models and hint for primary electron
  spectrum hardening}}, \bibinfo{journal}{Phys. Lett. B} \bibinfo{volume}{728}
  (\bibinfo{year}{2014}) \bibinfo{pages}{250--255}.

\bibtype{Article}%
\bibitem{Ibarra_2014}
\bibinfo{author}{A. Ibarra}, \bibinfo{author}{A.S. Lamperstorfer},
  \bibinfo{author}{J. Silk}, \bibinfo{title}{{Dark matter annihilations and
  decays after the AMS-02 positron measurements}}, \bibinfo{journal}{Phys. Rev.
  D} \bibinfo{volume}{89} (\bibinfo{year}{2014}) \bibinfo{pages}{063539}.

\bibtype{Article}%
\bibitem{Lin_2015}
\bibinfo{author}{S.-J. Lin}, \bibinfo{author}{Q. Yuan}, \bibinfo{author}{X.-J.
  Bi}, \bibinfo{title}{{Quantitative study of the AMS-02 electron/positron
  spectra: Implications for pulsars and dark matter properties}},
  \bibinfo{journal}{Phys. Rev. D} \bibinfo{volume}{91} (\bibinfo{year}{2015})
  \bibinfo{pages}{063508}.

\bibtype{Article}%
\bibitem{DiMauro_2016}
\bibinfo{author}{M.~Di~Mauro \emph{et al.}}, \bibinfo{title}{{Dark matter vs.
  astrophysics in the interpretation of AMS-02 electron and positron data}},
  \bibinfo{journal}{JCAP} \bibinfo{volume}{05} (\bibinfo{year}{2016})
  \bibinfo{pages}{031}.

\bibtype{Article}%
\bibitem{Chen_2016}
\bibinfo{author}{Y.-H. Chen}, \bibinfo{author}{K. Cheung},
  \bibinfo{author}{P.Y. Tseng}, \bibinfo{title}{{Dark matter with
  multiannihilation channels and the AMS-02 positron excess and antiproton
  data}}, \bibinfo{journal}{Phys. Rev. D} \bibinfo{volume}{93}
  (\bibinfo{year}{2016}) \bibinfo{pages}{015015}.

\bibtype{Article}%
\bibitem{Feng_2018}
\bibinfo{author}{J. Feng}, \bibinfo{author}{H.-H. Zhang}, \bibinfo{title}{{Dark
  matter search in space: Combined analysis of cosmic ray antiproton-to-proton
  flux ratio and positron flux measured by AMS-02}},
  \bibinfo{journal}{Astrophys. J.} \bibinfo{volume}{858} (\bibinfo{year}{2018})
  \bibinfo{pages}{116}.

\bibtype{Article}%
\bibitem{Bai_2018}
\bibinfo{author}{Y. Bai}, \bibinfo{author}{J. Berger}, \bibinfo{author}{S. Lu},
  \bibinfo{title}{{Supersymmetric resonant dark matter: A thermal model for the
  AMS-02 positron excess}}, \bibinfo{journal}{Phys. Rev. D}
  \bibinfo{volume}{97} (\bibinfo{year}{2018}) \bibinfo{pages}{115012}.

\bibtype{Article}%
\bibitem{Ghosh_2021}
\bibinfo{author}{S.~Ghosh \emph{et al.}}, \bibinfo{title}{{Leptophilic-portal
  dark matter in the light of AMS-02 positron excess}}, \bibinfo{journal}{Phys.
  Rev. D} \bibinfo{volume}{104} (\bibinfo{year}{2021}) \bibinfo{pages}{075016}.

\bibtype{Article}%
\bibitem{Navarro_1996}
\bibinfo{author}{J.F.~Navarro \emph{et al.}}, \bibinfo{title}{{A universal
  density profile from hierarchical clustering}}, \bibinfo{journal}{Astrophys.
  J.} \bibinfo{volume}{490} (\bibinfo{year}{1997}) \bibinfo{pages}{493--508}.

\bibtype{Article}%
\bibitem{Navarro_2010}
\bibinfo{author}{J.~Navarro \emph{et al.}}, \bibinfo{title}{{The diversity and
  similarity of simulated cold dark matter halos}}, \bibinfo{journal}{Mon. Not.
  R. Astron. Soc.} \bibinfo{volume}{402} (\bibinfo{year}{2010})
  \bibinfo{pages}{21--34}.

\bibtype{Article}%
\bibitem{Cirelli_2011}
\bibinfo{author}{M.~Cirelli \emph{et al.}}, \bibinfo{title}{{PPPC 4 DM ID: A
  poor particle physicist cookbook for dark matter indirect detection}},
  \bibinfo{journal}{JCAP} \bibinfo{volume}{03} (\bibinfo{year}{2011})
  \bibinfo{pages}{051}, \bibinfo{note}{[Erratum: JCAP 10, E01 (2012)]}.

\bibtype{Inproceedings}%
\bibitem{Pohl_2014}
\bibinfo{author}{M. Pohl}, \bibinfo{title}{{Particle detection technology for
  space-borne astroparticle experiments}}, in: \bibinfo{booktitle}{{Technology
  and Instrumentation in Particle Physics}}, \bibinfo{publisher}{Proceedings of
  Science} \bibinfo{year}{2014},
  \bibinfo{url}{\urlprefix\url{https://pos.sissa.it/213/013/pdf}}.

\bibtype{Inproceedings}%
\bibitem{Zhang_2014}
\bibinfo{author}{S.~N.~Zhang \emph{et al.}}, \bibinfo{title}{{The high energy
  cosmic-radiation detection (HERD) facility onboard China's space station}},
  in: \bibinfo{editor}{T.Takahashi}, \bibinfo{editor}{J.-W.A. den Herder},
  \bibinfo{editor}{M. Bautz} (Eds.), \bibinfo{booktitle}{Space Telescopes and
  Instrumentation 2014: Ultraviolet to Gamma Ray}, \bibinfo{comment}{vol.}
  \bibinfo{volume}{9144}, \bibinfo{organization}{International Society for
  Optics and Photonics}, \bibinfo{publisher}{SPIE} \bibinfo{year}{2014}, pp.
  \bibinfo{pages}{293 -- 301}.

\bibtype{Article}%
\bibitem{Kyratzis_2022}
\bibinfo{author}{D.~Kyratzis~(HERD Coll.)}, \bibinfo{title}{Overview of the
  {HERD} space mission}, \bibinfo{journal}{Physica Scripta}
  \bibinfo{volume}{97} (\bibinfo{year}{2022}) \bibinfo{pages}{054010}.

\bibtype{Article}%
\bibitem{Cagnoli_2024}
\bibinfo{author}{I. Cagnoli}, \bibinfo{author}{D. Kyratzis},
  \bibinfo{author}{D. Serini} (\bibinfo{collaboration}{HERD}),
  \bibinfo{title}{{HERD space mission: Probing the Galactic Cosmic Ray
  frontier}}, \bibinfo{journal}{Nucl. Instrum. Meth. A} \bibinfo{volume}{1068}
  (\bibinfo{year}{2024}) \bibinfo{pages}{169788}.

\bibtype{Article}%
\bibitem{Adriani_2019a}
\bibinfo{author}{O.~Adriani \emph{et al.}}, \bibinfo{title}{The {CALOCUBE}
  project for a space based cosmic ray experiment: design, construction, and
  first performance of a high granularity calorimeter prototype},
  \bibinfo{journal}{J. Inst.} \bibinfo{volume}{14} (\bibinfo{year}{2019})
  \bibinfo{pages}{P11004}.

\bibtype{Misc}%
\bibitem{Ambrosi_2025}
\bibinfo{author}{{G. Ambrosi \emph{et al.} (AMS-02 Coll.)}},
  \bibinfo{title}{The Silicon Tracker L0 Upgrade of the AMS-02 experiment on
  the ISS}, \bibinfo{howpublished}{Advances in Space AstroParticle Physics
  (ASAPP2025), to be published in the proceedings} \bibinfo{year}{2025}.

\bibtype{Article}%
\bibitem{Adriani_2022}
\bibinfo{author}{O.~Adriani \emph{et al.}}, \bibinfo{title}{Design of an
  Antimatter Large Acceptance Detector In Orbit (ALADInO)},
  \bibinfo{journal}{Instruments} \bibinfo{volume}{6} (\bibinfo{year}{2022})
  \bibinfo{pages}{19}.

\bibtype{Article}%
\bibitem{Schael_2019}
\bibinfo{author}{S.~Schael \emph{et al.}}, \bibinfo{title}{{AMS-100: The next
  generation magnetic spectrometer in space \textendash{} An international
  science platform for physics and astrophysics at Lagrange point 2}},
  \bibinfo{journal}{Nucl. Instrum. Meth. A} \bibinfo{volume}{944}
  (\bibinfo{year}{2019}) \bibinfo{pages}{162561}.

\bibtype{Article}%
\bibitem{Battiston_2021}
\bibinfo{author}{R.~Battiston \emph{et al.}}, \bibinfo{title}{{High precision
  particle astrophysics as a new window on the universe with an Antimatter
  Large Acceptance Detector In Orbit (ALADInO)}}, \bibinfo{journal}{Exper.
  Astron.} \bibinfo{volume}{51} (\bibinfo{year}{2021})
  \bibinfo{pages}{1299--1330}, \bibinfo{note}{[Erratum: Exper.Astron. 51,
  1331--1332 (2021)]}.

\bibtype{Article}%
\bibitem{Chung_2022}
\bibinfo{author}{C.~Chung \emph{et al.}}, \bibinfo{title}{{The development of
  SiPM-based fast time-of-flight detector for the AMS-100 experiment in
  space}}, \bibinfo{journal}{Instruments} \bibinfo{volume}{6}
  (\bibinfo{year}{2022}) \bibinfo{pages}{14}.

\end{thebibliography*}

\end{document}